\documentclass[pra,twocolumn,preprintnumbers,amsmath,amssymb]{revtex4}

\usepackage{graphicx}
\usepackage{dcolumn}
\usepackage{bm}
\usepackage{color}
\usepackage{epstopdf}
\usepackage{caption}

\def\bk{{\bm \kappa}}

\newcommand{\br}{ {\bm r}}

\def\EE{{\mathbb{E}}}
\def\bbeta{{\boldsymbol{\beta}}}
\def\bsigma{{\boldsymbol{\sigma}}}

\def\bPi{{\boldsymbol{\Pi}}}
\def\bGamma{\boldsymbol{\Gamma}}

\begin{document}



\title{Wave condensation with weak disorder vs beam self-cleaning in multimode fibers}

\author{J. Garnier$^{1}$, A. Fusaro$^{2}$, K. Baudin$^{2}$,  C. Michel,$^{3}$ K. Krupa$^{2}$, G. Millot$^{2}$, A. Picozzi$^{2}$}
\affiliation{$^{1}$ CMAP, CNRS, Ecole Polytechnique, 
Institut Polytechnique de Paris, 91128 Palaiseau Cedex, France}
\affiliation{$^{2}$ Laboratoire Interdisciplinaire Carnot de Bourgogne, CNRS, Universit\'e Bourgogne Franche-Comt\'e, Dijon, France}
\affiliation{$^{3}$Universit\'e C\^ote d'Azur, CNRS, Institut  de Physique de Nice, Nice, France}



\begin{abstract}
Classical nonlinear waves exhibit a phenomenon of condensation that results from the natural irreversible process of thermalization toward the Rayleigh-Jeans equilibrium spectrum.
Wave condensation originates in the divergence of the thermodynamic equilibrium Rayleigh-Jeans distribution, which is responsible for the macroscopic population of the fundamental mode of the system. 
Several recent experiments revealed a remarkable phenomenon of spatial organization of an optical beam that propagates through a graded-index multimode optical fiber (MMF), a phenomenon termed beam self-cleaning.
Our aim in this article is to provide some physical insight into the mechanism underlying optical beam self-cleaning through the analysis of wave condensation in the presence of structural disorder inherent to MMFs.
We consider  experiments of beam self-cleaning where long pulses are injected and populate many modes of a 10-20m MMF, for which the dominant contribution of disorder originates from polarization random fluctuations (weak disorder).
On the basis of the wave turbulence theory, we derive nonequilibrium kinetic equations describing the random waves in a regime where disorder dominates nonlinear effects.
The theory reveals that the presence of a conservative weak disorder introduces an effective dissipation in the system, which is shown to inhibit wave condensation in the usual continuous wave turbulence approach.
On the other hand, the experiments of beam-cleaning are described by a discrete wave turbulence approach, where the effective dissipation induced by disorder modifies the regularization of wave resonances, which leads to an acceleration of condensation that can explain the effect of beam self-cleaning.
By considering different models of weak disorder in MMFs, we show that a model where the modes experience a partially correlated noise is sufficient to accelerate the thermalization, whereas a fully mode-correlated noise does not lead to a dissipation-induced acceleration of condensation.
The simulations are in quantitative agreement with the theory, and evidence an effect of beam-cleaning even in a regime of moderate weak disorder.
At the leading order linear regime, random mode coupling among degenerate modes (strong disorder) can enforce thermalization and condensation.
The analysis also reveals that the effect of beam cleaning is characterized by a partial repolarization as a natural consequence of the condensation process.
In addition, the discrete wave turbulence approach explains why optical beam self-cleaning has not been observed in step-index multimode fibers.
\end{abstract}

\pacs{42.65.Sf, 05.45.a}

\maketitle

\section{Introduction}
\label{sec:intro}

Bose-Einstein condensation refers to a quantum process characterized by a thermodynamic transition into a single, macroscopically populated coherent state. 
This phenomenon has been observed in a variety of quantum systems, such as ultracold atoms and molecules \cite{stringari}, exciton polaritons \cite{carusotto13} and photons \cite{weitz} (also see \cite{fischer19}). 
On the other hand, several studies on wave turbulence predicted that completely classical waves can undergo a condensation process \cite{dyachenko92,Newell01,nazarenko11,Newell_Rumpf,berloff02,PRL05,onorato06,berloff07,PD09,salman09,proment12,
brachet11,PRA11b,Fleischer,suret,PR14,magnons15,nazarenko14,PRL18}.
The picture the reader may have in mind is the following. 
Considering an ensemble of  weakly nonlinear dispersive random waves, a redistribution of energy occurs among different modes, which is responsible for a self-organization process:  
While the (kinetic) energy is transferred to the small scales fluctuations (higher modes), an inverse cascade increases the `number of particles' into the lowest allowed mode, which leads to the emergence of a large scale coherent structure. 
This phenomenon refers to wave condensation.
It originates  in the natural thermalization of the wave system toward the thermodynamic Rayleigh-Jeans equilibrium distribution, whose divergence is responsible for the macroscopic occupation of the fundamental mode of the system \cite{Newell01,nazarenko11,
PRL05,PD09,salman09,PRA11b,zakharov92,shrira_nazarenko13,chiocchetta16,PR14,laurie12}.
We recall that this self-organization process takes place in a conservative (Hamiltonian) and formally reversible system:
The (`condensate') remains immersed in a sea of small-scale fluctuations (`uncondensed particles'), which store the information for time reversal.
In this respect, wave condensation is of different nature than other forms of condensation processes discussed in optical cavity systems, which are inherently nonequilibrium forced-dissipative systems \cite{PR14,conti08,fischer10,berloff13,fischer14,turitsyn12,turitsyn13,churkin15,fratalocchi16,sich18}.

The observation of wave thermalization with optical waves in a (cavity-less) free propagation is known to require very large propagation lengths, as discussed recently in different circumstances \cite{chiocchetta16,PRL18}.
The situation is different when the optical beam propagates in a waveguide configuration.
The finite number of modes supported by the waveguide significantly reduces the rate of thermalization, in particular by regularizing the ultra-violet catastrophe inherent to classical waves \cite{PRA11b}.
In this framework, a remarkable phenomenon of spatial beam self-organization, termed `beam self-cleaning', has been recently discovered in graded refractive index MMFs \cite{krupa16,wright16,krupa17}.
At variance with an apparently similar effect driven by the dissipative Raman effect in MMFs, known as Raman beam cleanup \cite{terry07}, 
this self-organization is due to a purely conservative Kerr nonlinearity \cite{krupa17}.

Recent works indicate that this phenomenon of beam self-cleaning can be interpreted as a consequence of wave thermalization and condensation \cite{PRL19,pod19,christodoulides19}.
In particular, in Ref.\cite{PRL19} a previously unrecognized mechanism of acceleration of condensation mediated by disorder has been reported, which can help to understand the effect of beam self-cleaning.
Indeed, light propagation in MMFs is known to be affected by a structural disorder of the material due to inherent imperfections and external perturbations \cite{kaminow13,ho14}, a feature which is attracting a growing interest, e.g., in image formation \cite{psaltis16,cao18}, or to study the dynamics of completely integrable Manakov equations \cite{menyuk,mecozzi12a,mecozzi12b,mumtaz13,xiao14,buch19}.



Our aim in this article is to provide some deeper physical insight into the mechanism underlying optical beam self-cleaning through the analysis of wave condensation in the presence of structural disorder of MMFs.
We pursue the work initiated in Ref.\cite{PRL19} along different lines:
(i) On the basis of the wave turbulence theory \cite{zakharov92,Newell01,nazarenko11,PR14} and related developments on finite size effects
\cite{Kartashova98,Zakharov05,Nazarenko06,kartashova08,Kartashova09,Kartashova10,Lvov10,Harris13,
Harper13,Bustamante14,kuksin,Mordant18}, 
we derive kinetic equations describing the nonequilibrium evolution of random waves in the regime where disorder effects dominate nonlinear effects.
Considering the dominant contribution of polarization random fluctuations (weak disorder) \cite{ho14}, the theory shows that a {\it conservative} disorder introduces an effective dissipation in the evolution of the moments equations. 
By following the conventional {\it continuous} wave turbulence approach describing very highly MMFs, the analysis reveals that the effective dissipation introduces a frequency broadening of four wave resonances, which inhibits the conservation of the (kinetic) energy and, consequently, the effect of condensation.
On the other hand, usual experiments of beam self-cleaning in MMFs are described by a {\it discrete} wave turbulence approach whereby dissipation induced by weak disorder modifies the regularization of wave resonances, which leads to a significant acceleration of the process of wave condensation.
(ii) By considering different models of weak disorder in MMFs, the theory shows that when all modes experience the same (correlated) noise, the effective dissipation induced by disorder vanishes and the system no longer exhibits a fast process of condensation. 
On the other hand, even a small decorrelation among the noise experienced by the modes is sufficient to re-establish a dissipation-induced acceleration of condensation mediated by disorder.
(iii) To improve our understanding of beam cleaning experiments, we have considered a regime where disorder effects are of the same order as nonlinear effects.
In this case the simulations reveal the existence of a mixed coherent-incoherent regime, which is still characterized by a relatively fast process of self-cleaning condensation that is consistent with the experiments.
(iv) By studying polarization effects \cite{krupa_polar},  we show that optical beam cleaning is responsible for an effective repolarization of the central part of the optical beam, a feature that can be explained by the macroscopic condensed population of the fundamental mode of the MMF.
(v) The discrete wave turbulence approach also explains why optical beam self-cleaning has not been observed in step-index MMFs.
This is due to a significant reduction of the number of resonances and their corresponding efficiencies, which leads to an effective freezing of the process of thermalization and condensation.
The same argument also explains why condensation is not observed in the recent experiments of beam cleaning under specific injection conditions \cite{deliancourt_lp11}, a feature that will be discussed in relation with the impact of a perturbation on the dispersion relations.


It is important to note that in this work we address the effect of optical beam self-cleaning in the quasi-continuous wave regime, where  long (sub-nanosecond) temporal pulses  are injected in the MMF.
Temporal dispersion effects such as modal group-velocity mismatch and chromatic dispersion can be neglected within the short MMF lengths considered in these experiments.
Accordingly, in this work we do not consider experiments of beam self-cleaning with high-power femto-second pulses \cite{liu16}, whose underlying mechanism should be related to complex spatio-temporal effects involving multimode solitons \cite{renninger13,wrightOE15,wright15_0,wright15,conforti17,agrawal19}.

\section{Theoretical model}

\subsection{Modal NLS equation}
\label{sec:modalNLSeq}

We consider the (2+1)D nonlinear Schr\"odinger (NLS) equation (or Gross-Pitaevskii equation), 
which is known to describe the transverse spatial evolution of an optical beam 
propagating along the $z-$axis of a waveguide modelled by a confining potential $V(\bm r)$ (with $\bm r=(x,y)$) \cite{horak12}.
Considering long temporal pulses (in the sub-nanosecond range), we neglect temporal effects related to first- and second-order chromatic dispersion.
The vector NLS equation accounting for the polarization degree of freedom can then be written
\begin{eqnarray}
i \partial_z {\bm \psi} =-\alpha \nabla^2 {\bm \psi} + V(\bm r){\bm \psi}  -\gamma_0{\cal P}({\bm \psi})  ,
\label{eq:nls}
\end{eqnarray}
where ${\bm \psi}({\bm r}, z)=(\psi_x, \psi_y)^T$ with $\psi_{x,y}({\bm r}, z)$ the vector field in the linear polarization basis, the superscript $^T$ denoting the transpose operation. The parameter $\alpha = 1/(2 k_0 n_0)$ is the diffraction coefficient where $k_0=2\pi/\lambda$ with $\lambda$ the vacuum laser wavelength and $n_0$ the  refractive index of the fiber core. 
The nonlinear term reads:
\begin{eqnarray}
{\cal P}({\bm \psi}) = \frac{1}{3}
{\bm \psi}^T {\bm \psi} {\bm \psi}^* +\frac{2}{3} {\bm \psi}^\dag  {\bm \psi} {\bm \psi},
\label{eq:nls_P}
\end{eqnarray}
where the nonlinear coefficient is $\gamma_0 = k_0 n_2$, 
with $n_2$ the nonlinear Kerr coefficient, and the superscripts $*,\dag$ stand for the conjugate, and conjugate transpose operations ($n_2>0$ for a focusing fiber nonlinearity in m$^2$/W, $|{\bm \psi}|^2$ in W/m$^2$).

We expand the random wave into the orthonormal basis ($\int u_p(\br) u_m^*(\br) d\br=\delta^K_{pm}$) of the eigenmodes $\{ u_p(\br) \}$ of the linearized NLS equation with the potential $V(\br)$ 
\begin{eqnarray}
{\bm \psi}(\br,z)=\sum_{p=0}^{N_*-1} {\bm B}_p(z) u_p(\br) \exp(-i\beta_p z),
\end{eqnarray}
where $N_*$ denotes the number of modes of the waveguide, ${\bm B}_p(z)=(B_{p,x},B_{p,y})^T$  refer to the linear polarization components of the $p$-th mode and $\beta_p$ are the corresponding eigenvalues verifying 
$\beta_p u_p(\br)=-\alpha \nabla^2 u_p(\br) + V(\bm r)u_p(\br)$.

In the following we focus the analysis on GRIN (graded-index) MMFs in which the effect of beam
self-cleaning has been observed experimentally, while the case of step-index MMFs will be discussed in Sec.~\ref{sec:step_index}.
For the ideal parabolic potential $V(\br)=q|\br|^2$ for $|\br| \le a$, $a$ being the fiber core radius, $u_p({\bm r})$ refer to the normalized Hermite-Gaussian functions with corresponding eigenvalues $\beta_p=\beta_{p_x,p_y}=\beta_0(p_x+p_y+1)$,
$u_{p_x,p_y}(x,y) = 
\kappa (\pi p_x! \, p_y! \, 2^{p_x+p_y})^{-1/2}
\, H_{p_x}(\kappa x) \, 
H_{p_y}(\kappa y) \, \exp[-\kappa^2 (x^2+y^2)/2]$,
where $\kappa= (q/\alpha)^{1/4}$, $q=k_0(n_0^2-n_1^2)/(2n_0 a^2)$, $\beta_0=2\sqrt{\alpha q}$, $n_1$ being the cladding refractive index. 
The number of modes (without including polarization degeneracy) is $N_* \simeq V_0^2/(2\beta_0^2)$, where $V_0$ denotes the depth of the parabolic potential.
Defining ${\bm A}_p(z)={\bm B}_p(z) \exp(-i\beta_p z)$, the evolutions of the modal components are governed by 
\begin{eqnarray}
i \partial_z {\bm A}_p = \beta_p {\bm A}_p 
- \gamma  {\bm P}_p({\bm A}).
\label{eq:modes_A_00}
\end{eqnarray}
The nonlinear term in (\ref{eq:modes_A_00}) reads
\begin{eqnarray}
{\bm P}_p ({\bm A})  = \sum_{l,m,n=0}^{N_*-1} S_{plmn} \Big(\frac{1}{3}
{\bm A}_l^T {\bm A}_m {\bm A}_n^* +\frac{2}{3} {\bm A}_n^\dag  {\bm A}_m {\bm A}_l\Big) ,
\label{def:nlterm0}
\end{eqnarray}
where 
$S_{lmnp} = \int u_l(\br) u_m(\br) u_n^*(\br) u_p^*(\br) d\br/\int |u_0|^4(\br) d\br$ so that $S_{0000}=1$, and $\gamma = \gamma_0/A_{eff}^0$, where $A_{eff}^0=1/\int |u_0|^4(\br) d\br$ is the effective area of the fundamental mode ($|{\bm A}_p|^2$ in Watts).

The modal NLS Eq.(\ref{eq:modes_A_00}) has a form analogous to the basic equation usually considered to study light propagation in MMFs in the absence of disorder \cite{horak12,poletti08,agrawal_book,mumtaz13,xiao14}.
It conserves the total power $N=\int |{\bm \psi} |^2 d\br=\sum_{p=0}^{N_*-1} |{\bm A}_p|^2$, and the Hamiltonian $H=E+U$, which has a linear  contribution $E=\alpha \int |\nabla {\bm \psi} |^2 d\br+\int V(\br) |{\bm \psi}|^2d\br=\sum_{p=0}^{N_*-1} \beta_p |{\bm A}_p|^2$, and a nonlinear contribution:
\begin{eqnarray}
U=\frac{\gamma}{4} \sum_{l,m,n,p=0}^{N_*-1}  S_{lmnp} {\bar U}_{lmnp}
+ c.c.
\label{eq:ham}
\end{eqnarray}
where 
${\bar U}_{lmnp}=\frac{1}{3}
({\bm A}_l^T {\bm A}_m)( {\bm A}_n^\dag  {\bm A}_p^*) +\frac{2}{3} ({\bm A}_n^\dag  {\bm A}_m)( {\bm A}_l^T {\bm A}_p^*) $
and $c.c.$ denotes the complex conjugate.

The characteristic length $L_0=1/(\gamma N)=1/(\gamma_0 N/A_{eff}^0)$ corresponds to the nonlinear propagation length when all the power is confined in the fundamental mode. 
Usually the optical beam populates many higher-order modes and the nonlinear length is $L_{nl}=1/(\gamma_0 P/S_{eff})$ where $S_{eff}$ is the effective surface section of the beam.
We also define the linear characteristic length from Eq.(\ref{eq:modes_A_00}) as $L_{lin}=1/\delta \beta_{eff}$, where $\delta \beta_{eff}$ denotes the effective (`spectral') bandwidth of the optical beam in the mode space.
In usual experiments of beam self-cleaning in GRIN fibers we have $\beta_0 \sim 10^3$m$^{-1}$, so that the optical wave evolves in the weakly nonlinear regime
\begin{eqnarray}
L_{lin} \ll L_{nl}, 
\end{eqnarray}
where $L_{lin} \lesssim 1$mm while $L_{nl}$ is typically  larger than 10cm. 

\subsection{Coherent modal regime: Impact of a confining parabolic potential on wave resonances}
\label{sec:coherent_regime}

\begin{figure}[]
\begin{center}
\includegraphics[width=1\columnwidth]{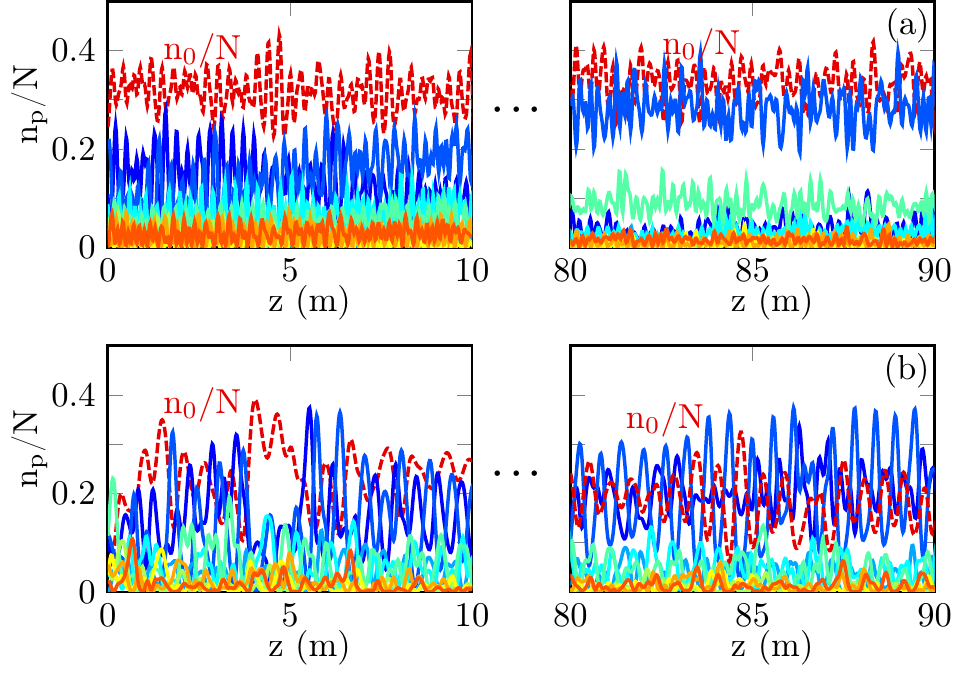}
\caption{
Coherent regime of mode interaction:
Numerical simulations of the NLS Eq.(\ref{eq:modes_A_00})  showing the evolutions of the modal components $n_p(z)=|{\bm A}_p(z)|^2$ (in the absence of disorder), starting from a coherent initial condition (a), starting from a `speckle' beam with random phase among the modes (b). 
Evolutions of the modal populations $n_p/N$: fundamental mode $p=0$ (red dashed), $p=1$ (dark blue solid), $p=2$ (blue solid), $p=3$ (light blue solid), $p=4$ (cyan solid), $p=5$ (light green solid), $p=6$ (green solid), $p=7$ (yellow solid), $p=8$ (orange solid).
The system exhibits a persistent oscillatory dynamics among the low-order modes.
At complete thermal equilibrium the system would reach the condensate fraction $n_0^{eq}/N \simeq 0.68$ in (a), and $n_0^{eq}/N \simeq 0.58$ in (b).
The power is $N=47.5$kW, $N_*=120$ modes, $a=26\mu$m.
}
\label{fig:0}
\end{center}
\end{figure}

In a typical experiment of optical beam self-cleaning a laser beam featured by a coherent transverse phase front is launched into the MMF.
Under these conditions, the modal components excited at the fiber input exhibit a strong phase-correlation among each other.
We remark that this is in contrast to usual simulations in wave turbulence where one imposes a random phase among the modes in the initial condition \cite{zakharov92,nazarenko11,shrira_nazarenko13,PR14,nazarenko14}.
The numerical simulations of the NLS Eq.(\ref{eq:modes_A_00}) reveal a remarkable fact:
The strong phase-correlation among the low-order modes is preserved during the propagation, thus leading to a phase-sensitive mode interaction, as illustrated in Fig.~\ref{fig:0}(a).
In this coherent regime, the phase relation-ship among the modes plays an important role in the dynamics.
The low-order modal components ${\bm A}_p(z)$ that are strongly populated then experience a quasi-reversible exchange of power with each other. 
The corresponding intensity pattern $|{\bm \psi}|^2(\br,z)$ exhibits itself an oscillatory dynamics during the propagation: The  beam populates several modes of the potential $V(\br)$ and does not  exhibit an enhanced brightness characterizing a stable self-cleaning effect.
As already discussed in Ref.\cite{PRL19}, this coherent regime of mode interactions freezes the thermalization process.
In addition, we report in Fig.~\ref{fig:0}(b) a simulation of the NLS Eq.(\ref{eq:modes_A_00}) starting from a speckle beam.
Although the initial random phases modify the regular oscillatory behavior shown in Fig.~\ref{fig:0}(a), the low-order modal amplitudes still exhibit a rapid and significant power exchange among each other. 
This indicates that the phase relation-ship among the modes still plays a non-trivial role thus leading to an effective freezing of the process of thermalization.
As will be discussed throughout this paper, these simulations reflect the discrete nature of the resonance manifold underlying wave turbulence in MMFs. 
We note that understanding the mechanisms that can freeze the process of thermalization is an important problem that is currently analyzed in various systems, such as, e.g., finite size effects in discrete or mesoscopic wave turbulence \cite{Kartashova98,Zakharov05,Nazarenko06,kartashova08,Kartashova09,Kartashova10,Lvov10,Harris13,
Harper13,Bustamante14,kuksin,Mordant18}, in Fermi-Pasta-Ulam chains \cite{gallavotti,benettin13,onorato15,onorato18,PRX17}, or in nonlinear disordered systems \cite{ermann13,ermann15}.

The phase-sensitive regime of interaction among the highly populated modes of the MMF originates in the small number of modes confined in the parabolic potential (typically $\sim 15$ groups of degenerate modes, i.e., $\sim 120$ modes), which significantly reduces the effective number of modes that can interact efficiently.
To illustrate this aspect, one can compare the number of resonances in a GRIN MMF used in the experiments of beam cleaning and the corresponding number of resonances in a 2D box.
Let consider such two systems with approximately the same number of modes, say $N_*=120$ for the GRIN fiber and $N_*=121$ for the 2D box.
In a 2D box with periodic boundary conditions the resonances verify energy conservation ($\omega_{{\bm k}_1}+\omega_{{\bm k}_2}=\omega_{{\bm k}_3}+\omega_{{\bm k}_4}$, $\omega_{{\bm k}}=\alpha |{\bm k}|^2$), and momentum conservation (${\bm k}_1+{\bm k}_2={\bm k}_3+{\bm k}_4$).
In a GRIN fiber, the resonances verify energy conservation $\beta_p+\beta_l=\beta_m+\beta_n$, while the tensor $S_{lmnp}$ reflects the efficiency of the corresponding resonance.
Because the eigenvalues $\beta_p$ are regularly spaced for a parabolic potential, there are much more resonances in a GRIN fiber than in a 2D box.
However, most of the resonances are not efficient: only 41 (1067) resonances are characterized by a coefficient $S_{lmnp} > 0.4$ ($S_{lmnp} > 0.2$), where we recall that $0<S_{lmnp} \le 1$.
In contrast,  the 2D box exhibits $\sim 10^5$ exact resonances for energy and momentum conservation (the computation being independent on the box size).
The analysis indicates that, as compared to the 2D box, the GRIN fiber effectively behaves as a dynamical system with a relative small number of degrees of freedom, 
which provides an intuitive interpretation for the persistent coherent regime of mode interaction discussed above. 
In the following we will see that the introduction of a structural disorder of the MMF breaks the coherent phase-sensitive regime of modal interaction.

\section{Weak disorder: Kinetic equations and simulations}
\label{sec:weak_disorder}

The modal NLS Eq.(\ref{eq:modes_A_00}) assumes that the MMF is ideal, in the sense that the model does not account for any form of disorder. 
However, in practice light propagation in MMFs is known to be affected by random fluctuations of the longitudinal and transverse profiles of the index of refraction as a consequence of external factors such as bending, twisting, tensions, kinks, or core-size variations in the process of fabrication of the fiber.
Such multiple physical perturbations introduce random polarization fluctuations as well as a random coupling among the modes of the fiber.
The specific mechanisms and models that describe how fiber imperfections impact light propagation in MMFs still remains an active topic of research \cite{kaminow13,mecozzi12a,mecozzi12b,mumtaz13,xiao14,buch19,ho14}.
Usually, one considers that for relative short propagation lengths ($\sim$10m) as those considered in optical beam self-cleaning experiments, the dominant contribution of noise arises from polarization modal fluctuations (weak disorder), while for larger propagation lengths mode coupling occurs among degenerate modes (mode group coupling state of strong disorder), and for even larger fiber lengths mode coupling takes place for non-degenerate modes (strong disorder) \cite{ho14}.
In the following we consider the dominant contribution of weak disorder \cite{ho14}, while strong disorder will be considered in Sec.~\ref{sec:strong_disorder}.
We will see that the source of noise introduced by weak disorder is sufficient to break the coherent regime of modal interaction discussed here above in Sec.~\ref{sec:coherent_regime}. 
The resulting random phase dynamics then leads to a turbulent {\it incoherent regime} of the modal components that we describe in the framework of the wave turbulence theory.

In the following we discuss in detail the impact of a weak disorder on the derivation of the wave turbulence kinetic equation for MMFs.
We first discuss the conventional continuous wave turbulence regime relevant for very highly multimode fibers, and then the discrete wave turbulence regime relevant to the experiments of beam self-cleaning.
We provide a detailed derivation of the discrete wave turbulence kinetic equation accounting for weak disorder -- a sketch of the derivation was reported in Ref.\cite{PRL19}. 
We first consider a model of weak disorder in which the modes experience an independent (decorrelated) noise. 
We note that, in the opposite limit where {\it the modes experience the same disorder}, the analysis reveals that the optical wave recovers a quasi-coherent regime of mode interaction.
In other words, this fully mode-correlated model of disorder appears ineffective and in this limit the optical wave does not exhibit a fast process of condensation.
Accordingly, we consider an intermediate model of partially correlated disorder where degenerate modes experience the same noise.
The theory shows that this partially mode-correlated disorder is sufficient to re-establish an efficient acceleration of condensation.

\subsection{Decorrelated model of disorder}

\subsubsection{Model}
\label{sec:model}

As discussed above, in the weak disorder regime different spatial modes do not couple, but the two polarization states of each spatial mode exhibit a random coupling \cite{mumtaz13,xiao14}. 
The evolutions of the modal components are governed by 
\begin{eqnarray}
i \partial_z {\bm A}_p = \beta_p {\bm A}_p + {\bf D}_p(z) {\bm A}_p 
- \gamma  {\bm P}_p({\bm A}) .
\label{eq:nls_Ap}
\end{eqnarray}
Let us recall that longitudinal and transverse fluctuations in the refractive index profile of the MMF lead to a random coupling among the modes, as expressed by the coupled-mode theory \cite{kaminow13,mumtaz13}. 
However, by following this procedure one obtains a scalar perturbation that does not introduce coupling among the polarization components, a feature commented in particular in Ref.\cite{xiao14}. 
Accordingly, random polarization fluctuations are usually introduced in a phenomenological way, see e.g.  \cite{menyuk,mumtaz13,xiao14}. 
Here, we consider the most general form of polarization disorder that conserves the power of the optical beam.
The Hermitian matrices ${\bf D}_p(z)$ are expanded into the Pauli matrices, which are known to form a basis for the vector space of 2$\times$2 Hermitian matrices.
The matrices then have the form
\begin{eqnarray}
\label{model:Dp0a}
{\bf D}_p(z) = \sum_{j=0}^3 \nu_{p,j} (z) \bsigma_j  ,
\end{eqnarray}
where $\bsigma_j$ ($j=1,2,3$) are the Pauli matrices and $\bsigma_0$ is the identity matrix.
The functions $\nu_{p,j}(z)$ are independent and identically distributed Gaussian real-valued  random processes, with 
\begin{eqnarray}
\left< \nu_{p,j}(z) \nu_{p',j'}(z') \right>
=  \sigma^2_\beta \delta^K_{pp'} \delta^K_{jj'} {\cal R}\Big( \frac{z-z'}{l_\beta}\Big) .
\label{eq:corr_nu}
\end{eqnarray}
Here $l_\beta$ is the correlation length of the random process and $\sigma^2_\beta$ is its variance.
The normalized correlation function is such that ${\cal R}(0)=O(1)$,  $\int_{-\infty}^\infty {\cal R}(\zeta)d\zeta=O(1)$, and ${\cal R}( z/l_\beta) \to l_\beta \delta(z)$ in the limit $l_\beta \to 0$.
We will consider Ornstein-Uhlenbeck processes:
$$
d \nu_{p,j} = - \frac{1}{l_\beta} \nu_{p,j} dz + \frac{\sigma_\beta}{\sqrt{l_\beta}} dW_{p,j}(z)
$$
where $W_{p,j}$ are independent Brownian motions. This means that $\nu_{p,j}$ are  Gaussian processes with mean zero and covariance 
function of the form (\ref{eq:corr_nu}) with ${\cal R}(\zeta) = \exp(-|\zeta|) /2$. 
We introduce an effective parameter of disorder 
$\Delta \beta=\sigma^2_\beta l_\beta$ and the associated length scale 
\begin{eqnarray}
L_d=1/\Delta \beta.
\label{eq:L_d}
\end{eqnarray}
We assume that disorder effects dominate nonlinear effects 
\begin{eqnarray}
L_d \ll L_{nl},
\end{eqnarray}
and that $l_\beta \ll L_d$ (or $\sigma_\beta l_\beta \ll 1$).
We recall that usual experiments of beam self-cleaning refer to a weakly nonlinear regime $L_{lin} \ll L_{nl}$, where we have typically $\beta_0 \sim 10^3$m$^{-1}$, so that $\beta_0^{-1} \ll L_d$ (or $\Delta \beta \ll \beta_0$).
Aside from beam-cleaning experiments, in the following we will also consider a regime where the mode-spacing can be very small $\beta_0 \ll \Delta \beta$ (MMFs with huge number of modes $N_* \simeq V_0^2/(2\beta_0^2) \gg 1$) so as to address a regime described by a continuous wave turbulence approach. 
Finally note that since the disorder is (`time') $z-$dependent, our system is of different nature than those studying the interplay of thermalization and Anderson localization \cite{cherroret15}.

\subsubsection{Vanishing correlations among modes}
\label{sec:wd_vanish_corr}

First of all, we study the correlations among the modes through the analysis of the evolution of the second-order moments in the 2$\times$2 matrix $\left< {\bm A}_p^* {\bm A}_q^T\right>(z)$. 
The impact of disorder is treated by making use of the Furutsu-Novikov theorem.
This reveals that the conservative disorder introduces an effective dissipation in the system.
In this respect, we recall that in principle the laser beam excites strongly correlated modes at the fiber input, as discussed above through the coherent modal regime of interaction.
It is the effective dissipation due to disorder that breaks such a strong modal phase-correlation.

The analysis developed in Appendix~A (section~\ref{appA1}) completes that reported in the Supplemental of Ref.\cite{PRL19}, which was focused on correlations between different modes, $\left< {\bm A}_p^* {\bm A}_q^T\right>$ for $p \neq q$.
The correlations within a single mode (i.e., among the orthogonal polarization components for $p=q$) is more delicate and it is reported in detail in Appendix~A (section~\ref{appA1}).
The theory reveals that in the regime discussed in Sec.~\ref{sec:model}, the correlations among different modes ($p \neq q$), or within a single mode ($p=q$), both have a form 
$\sim  \gamma {\bm G}_{pq}({\bm A}(z)) /\big(4 \Delta \beta -i(\beta_p-\beta_q)\big)$,  
where ${\bm G}_{pq}({\bm A}(z))={\bm P}_p({\bm A}(z))^* {\bm A}_q^T(z)
- {\bm A}_p^*(z) {\bm P}_q({\bm A}(z))^T $.
Recalling that $L_d \ll L_{nl}$ we see that, at leading order, modal correlations vanish for propagation lengths larger than the nonlinear length, $\left< {\bm A}_p^* {\bm A}_q^T\right>(z) \simeq 0$ for $z \gg L_{nl}$.

\subsubsection{Closure of the moments equations}

In the following we derive the kinetic equation by following the wave turbulence perturbation expansion procedure in which linear dispersive effects dominate nonlinear effects $L_{lin} \ll L_{nl}$. 
Accordingly, an effective large separation of the linear and the nonlinear lengths scales takes place \cite{Newell01,nazarenko11}.
When combined with disorder effects, the modes exhibit random phases with quasi-Gaussian statistics, which allows one to achieve the closure of the infinite hierarchy of moment equations.
More precisely, because of the nonlinear character of the NLS equation, the evolution of the second-order moment of the field depends on the fourth-order moment, while the equation for the fourth-order moment depends on the sixth-order moment, and so on. In this way, one obtains an infinite hierarchy of moment equations, in which the $n-$th order moment depends on the $(n+2)-$th order moment of the field. This makes the equations impossible to solve unless some way can be found to truncate the hierarchy. The closure of the infinite hierarchy of moment equations can be realized in the weakly nonlinear regime by virtue of the Gaussian moment theorem. We remark in this respect that the key assumption underlying the wave turbulence approach is the existence of a random phase among the modes rather than a genuine Gaussian statistics, as recently discussed in detail in Refs.\cite{nazarenko11,chibbaro17,chibbaro18}.
We emphasize that in the present work the random phase of the modes is induced by the structural disorder of the medium that dominates nonlinear effects ($L_d \ll L_{nl}$).

As discussed here above, the nondiagonal components of the 2$\times$2 matrix $\left< {\bm A}_p^* {\bm A}_p^T\right>(z)$ vanish. 
Then our aim  is to derive an equation governing the evolutions of the diagonal components $w_p(z)=\frac{1}{2}\left< |{\bm A}_p(z)|^2\right>$.
Starting from the modal NLS Eq.(\ref{eq:nls_Ap}), we have 
\begin{eqnarray}
\partial_z w_p &=&  \frac{1}{3}  \gamma \left< X_p^{(1)}\right> 
+ \frac{2}{3}  \gamma \left<  {X}_p^{(2)}\right> ,
\label{eq:pzwp1}\\
X_p^{(1)} &=&{\rm Im}\Big\{ \sum_{l,m,n} S_{lmnp}^* ({\bm A}_l^\dag {\bm A}_m^*) ({\bm A}_n^T {\bm A}_p) \Big\}  ,
\label{def:Xp1}\\
{X}_p^{(2)} &=&{\rm Im}\Big\{ \sum_{l,m,n} S_{lmnp}^* ({\bm A}_n^T {\bm A}_m^*) ({\bm A}_l^\dag {\bm A}_p) \Big\}  .
\label{def:Xp2}
\end{eqnarray}
The detailed derivation of the equations for the fourth order correlators $J_{lmnp}^{(1)}(z)=\left<({\bm A}_l^\dag {\bm A}_m^*) ({\bm A}_n^T {\bm A}_p)\right>$ and ${J}_{lmnp}^{(2)}(z)=\left<({\bm A}_n^T {\bm A}_m^*) ({\bm A}_l^\dag {\bm A}_p)\right>$ is given in the Appendix~A (section~\ref{appA2}).
As already noticed, as a result of the Furutsu-Novikov theorem, the conservative disorder introduces an effective dissipation in the system, so that the evolutions of the fourth-order moments have the form of a forced-damped oscillator equation:
\begin{eqnarray}
\partial_z J_{lmnp}^{(j)} &=&
(-8\Delta \beta +i  \Delta \omega_{lmnp} ) J_{lmnp}^{(j)}
+ i \gamma \big<{Y}_{lmnp}^{(j)}\big> 
\label{eq:4th_order_moment}
\end{eqnarray}
where $\Delta \omega_{lmnp}  = \beta_l+\beta_m-\beta_n-\beta_p$ is the frequency resonance, while $\big<{Y}_{lmnp}^{(j)}\big>$ denote the six-th order moments that have been split into products of second-order moments by virtue of the factorizability property of statistical Gaussian fields (see the detailed expressions of $\big<{Y}_{lmnp}^{(j)}\big>$ for $j=1,2$ in Eq.(\ref{eq:Y_lmnp_1}) and Eq.(\ref{eq:Y_lmnp_2}) in the Appendix~A, section~\ref{appA2}).
The solution to Eq.(\ref{eq:4th_order_moment}) reads
\begin{eqnarray}
J_{lmnp}^{(j)}(z) &=& 
J_{lmnp}^{(j)}(0) G_{lmnp}(z) + {\cal I}_{lmnp}^{(j)}(z)
\label{eq:mom4pq}
\end{eqnarray}
where the convolution integral reads
\begin{eqnarray}
{\cal I}_{lmnp}^{(j)}(z) = i \gamma \int_0^z\left< {Y}_{lmnp}^{(j)}\right> (z-z') G_{lmnp}(z') dz'  
\label{eq:conv_int}
\end{eqnarray}
with the Green function 
\begin{eqnarray}
G_{lmnp}(z)=H(z) \exp(i \Delta \omega_{lmnp} z -8\Delta \beta z),
\label{eq:green}
\end{eqnarray}
and $H(z)$ the Heaviside function.  
We now arrive at the key point of our analysis, in which one distinguishes the continuous from the discrete wave turbulence approaches.

\subsubsection{Continuous wave turbulence}
\label{sec:cont_wt}

In general, an exchange of power among four different modes does not need to satisfy the exact resonance condition $\Delta \omega_{lmnp}=0$. 
Indeed, it is sufficient for a mode quartet to verify the quasi-resonant condition 
\begin{eqnarray}
|\Delta \omega_{lmnp}| \lesssim 1/L_{kin}^{disor}
\label{eq:cont_turb}
\end{eqnarray}
to provide a non-vanishing contribution to the convolution integral (\ref{eq:conv_int}), where $L_{kin}^{disor}$ denotes the characteristic evolution length scale of the moments of the field $\left< {Y}_{lmnp}^{(j)}\right>(z)$ (see Eq.(\ref{eq:L_kin_disor}) below).
In the usual {\it continuous wave turbulence approach}, there are a large number of non-resonant mode quartets $\Delta \omega_{lmnp} \neq 0$ that verify the quasi-resonant condition (\ref{eq:cont_turb}) and that contribute significantly to the evolution of the moments equations.

We anticipate that the continuous wave turbulence regime does not correspond to the usual experiments of optical beam self-cleaning \cite{wright16,krupa17}, since we have typically $\beta_0 \sim 10^3$m$^{-1}$, so that $\beta_0 \gg 1/L_{kin}^{disor}$.
In other words, since any non-resonant mode quartet verifies $|\Delta \omega_{lmnp}| \geq \beta_0 \gg 1 /L_{kin}^{disor}$, the system does not exhibit quasi-resonances, but solely exact resonances $\Delta \omega_{lmnp}=0$.
Actually, the experiments of beam self-cleaning are described by the discrete wave turbulence approach that will be discussed in detail in the next Sec.~\ref{sec:disc_wt}.
Here, for the sake of clarity, we first discuss the conventional continuous wave turbulence regime \cite{nazarenko11}, which can be relevant for MMFs characterized by a huge number of modes ($N_* \simeq V_0^2/(2\beta_0^2) \gg 1$ such that $\beta_0 \ll \Delta \beta$).
The discussion of the usual continuous regime is also important in that it enlightens the effect of acceleration of thermalization due to the presence of structural disorder in the system.

Recalling that $L_d=1/\Delta \beta \ll L_{nl}$, then the Green function decays on a length scale much smaller than the evolution length of $\left<Y_{lmnp}^{(j)}\right>(z)$, so that the convolution integral can be approximated for $z \gg L_d$
\begin{eqnarray}
{J}_{lmnp}^{(j)} \simeq \gamma \left< {Y}_{lmnp}^{(j)}\right> 
\frac{i8\Delta \beta-\Delta \omega_{lmnp}}{\Delta \omega_{lmnp}^2+(8\Delta \beta)^2}.
\label{eq:I_lmnp_damp}
\end{eqnarray}
By substitution of Eq.(\ref{eq:I_lmnp_damp}) in the fourth-order moments (\ref{eq:mom4pq}), one obtains the expressions of the averaged moments $\left< X_p^{(j)}\right>$ given in (\ref{def:Xp1}-\ref{def:Xp2}).
Collecting all terms in (\ref{eq:pzwp1}) gives the equation for the modal amplitudes $n_p(z)=2w_p(z)$.
Since the MMF supports a  large number of modes $N_* \simeq V_0^2/(2\beta_0^2) \gg 1$, we consider the continuous limit where the discrete sums in (\ref{def:Xp1}-\ref{def:Xp2}) are replaced by continuous integrals ($\beta_0 /V_0 \ll 1$). We obtain the continuous kinetic equation 
\begin{eqnarray}
\nonumber 
\partial_z \tilde{n}_{\bk_4}(z) =
 \frac{4 \gamma^2}{3\beta_0^6} \iiint d\bk_{1,2,3}  \frac{\overline{\Delta \beta}}{\Delta {\tilde \omega}_{\bk_1 \bk_2 \bk_3 \bk_4}^2+\overline{\Delta \beta}^2} \\
\nonumber
\times |\tilde{S}_{\bk_1 \bk_2 \bk_3 \bk_4}|^2 
\tilde{M}_{\bk_1 \bk_2 \bk_3 \bk_4}
\\
 +\frac{32 \gamma^2}{9\beta_0^2} \int d\bk_1  \frac{\overline{\Delta \beta}}{\Delta \tilde{\omega}_{\bk_1 \bk_4}^2+\overline{\Delta \beta}^2} |\tilde{s}_{\bk_1 \bk_4}(\tilde{\bm n})|^2 ( \tilde{n}_{\bk_1}-\tilde{n}_{\bk_4})  
\label{eq:kin_contin_dis}
\end{eqnarray}
where $d\bk_{1,2,3}=d\bk_1 d\bk_2 d\bk_3$, $\overline{\Delta \beta}=8\Delta \beta$ and
\begin{equation}
{\tilde s}_{\bk_1 \bk_4} (\tilde{\bm n})= \frac{1}{\beta_0^2}\int d\bk' \, {\tilde S}_{\bk_1 \bk' \bk' \bk_4} \, {\tilde n}_{\bk'}.
\label{eq:kin_np_cont}
\end{equation}
The functions with a tilde refer to the natural continuum extension of the corresponding discrete functions, i.e., $\tilde{n}_{\bm k}(z)=n_{[\bm k/\beta_0]}(z)$, 
$\tilde{\beta}_\bk = \beta_{[\bk/\beta_0]}$, $\tilde{S}_{\bk_1 \bk_2\bk_3 \bk_4}= S_{[\bk_1/\beta_0][\bk_2/\beta_0][\bk_3/\beta_0][\bk_4/\beta_0]}$ and so on, where $[x]$ denotes the integer part of $x$.
With these notations we have  
$\tilde{M}_{\bk_1 \bk_2 \bk_3 \bk_4}=\tilde{n}_{\bk_1} \tilde{n}_{\bk_2}\tilde{n}_{\bk_3}\tilde{n}_{\bk_4}
\big( \tilde{n}_{\bk_4}^{-1}+ \tilde{n}_{\bk_3}^{-1}- \tilde{n}_{\bk_1}^{-1}-\tilde{n}_{\bk_2}^{-1} \big)$,
$\Delta {\tilde \omega}_{\bk_1 \bk_2 \bk_3 \bk_4}=\tilde{\beta}_{\bk_1}+\tilde{\beta}_{\bk_2}-\tilde{\beta}_{\bk_3}-\tilde{\beta}_{\bk_4}$, $\Delta \tilde{\omega}_{\bk_1 \bk_2}=\tilde{\beta}_{\bk_1}-\tilde{\beta}_{\bk_2}$, and 
$\tilde{\beta}_\bk= \kappa_x + \kappa_y + \beta_0$, with ${\bm \kappa}=\beta_0 (p_x, p_y)$ \cite{PRA11b}.

The main novelty of the kinetic Eq.(\ref{eq:kin_np_cont}) with respect to the previous work without any structural disorder \cite{PRA11b}, is that the mechanism of disorder-induced dissipation introduces a finite bandwidth into the four-wave resonances among the modes.
Accordingly, instead of the Dirac $\delta$ distribution that guarantees energy conservation at each four-wave interaction, here the kinetic equation involves a Lorentzian distribution.
We remark that this aspect was already discussed in different circumstances \cite{dyachenko92}, in particular in the recent work \cite{churkin15} dealing with random fiber lasers in the presence of gain and losses.
Here, the originality is that the finite bandwidth of the Lorentzian distribution and the associated effective dissipation $\Delta \beta$ originate in the {\it conservative} structural disorder of the material.

Taking the formal limit $\Delta \beta \to 0$:
$\overline{\Delta \beta}/\big(\Delta {\tilde \omega}_{\bk_1 \bk_2 \bk_3 \bk_4}^2+\overline{\Delta \beta}^2 \big)
\rightarrow
\pi \delta(\Delta {\tilde \omega}_{\bk_1 \bk_2 \bk_3 \bk_4}),
$
we recover a continuous kinetic equation with a form similar to that derived in \cite{PRA11b} in the absence of disorder ($\Delta \beta = 0$) and in the absence of polarization effects (scalar limit ${\bm A}_p \to A_{p,x}$).
However the above limit $\Delta \beta \to 0$ is not physically relevant here since we have assumed $L_d=1/\Delta \beta \ll L_{nl}$ to derive Eq.(\ref{eq:4th_order_moment}).
Here, we consider the regime $\beta_0 \ll \Delta \beta$, so that it is the opposite limit that is physically relevant
$\overline{\Delta \beta}/\big(\Delta {\tilde \omega}_{\bk_1 \bk_2 \bk_3 \bk_4}^2+\overline{\Delta \beta}^2 \big)
\rightarrow 1/\overline{\Delta \beta}
$, and the kinetic equation takes the reduced form:
\begin{eqnarray}
\nonumber 
\partial_z \tilde{n}_{\bk_4}(z) =
 \frac{\gamma^2}{6 \Delta \beta \beta_0^6} \iiint d\bk_{1,2,3}   
 |\tilde{S}_{\bk_1 \bk_2 \bk_3 \bk_4}|^2 
\tilde{M}_{\bk_1 \bk_2 \bk_3 \bk_4}
\\
 +\frac{4 \gamma^2}{9 \Delta \beta \beta_0^2 } \int d\bk_1   |\tilde{s}_{\bk_1 \bk_4}(\tilde{\bm n})|^2 ( \tilde{n}_{\bk_1}-\tilde{n}_{\bk_4})  
\label{eq:kin_contin_dis2}
\end{eqnarray}
The second term in the right-hand side of (\ref{eq:kin_contin_dis2}) enforces the isotropization of the mode occupancies $\tilde{n}_{\bk}(z)$ among the degenerate modes, while the first term enforces the mode occupancies to reach the most disordered equilibrium distribution. 
The kinetic equation (\ref{eq:kin_contin_dis2}) conserves the power, $N=\beta_0^{-2}\int \tilde{n}_{\bk} d\bk$, and exhibits a $H-$theorem of entropy growth, $\partial_z {\cal S}(z) \ge 0$, where the nonequilibrium entropy is defined by ${\cal S}(z) = \beta_0^{-2} \int \log(\tilde{n}_{\bk}) d\bk$. 
However, at variance with the conventional wave turbulence kinetic equation, the kinetic Eq.(\ref{eq:kin_contin_dis2}) does not conserve the energy, $E=\beta_0^{-2} \int {\tilde \beta}_{\bk} \tilde{n}_{\bk} d\bk$.
Then the equilibrium distribution maximizing the entropy given the constraint on the conservation of the power $N$ is given by the uniform distribution 
\begin{eqnarray}
\tilde{n}^{eq}_{\bk}= {\rm const}.
\label{eq:rj_cont_disor}
\end{eqnarray}
This equilibrium state denotes an equipartition of (`particles') power among all modes.
Recalling that $\beta_0 \ll \Delta \beta$, the Lorentzian distribution in the kinetic Eq.(\ref{eq:kin_contin_dis}) is dominated by disorder, so that the length scale characterizing the rate of thermalization toward the equilibrium distribution (\ref{eq:rj_cont_disor}) is given by
\begin{eqnarray}
L_{kin}^{disor} \sim \Delta \beta L_{nl}^2/{\bar S_{lmnp}^2},
\label{eq:L_kin_disor}
\end{eqnarray}
where ${\bar S_{lmnp}^2}$ denotes the average square of the tensor $S_{lmnp}$ involving non-trivial resonances among nondegenerate modes.

The main conclusion is that the kinetic Eq.(\ref{eq:kin_np_cont}) does not describe a process of condensation characterized by a macroscopic occupation of the fundamental mode.
This indicates that weak disorder should prevent an effect of beam self-cleaning in MMFs featured by huge number of modes. 


\subsubsection{Discrete wave turbulence}
\label{sec:disc_wt}

We have seen that in the continuous wave turbulence regime quasi-resonances verifying $\Delta \omega_{lmnp} \lesssim 1/L_{kin}^{disor}$ contribute to the convolution integral (\ref{eq:conv_int}).
At variance with the continuous regime, in the discrete case the non-vanishing minimum value of $\Delta \omega_{lmnp}$ is such that
\begin{eqnarray}
{\rm min}(|\Delta \omega_{lmnp}|) = \beta_0 \gg 1/L_{kin}^{disor}.
\label{eq:disc_turb}
\end{eqnarray}
Accordingly, only exact resonances $\Delta \omega_{lmnp} =0$ contribute to the integral (\ref{eq:conv_int}), while non-resonant mode quartets lead to a vanishing integral \cite{nazarenko11}.
Note that this procedure can also be justified by a homogenization procedure, as reported in \cite{kuksin} in the presence of a non-conservative disorder accounting for gain and losses in the system.
As discussed above, usual experiments of optical beam self-cleaning are described by the discrete wave turbulence approach since $\beta_0^{-1} \simeq 10^{-3}$m and the above condition is well verified in the experiments \cite{wright16,krupa17}.

In the discrete wave turbulence regime we need to consider separately the cases of resonant and non-resonant mode interactions.
For mode quartets verifying $\Delta \omega_{lmnp}=0$, the Green function (\ref{eq:green}) decays on a length scale much smaller than the evolution of $n_p(z)$, because $L_{d}=(\Delta \beta)^{-1} \ll L_{kin}$, so that  (\ref{eq:conv_int}) can be approximated by $J_{lmnp}^{(j)}(z)  \simeq \frac{i\gamma}{8\Delta \beta } \left< {Y}_{lmnp}^{(j)}\right>(z)$.
For $\Delta \omega_{lmnp} \neq 0$, the Green function oscillates rapidly over a length scale smaller than $L_{lin}=\beta_0^{-1} \ll L_{kin}$.
Such a rapid phase rotation combined to the fast decay of the Green function over a length $L_d \ll L_{nl}$ leads to a vanishing convolution integral in (\ref{eq:conv_int}).
As a result, the fourth-order moment can be written in the form
\begin{eqnarray}
J_{lmnp}^{(j)}(z)  \simeq
\frac{i\gamma}{8\Delta \beta }\left< {Y}_{lmnp}^{(j)}\right>(z) \delta^K(\Delta \omega_{lmnp}),
\label{eq:mom4_disc_dis}
\end{eqnarray}
where $\delta^K(\Delta \omega_{lmnp})=1$ if $\Delta \omega_{lmnp}=0$, and zero otherwise.
Note that the discrete regime discussed here does not exactly correspond to the discrete regimes due to finite size effects in homogeneous wave turbulence \cite{nazarenko11} -- here the system is non-homogeneous ($V(\br) \neq $const) and the resonance for the momentum reflected by the tensor
$S_{lmnp}$ is not as rigid as the usual one involving the Dirac $\delta$ distribution in homogeneous turbulence.

We provide a detailed computation of the fourth-order moments $J_{lmnp}^{(j)}(z)$ and corresponding moments $\left< X_p^{(j)}\right>$ defined in (\ref{def:Xp1}-\ref{def:Xp2}) in the Appendix~A (section~\ref{appA2}), see Eq.(\ref{eq:momX_p1}) and (\ref{eq:momX_p2}).
Then collecting all terms in (\ref{eq:pzwp1}) give the discrete kinetic equation for the modal amplitudes $n_p(z)=2w_p(z)$
\begin{eqnarray}
\nonumber
\partial_z n_p(z) &=&  \frac{ \gamma^2}{6\Delta \beta} \sum_{l,m,n} |S_{lmnp}|^2 \delta^K(\Delta\omega_{lmnp}) M_{lmnp}({\bm n}) \\
&&
+  \,  \frac{4\gamma^2}{9 \Delta \beta}  \sum_l  
 |  s_{lp}({\bm n}) |^2 \delta^K(\Delta\omega_{lp})   (n_l-n_p), \quad \quad
\label{eq:kin_np_disc}
\end{eqnarray}
with $s_{lp}({\bm n})=\sum_{m'} S_{lm'm'p} n_{m'}$, and $M_{lmnp}({\bm n})=  n_l n_m n_p+n_l n_m n_n -  n_n n_p n_m -n_n n_p n_l$, with `$n_m$' for `$n_m(z)$', $\Delta \omega_{lp}=\beta_l-\beta_p$.
According to the kinetic equation (\ref{eq:kin_np_disc}), the length scale characterizing the rate of thermalization is the same as that obtained in the continuous wave turbulence regime (\ref{eq:L_kin_disor}), namely 
$L_{kin}^{disor} \sim \Delta \beta L_{nl}^2/{\bar S_{lmnp}^2}$.

Aside from its discrete form, the kinetic Eq.(\ref{eq:kin_np_disc}) has a structure analogous to the conventional wave turbulence kinetic equation, so that it describes a process of wave condensation that occurs irrespective of the sign of the nonlinearity $\gamma$ (see the factor $\gamma^2$ in Eq.(\ref{eq:kin_np_disc}))\cite{PRL05,nazarenko11}.
We refer the reader to Refs.\cite{PRL05,PD09,nazarenko11} for details on condensation in the homogeneous case ($V(\br)=0$), and to \cite{PRA11b} for the non-homogeneous case in a waveguide potential ($V(\br) \neq 0$).
It is important to note that, in contrast to the kinetic Eq.(\ref{eq:kin_contin_dis2}) derived in the continuous wave turbulence regime, here Eq.(\ref{eq:kin_np_disc}) also conserves the energy $E=\sum_p \beta_p n_p(z)$ --  despite the presence of the dissipation effect $\Delta \beta$.
The reason for this is that only exact resonances contribute to the discrete turbulence regime, so that the discrete kinetic equation is not affected by the dissipation-induced resonance broadening that inhibits energy conservation in the continuous turbulence regime.
The kinetic Eq.(\ref{eq:kin_np_disc}) also conserves the `number of particles' $N=\sum_p n_p(z)$ and exhibits a $H-$theorem of entropy growth for the nonequilibrium entropy ${\cal S}(z)=\sum_p \log\big(n_p(z)\big)$.
Accordingly, it describes an irreversible evolution to the (maximum entropy) thermodynamic Rayleigh-Jeans equilibrium 
\begin{eqnarray}
n^{eq}_p=T/(\beta_p - \mu).
\label{eq:wd_rj}
\end{eqnarray}
The system exhibits a phase transition to condensation when $\mu \to \beta_0$ \cite{PRA11b}: for $E \ge E_{\rm crit}=\frac{NV_0}{2}(1+2\beta_0/V_0)$ there is no condensation $n_0^{eq}/N=0$, while for $E < E_{\rm crit}$ the fundamental mode of the MMF gets macroscopically populated
\begin{eqnarray}
\frac{n_0^{eq}}{N}=1-\frac{E-E_0}{E_{\rm crit}-E_0},
\label{eq:n_eq_vs_E}
\end{eqnarray} 
where $E_0=N\beta_0$ is the minimum energy (all particles are in the fundamental mode).
Note that Eq.(\ref{eq:n_eq_vs_E}) only provides an approximation of the condensation curve $n_0^{eq}$ vs $E$, see Ref.\cite{PRA11b} for a more detailed discussion.
As discussed in \cite{PRL19}, a stable self-cleaned shape of the intensity pattern $|\psi|^2(\br)$ can be interpreted as a consequence of the macroscopic population of the fundamental mode of the MMF (see the movie published in the Supplemental of \cite{PRL19}).

\begin{figure}[]
\begin{center}
\includegraphics[width=1\columnwidth]{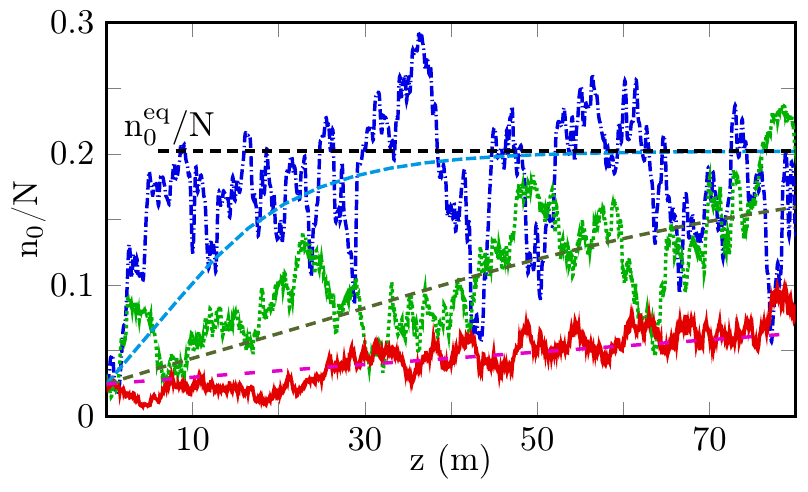}
\caption{
Scaling of acceleration of thermalization with mode-decorrelated disorder:
Numerical simulations of the NLS Eq.(\ref{eq:nls_Ap}) showing the evolutions of the fundamental mode $n_0(z)$, for different amounts of disorder $\Delta \beta$.
The dashed lines show the corresponding simulations of the discrete kinetic Eq.(\ref{eq:kin_np_disc}), starting from the same initial condition as the NLS simulations.
Parameters are: 
$\Delta \beta \simeq 2.6$m$^{-1}$ ($2\pi/\sigma_\beta=2.1$m, $l_\beta= 30$cm) blue (dashdotted);
$\Delta \beta \simeq 10.5$m$^{-1}$, ($2\pi/\sigma_\beta=26$cm, $l_\beta= 1.88$cm) green (dotted);
$\Delta \beta \simeq 42$m$^{-1}$, ($2\pi/\sigma_\beta=6.6$cm, $l_\beta= 0.47$cm) red (solid);
while the power is $N=47.5$kW ($N_*=120$ modes, $a=26\mu$m).
The curves eventually relax to the common theoretical equilibrium value $n_0^{eq}/N \simeq 0.2$ (dashed black line from Eq.(\ref{eq:n_eq_vs_E})) with different rates, confirming the scaling of acceleration of condensation predicted by the theory in Eq.(\ref{eq:accel_thermal}),  without using adjustable parameters.
}
\label{fig:1}
\end{center}
\end{figure}

\subsubsection{Acceleration of thermalization mediated by disorder}
\label{sec:acc_thermal}

The length scale characterizing the rate of thermalization in the absence of structural disorder is obtained from the continuous kinetic equation that was derived in Ref.\cite{PRA11b}, 
$L_{kin}^{ord} \sim \beta_0 L_{nl}^2/{\bar S_{lmnp}^2}$.
As discussed in Appendix~B, a similar scaling behavior is expected in the discrete wave turbulence regime.
Hence, in usual regimes of beam self-cleaning experiments where $\beta_0 \gg \Delta \beta$, the weak disorder is responsible for a significant acceleration of the rate of thermalization and condensation 
\begin{eqnarray}
L_{kin}^{disor}/L_{kin}^{ord} \sim \Delta \beta / \beta_0.
\label{eq:accel_thermal}
\end{eqnarray}
Considering typical values of $\beta_0 \sim 10^3$m$^{-1}$ and $L_d =1/\Delta \beta$ larger than tens of centimetres, we see that we always have $L_{kin}^{disor}/L_{kin}^{ord} \ll 1$.
Accordingly, weak disorder is responsible for a significant acceleration of the rate of thermalization and condensation \cite{PRL19}.
Note that the presence of a perturbation on the dispersion relation can modify the above scaling, a feature that will be discussed later.

We have confirmed the scaling (\ref{eq:accel_thermal}) by performing numerical simulations of the NLS Eq.(\ref{eq:nls_Ap}) for different amounts of disorder $\Delta \beta$.
We refer the reader to the Supplemental of Ref.\cite{PRL19} for the numerical scheme used to solve the  NLS Eq.(\ref{eq:nls_Ap}) in the presence of disorder.
The results are reported in Fig.~\ref{fig:1} and confirm the scaling of the rate of acceleration of thermalization and condensation predicted by the theory.
More precisely, a remarkable quantitative agreement has been obtained between the simulations of the NLS Eq.(\ref{eq:nls_Ap}) and those of the discrete kinetic Eq.(\ref{eq:kin_np_disc}), without using any adjustable parameter.
Note that the equilibrium value of the condensate fraction $n_0^{eq}/N$ vs the energy $E$ is obtained from a rather simple analysis of the RJ equilibrium distribution \cite{PRL05,PRA11b,nazarenko11}.
In our case, the condensation curve $n_0^{eq}/N$ vs the energy $E$ has been reported in explicit form in Ref.\cite{PRL19} (see Eqs.(17-18) in the Supplemental). 
Also note that we have deliberately chosen a small value of the condensate fraction $n_0/N \simeq 0.2$ so as to avoid large deviations from Gaussianity for the fundamental mode -- though the theory has been validated even for large condensate fractions in \cite{PRL19}.
We recall here the recent works showing that a key assumption of the wave turbulence approach is the existence of a random phase among the modes rather than a genuine Gaussian statistics \cite{chibbaro17,chibbaro18}.

\begin{figure}[]
\begin{center}
\includegraphics[width=1\columnwidth]{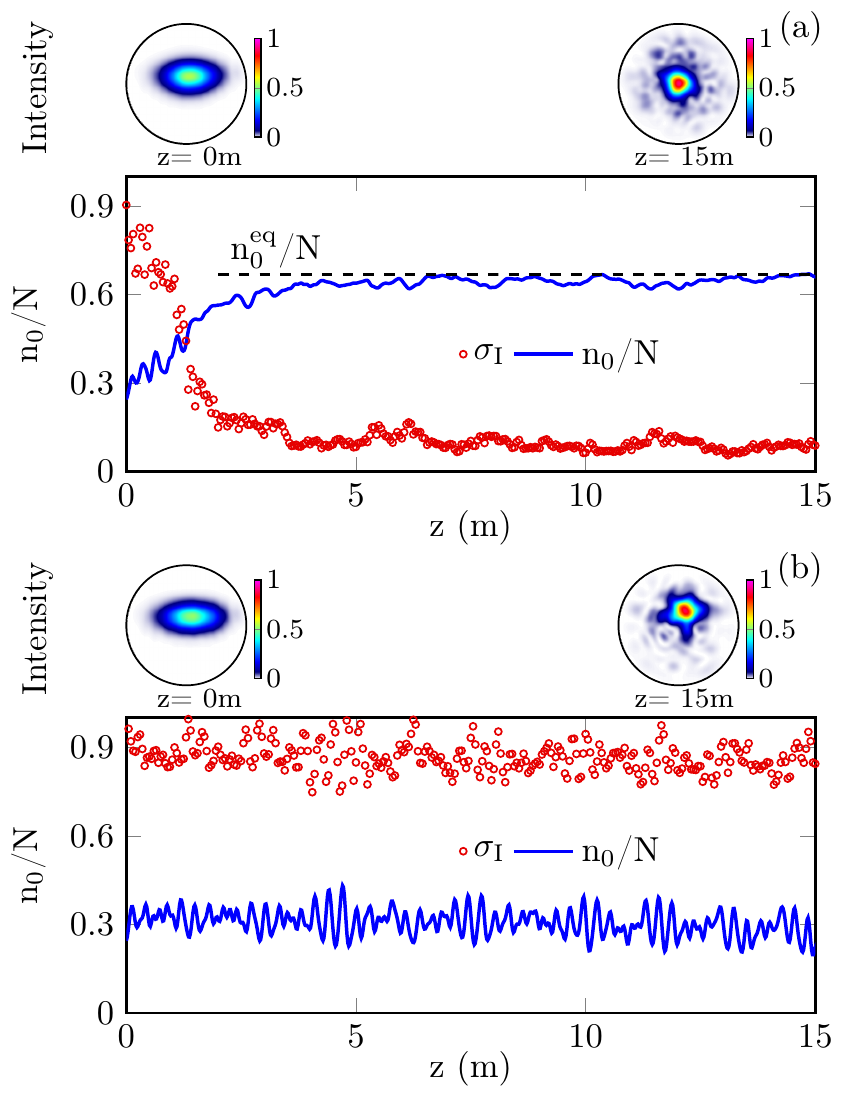}
\caption{
Disorder-induced beam cleaning:
Numerical simulations of the NLS Eq.(\ref{eq:nls_Ap}) showing the evolutions of the relative variance of intensity fluctuations $\sigma_{I}(z)$ from Eq.(\ref{eq:variance}) (red circles), and condensate fraction $n_0(z)/N$ (blue line), in the presence of disorder (a), and in the absence of disorder (b).
The intensity patterns $I(\br)$ inside the fiber core (circles) are shown at $z=0$ and $z=15$m.
The horizontal dashed black line in (a) shows the condensate fraction at thermal equilibrium, $n_0^{eq}/N \simeq 0.68$ (from Eq.(\ref{eq:n_eq_vs_E})).
The disorder induces a stable beam cleaning condensation characterized by a significant reduction of intensity fluctuations down to $\sigma_{I} \simeq 0.1$ (a), which is in contrast with the case without disorder where the variance of intensity fluctuations remain almost constant throughout propagation with $\sigma_{I} \simeq 0.9$.
Parameters are: $\Delta \beta=2.6$m ($2\pi/\sigma_\beta=2.13$m, $l_\beta= 30$cm), $N_*=120$ modes, $a=26\mu$m, $N=47.5$kW.}
\label{fig:variance}
\end{center}
\end{figure}

\subsubsection{Disorder-induced beam cleaning}
\label{sec:dis_ind_bclean}

The amount of `beam cleaning' can be quantified through the analysis of the fluctuations of the intensity during the propagation in the MMF.
To this aim we consider the relative variance of intensity fluctuations relevant for a spatially non-homogeneous incoherent beam
\begin{eqnarray}
\sigma_{I}^2(z)= \frac{\int \left< I^2(\br,z) \right> - \left< I(\br,z) \right>^2 d\br }{\int \left< I(\br,z) \right>^2 d\br},
\label{eq:variance}
\end{eqnarray}
where the intensity is $I(\br,z)=|{\bm \psi}|^2(\br,z)$.
Note that for a beam with Gaussian statistics we have $\sigma_{I}^2=1$.
We report in Fig.~\ref{fig:variance} the evolutions of the variance of intensity fluctuations in the presence and the absence of disorder.
The brackets $\left<.\right>$ in Eq.(\ref{eq:variance}) refers to an averaging over the propagation length $\Delta z=10$mm, which is larger than the mode-beating length scale $\sim \beta_0^{-1}$($\simeq 0.2$mm in the example of Fig.~\ref{fig:variance}). 
The simulations in Fig.~\ref{fig:variance} clearly show that the presence of disorder induces a rapid condensation process, which in turn leads to a significant reduction of the relative standard deviation of intensity fluctuations $\sigma_{I} \simeq 0.1$.
Conversely, in the absence of disorder the multimode beam can exhibit an enhanced brightness at some propagation lengths (see the intensity pattern at $z=15$m in Fig.~\ref{fig:variance}(b)), however its oscillatory multimode nature prevents a stable beam-cleaning propagation, as evidenced by the relative standard deviation of intensity fluctuations that only slowly decreases below $\sigma_I \simeq 0.9$.

\subsection{Correlated and partially correlated disorder}
\label{sec:corr_part_corr_dis}

In the previous section~\ref{sec:model} we have considered a model of disorder that can be termed `mode-decorrelated', in the sense that each individual mode of the MMF experiences a different noise, i.e., the functions $\nu_{p,j}(z)$ in (\ref{model:Dp0a}) are independent of each other.
Although this approach can be considered as justified in different circumstances \cite{mumtaz13,xiao14,cao18}, 
one may question its validity for a MMF featured by a large number of modes. 
In the following we address this question by considering two different models of disorder, namely the fully mode-correlated disorder model, and the partially mode-correlated disorder model.

\subsubsection{Mode-correlated noise}

We first consider the fully mode-correlated model that can be considered as the opposite limit of the `mode-decorrelated' one, in the sense that all  modes experience the same noise (more precisely, the same realization of the noise).
In this limit, the 2$\times$2 matrices describing the modal noise in (\ref{model:Dp0a}) reduce to ${\bf D}_p={\bf D}$ with
${\bf D} = \sum_{j=0}^3 \nu_j \bsigma_j$.

\begin{figure}[]
\begin{center}
\includegraphics[width=1\columnwidth]{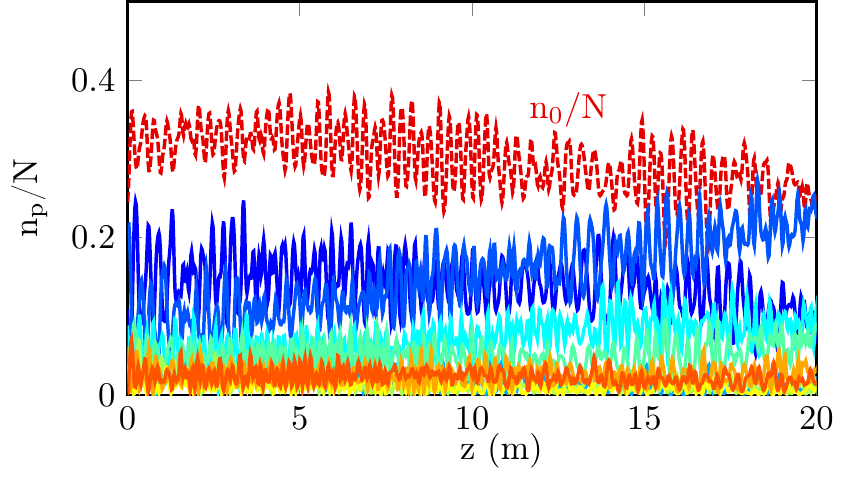}
\caption{
Mode-correlated disorder:
Numerical simulations of the NLS Eq.(\ref{eq:nls_Ap}) showing the evolutions of the modal components $n_p(z)$, starting from a coherent initial condition: 
fundamental mode $p=0$ (red dashed), $p=1$ (dark blue solid), $p=2$ (blue solid), $p=3$ (light blue solid), $p=4$ (cyan solid), $p=5$ (light green solid), $p=6$ (green solid), $p=7$ (yellow solid), $p=8$ (orange solid).
The mode-correlated disorder does not introduce an effective dissipation in the system ($\Delta \beta=0$): phase-correlations among the low-order modes are preserved and lead to an oscillatory dynamics similar to that in the absence of disorder (see Fig.~\ref{fig:0}).
At complete thermal equilibrium the system would reach the condensate fraction $n_0^{eq}/N \simeq 0.68$ (from Eq.(\ref{eq:n_eq_vs_E})).
Parameters are: 
$\Delta \beta \simeq 2.6$m$^{-1}$ ($2\pi/\sigma_\beta=2.1$m, $l_\beta= 30$cm), 
the power is $N=47.5$kW, $N_*=120$ modes, $a=26\mu$m.
}
\label{fig:mode_correl}
\end{center}
\end{figure}

The theory developed above for the decorrelated model of disorder can be extended to this fully correlated model.
The theory reveals in this case a remarkable result, namely that the equations for the fourth-order moments $J_{lmnp}^{(j)}(z)$ do not exhibit an effective damping, i.e., $\Delta \beta=0$, see Appendix~A (section~\ref{appA3}). 
This is in marked contrast with the decorrelated model of disorder discussed above, see Eqs.(\ref{eq:4th_order_moment}).
This result has a major consequence: the fully mode-correlated noise does not modify the regularization of wave resonances and the system recovers a dynamics analogous to that obtained in the absence of any disorder. 
We illustrate this in Fig.~\ref{fig:mode_correl} that reports the evolution of the modal components $n_p(z)$ obtained by simulation of the modal NLS Eq.(\ref{eq:nls_Ap}) in the presence of a mode-correlated disorder.
The initial condition is coherent (all modes are correlated with each other) and we see in Fig.~\ref{fig:mode_correl} that the low-order modes recover an oscillatory dynamics reflecting the presence of strong phase-correlations, as for the coherent regime discussed in the absence of disorder in Fig.~\ref{fig:0}.
This coherent dynamics is consistent with the intuitive idea that, in the absence of an effective dissipation ($\Delta \beta=0$), the phase-correlations are not forgotten during the evolution.

\subsubsection{Partially mode-correlated noise}

We have seen that the fully mode-correlated disorder does not introduce an effective dissipation ($\Delta \beta=0$)  and thus leads to a coherent dynamics analogous to that obtained in the absence of disorder.
We have thus  considered a `partially correlated' model of disorder in which modes that belong to different groups of degenerate modes experience a decorrelated noise, while degenerate modes of the same group experience the same noise.
This artificial model for MMFs can be considered as an intermediate model between the two limits of correlated and decorrelated models. It is mathematically tractable and it illustrates the conjecture that the discrete kinetic Eq.(\ref{eq:kin_np_disc}) is robust as soon as disorder is not fully mode-correlated.

The theory developed above for the model of decorrelated disorder has been extended to address the partially correlated model, see Appendix~A (section~\ref{appA4}).
The theory reveals that (second-order) correlations among non-degenerate modes are vanishing small and can be neglected, as it was shown for the model of decorrelated disorder.
However, the computation of the fourth-order moments $J_{lmnp}^{(j)}$ is more delicate because the mode-correlated noise introduces more terms in the calculations of the equations for the moments.
We obtain different results for the fourth-order moments that depend on the specific modes involved in the moments. 
Almost all of the fourth-order moments satisfy an evolution equation with a dissipation that is proportional to $\Delta \beta$.
This result is analogous to that obtained for the model of decorrelated disorder considered above, though the coefficients in front of $\Delta \beta$ are different and their values depend on the specific modes involved in the fourth-order moment. 
In addition, at variance with the model of decorrelated disorder, here there are also particular cases where the fourth-order moments do not exhibit any dissipation.
Such special cases do not contribute to the fast thermalization process described by the effective dissipation $\Delta \beta$, but instead they induce a reversible exchange of power within a group of degenerate modes, see Appendix~A (section~\ref{appA4}).

\begin{figure}[]
\begin{center}
\includegraphics[width=1\columnwidth]{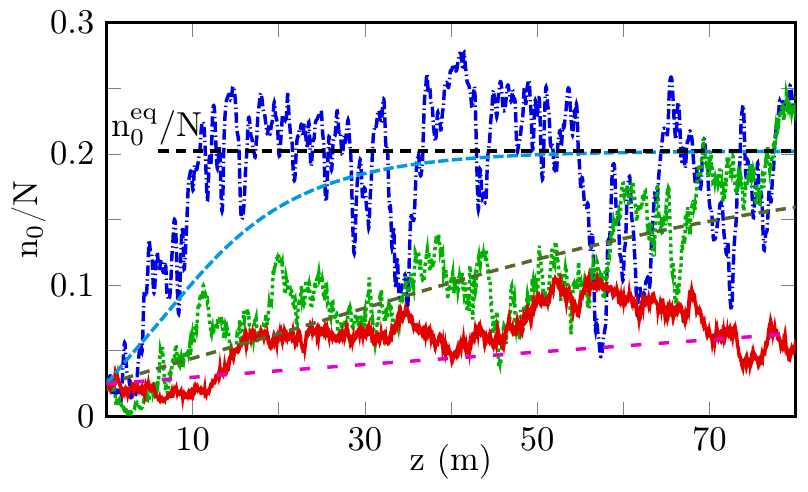}
\caption{
Scaling of acceleration of thermalization with a partially mode-correlated disorder:
Numerical simulations of the NLS Eq.(\ref{eq:nls_Ap}) showing the evolutions of the fundamental mode $n_0(z)$, for different amounts of disorder $\Delta \beta$.
The dashed lines show the corresponding simulations of the discrete kinetic Eq.(\ref{eq:kin_np_disc}), starting from the same initial condition as the NLS simulations.
Parameters are: 
$\Delta \beta \simeq 2.6$m$^{-1}$ ($2\pi/\sigma_\beta=2.1$m, $l_\beta= 30$cm) blue (dashdotted);
$\Delta \beta \simeq 10.5$m$^{-1}$, ($2\pi/\sigma_\beta=26$cm, $l_\beta= 1.88$cm) green (dotted);
$\Delta \beta \simeq 42$m$^{-1}$, ($2\pi/\sigma_\beta=6.6$cm, $l_\beta= 0.47$cm) red (solid);
while the power is $N=47.5$kW ($N_*=120$ modes, $a=26\mu$m).
The curves eventually relax to the common theoretical equilibrium value $n_0^{eq}/N$ (dashed black line from Eq.(\ref{eq:n_eq_vs_E})) with different rates,  and confirm the theoretical scaling of acceleration of condensation in Eq.(\ref{eq:accel_thermal}) for a partially mode-correlated disorder.
}
\label{fig:2}
\end{center}
\end{figure}

To summarize, 
the theoretical developments reported in Appendix~A (section~\ref{appA4}) allow us to infer that the kinetic equation is still of the form given by the discrete kinetic Eq.(\ref{eq:kin_np_disc}) and that the scaling of the rate of thermalization is still given by $L_{kin}^{disor} \sim  \Delta \beta L_{nl}^2/{\bar S_{lmnp}^2}$ in Eq.(\ref{eq:L_kin_disor}).
This scaling of the rate of thermalization has been confirmed by the numerical simulations of the NLS Eq.(\ref{eq:nls_Ap}).
The results are reported in Fig.~\ref{fig:2} for the same parameters of disorder as those considered in Fig.~\ref{fig:1}, except that a model of partial mode-correlated disorder has been considered. 
The  agreement with the discrete kinetic Eq.(\ref{eq:kin_np_disc}) confirms the scaling (\ref{eq:accel_thermal}) of the rate of acceleration of thermalization for the `partially correlated' model of disorder.

\section{Strong  disorder}
\label{sec:strong_disorder}

In this section we discuss the impact of a strong structural disorder, i.e., a noise that couples distinct modes on the evolution of the averaged modal components.
More precisely, here we complete the discussion of Ref.\cite{PRL19} on the impact of strong disorder in several respects.
We show that a general form of conservative disorder introduces an effective dissipation in the system, which is responsible for an irreversible decay of the first-order moments, as well as an irreversible relaxation of the second-order moments toward different forms of power equipartition among the modes.
In this way, we provide a detailed derivation of the kinetic equation accounting for strong disorder at the leading order linear regime, see Appendix~C.
We also discuss the different regimes described by the considered model of strong disorder, namely the random coupling among polarization components, as well as random coupling among degenerate and non-degenerate modes.
In this way, we relate our approach with the corresponding limits described by the so-called Manakov approximation \cite{mecozzi12a,mecozzi12b,mumtaz13,xiao14}.

\subsection{Model} 

Random coupling among the modes can be described by a generalized form of the modal NLS Eq.(\ref{eq:nls_Ap}).
The random field composed of $N_*$ spatial modes is represented by a $2N_*$-dimensional
complex valued vector ${\bm A}(z)=(A_p(z))_{p=0}^{2N_*-1}$, which is obtained by stacking the $2N_*-$modal components that include the polarization degrees of freedom.
The vector ${\bm A}(z)$ is governed by the modal NLS equation:
\begin{eqnarray}
i \partial_z {\bm A}= \bbeta {\bm A} + {\bf D}^{\rm sd}(z) {\bm A} -  \gamma {\bm P}({\bm A}),
\label{eq:nlsmulti}
\end{eqnarray}
where $\bbeta$ is a diagonal matrix with diagonal terms $\beta_p$.
Random linear coupling between the modes is modelled by the random matrix-valued process ${\bf D}^{\rm sd}(z)$.
It consists of a 2$N_*$ dimensional generalization of the 2-dimensional model of weak disorder discussed in section~\ref{sec:model}.
The $2N_* \times 2N_*$ Hermitian matrix ${\bf D}^{\rm sd}(z)$ can be written as
\begin{eqnarray}
{\bf D}^{\rm sd}(z) &=&  \sum_{q<r}
g_{qr} \big(  {\nu}_{qr}(z)  {\bf H}^{qr} 
+ {\mu}_{qr}(z)  {\bf K}^{qr} 
\big) \nonumber \\
&&+
\sum_q g_q {\eta}_q(z) {\bf J}^q  ,
\label{eq:D_sd}
\end{eqnarray}
where the matrices ${\bf H}^{qr} $, $ {\bf K}^{qr} $, $ {\bf J}^{q} $ refer to a $2N_*$ dimensional generalization of the Pauli matrices $\bsigma_j$ \cite{mecozzi12c}:
The matrices $ {\bf H}^{qr} $, $ {\bf K}^{qr} $, $ {\bf J}^{q} $ also form a basis for the vector space of $2N_* \times 2N_*$ Hermitian matrices. 
In this respect, the model (\ref{eq:D_sd}) can be considered as a general model of strong disorder.
For definiteness, the three matrices are given in explicit form in the Appendix~C.
Following the generalization of the weak disorder model of section~\ref{sec:model}, in the strong disorder model given by Eq.(\ref{eq:D_sd}) the functions  $\eta_q$ for  $0\leq q \leq 2N_*-1$,  and $\nu_{qr}$, $\mu_{qr}$ for   $0\leq q<r\leq 2N_*-1$, are zero mean independent and identically distributed Gaussian real-valued  random processes, with 
$\left< \eta_q(z) \eta_q (z') \right>
=
\left< \nu_{qr}(z) \nu_{qr} (z') \right>=
\left< \mu_{qr}(z) \mu_{qr} (z') \right>
=  \sigma_\beta^2   {\cal R}\Big( \frac{z-z'}{l_{\beta}}\Big)
$.
We consider Ornstein-Uhlenbeck processes for $\eta_q$, $\nu_{q,r}$ and $\mu_{q,r}$, with ${\cal R}(\zeta) = \exp(-|\zeta|) /2$.
The parameters $g_{qr}$ and $g_q$ are mode coupling coefficients of order one.

As will be discussed below, the model of disorder (\ref{eq:D_sd}) allows us to distinguish three different regimes of random mode coupling, namely polarization coupling, coupling among degenerate modes and coupling among non-degenerate modes.


\subsection{Kinetic equation}

In the Appendix~C, we derive the equations that govern the evolutions of the first-order moments of the field.
The main result obtained by using the Furutsu-Novikov theorem is that the conservative strong disorder introduces an effective dissipation for the evolutions of the average $\left< {\bm A} \right>(z)$, which thus exhibits an exponential decay during the propagation, see Eqs.(\ref{eq:sd_avU}-\ref{eq:sd_avU_Q}).
This result allows us to study the dynamics of the second-order moments with strong disorder in the regime where linear effects dominate nonlinear effects, as for the weak disorder case considered in section \ref{sec:weak_disorder}.
However, at variance with weak disorder where the linear dynamics of the second-order moments is trivial (i.e., vanishing correlations, see section \ref{sec:wd_vanish_corr}), the presence of strong disorder introduces a non-trivial linear dynamics that dominates nonlinear effects.
In the following we study strong disorder at the leading order linear regime.

In Appendix~C we obtain the following equation governing the evolution of the modal components:
\begin{eqnarray}
\partial_z w_p &=& 2\sigma_\beta^2 \sum_{m=0}^{2N_*-1}  g_{pm}^2 \int_0^z \big( {w_m}(z') - {w_p}(z')\big) \nonumber \\
&& \times {\cal R}\big((z-z')/l_{\beta}\big) \cos\big( (\beta_p-\beta_m)(z-z')\big) dz' \quad \quad
\label{eq:sd_w_p}
\end{eqnarray}
with $w_p(z)=\left<|A_p|^2(z)\right>$, $p=0,...,2N_*-1$.
Considering that the coupling coefficients do not depend on polarization 
the $2N_* \times 2N_*$ matrix $g_{pm}$ can be reduced to a $N_* \times N_*$ matrix $\rho_{pm}$. 
Furthermore, since $l_{\beta} \ll z,L_{kin}$, the additional term in the kinetic equation takes the simplified form 
\begin{eqnarray}
\partial_z n_p = \Delta \beta \sum_{m=0}^{N_*-1}  \Gamma_{pm}  {\hat {\cal R}}\big((\beta_m-\beta_p) l_{\beta}\big) ( n_m - n_p)
\label{eq:kin_strong_dis}
\end{eqnarray}
where $n_p(z)=w_{2p}(z)+w_{2p+1}(z)$ is the power in the mode $p(=0,...,N_*-1)$, and $\Gamma_{pm}=\rho_{pm}^2$ denotes the mode coupling matrix.
The function ${\hat {\cal R}}(\kappa)$ denotes the Fourier transform of the correlation function ${\cal R}(x)$,  which reads ${\hat {\cal R}}(\kappa)=1/(1+\kappa^2)$ for the considered Ornstein-Uhlenbeck process.
Note that Eq.(\ref{eq:kin_strong_dis}) has a form similar to that considered to model power cross-talk among different modes in optical telecommunications \cite{kaminow13}.

\subsection{Characteristic length scales }

We first note that by adding the extra term (\ref{eq:kin_strong_dis}) in the kinetic equation derived above with weak disorder (see Eq.(\ref{eq:kin_np_disc})), the conservation of the energy $E$ is no longer verified, so that a $H-$theorem of the complete kinetic Eq.(\ref{eq:kin_np_disc}) and (\ref{eq:kin_strong_dis}) would describe an irreversible evolution toward a maximum entropy equilibrium state characterized by an equipartition of power (`number of particles') among the modes.
Accordingly, strong disorder would inhibit the self-cleaning condensation process discussed through weak disorder in section~\ref{sec:weak_disorder}.
As will be discussed below (section \ref{sec:appl_bsc}), this conclusion is not correct when one considers the usual regime of optical beam self-cleaning.
In this view, we now discuss the physical meaning of Eq.(\ref{eq:kin_strong_dis}) through the analysis of the model of disorder considered in Eq.(\ref{eq:D_sd}) and the corresponding different length scales of random mode coupling.

\noindent
(i) {\it Polarization coupling:} 
The model (\ref{eq:D_sd}) describes random coupling among the polarizations of a single mode that occurs over the propagation length $L_{d} \sim 1/(\Delta \beta \Gamma_{pp})$.
Considering a moderate impact of disorder, we can have $\Gamma_{pp} \sim 1$, so that this length scale of polarization random coupling is the same as that considered above in Eq.(\ref{eq:L_d}). 

\noindent
(ii) {\it Coupling among degenerate modes:} 
The model (\ref{eq:D_sd}) describes random coupling within a group of $M_g$ degenerate modes over the characteristic propagation length 
\begin{eqnarray}
L_{sd}^{deg} \sim 1/(M_g \Delta \beta {\bar \Gamma}_{g})
\label{eq:L_sd_deg}
\end{eqnarray}
where ${\bar \Gamma}_{g}$ is determined by an average of the coupling coefficients $\Gamma_{mp}$ ($m \neq p$) among the $M_g$ modes of the $g$th mode group.
Considering a moderate coupling among the modes $\Gamma_{mp} < \Gamma_{pp}$ for $m\neq p$, we have $L_{d} < L_{sd}^{deg}$.
Physically, the length scale $L_{sd}^{ndeg}$ represents the typical propagation length such that strong disorder achieves an equipartition of power within each group of degenerate modes.
When one considers temporal propagation effects through generalized coupled NLS equations, this effect is known as the Manakov limit of strongly coupled groups of modes, see \cite{mecozzi12b}.

\noindent
(iii) {\it Coupling among non-degenerate modes ($\beta_p \neq \beta_m$):} By Perron-Frobenius theorem, the symmetric  matrix $\widetilde{\bGamma}$ defined by
\begin{eqnarray}
\widetilde{\Gamma}_{mp} &=& \Gamma_{mp} \hat{\cal R}\big( (\beta_p-\beta_m)l_\beta \big) (1-\delta^K_{mp}) \nonumber \\
&& - \Big( \sum_{p'\neq m}  \Gamma_{mp'} \hat{\cal R}\big( (\beta_{p'}-\beta_m)l_\beta \big)\Big) \delta^K_{mp}
\end{eqnarray}
has a simple zero eigenvalue with the associated unit eigenvector, and all other eigenvalues are negative.
As a result, the additional linear coupling terms in the kinetic Eq.(\ref{eq:kin_strong_dis}) tend to redistribute the power amongst all modes, at an exponential rate that can be determined by the second eigenvalue $\lambda_2 (\widetilde{\bGamma})$ of the  matrix $\widetilde{\bGamma}$.
The corresponding characteristic length scale is given by 
\begin{eqnarray}
L_{sd}^{ndeg} \sim 1/\big( \Delta \beta | \lambda_2 (\widetilde{\bGamma})| \big).
\label{eq:L_sd_ndeg}
\end{eqnarray}
We recall that ${\cal R}(\kappa)$ decays to zero as $\kappa$ goes to infinity, so that ${\cal R}((\beta_p-\beta_m)l_\beta)$ is much smaller than ${\cal R}(0)=1$ and $L_{sd}^{deg} < L_{sd}^{ndeg}$.
In other words, the length scale $L_{sd}^{ndeg}$ represents the typical propagation length such that strong disorder achieves an equipartition of power among the modes.
Considering temporal effects through generalized coupled NLS equations, this effect of power equipartition among all fiber modes is known as the strong coupling regime in the Manakov limit \cite{mecozzi12a,mumtaz13}.

We finally note that typical values of the three length scales discussed here are estimated in Ref.\cite{ho14}, where complete polarization random coupling is expected to occur over several meters, random mode coupling among degenerate modes over tens meters, and mode coupling among non-degenerate modes over hundreds meters.

\subsection{Application to beam cleaning: Acceleration of thermalization}
\label{sec:appl_bsc}

The additional term in the kinetic Eq.(\ref{eq:kin_strong_dis}) provides the characteristic length scales  due to strong mode coupling $L_{sd}^{deg}$ and $L_{sd}^{ndeg}$. 
In the usual experiments of beam cleaning we have $\beta_0 l_\beta \gg 1$ (since $\beta_0 \sim 10^3$m$^{-1}$), so that mode coupling among non-degenerate modes is quenched by the Fourier transform of the correlation function ${\cal {\hat R}}(\beta_0 l_\beta ) \ll 1$, i.e., $L_{sd}^{deg} \ll L_{sd}^{ndeg}$.
However, as discussed here above, mode coupling among degenerate modes leads to an exponential relaxation to an {\it equipartition of power within groups of degenerate modes}.
Unexpectedly, this is a property of the Rayleigh-Jeans distribution, since this equilibrium only depends on the eigenvalue $\beta_p$, see the expression of $n_p^{eq}$ in Eq.(\ref{eq:wd_rj}).
This shows that the impact of strong disorder is not detrimental to achieve wave condensation, but instead it enforces the process of thermalization to the Rayleigh-Jeans distribution, though such an acceleration of thermalization is negligible with respect to the acceleration mediated by weak disorder, see Eq.(\ref{eq:accel_thermal}).
In addition, considering the short fiber lengths typically used in the experiments of beam self-cleaning ($L \sim 10-20$m), mode coupling among degenerate modes is expected to play a negligible role, $L, L_{kin}^{disor} \lesssim L_{sd}^{deg}$.
In the remainder of this article, we will neglect the impact of random mode coupling among degenerate modes.




\section{Reduced impact of disorder}

The kinetic equations describing the evolution of the modal components $n_p(z)$ have been derived under the assumption that disorder effects dominate nonlinear effects $L_d \ll L_{nl}$.
Although the parameters that characterize the disorder, namely the correlation length $l_\beta$ and the effective `beating length' $2\pi/\sigma_\beta$ (reflecting the `strength' of disorder) are not precisely known, accurate measurements in {\it single mode} optical fibers indicate that such length scales can be of few to several meters \cite{galtarossa01}. 

\begin{figure}[]
\begin{center}
\includegraphics[width=1\columnwidth]{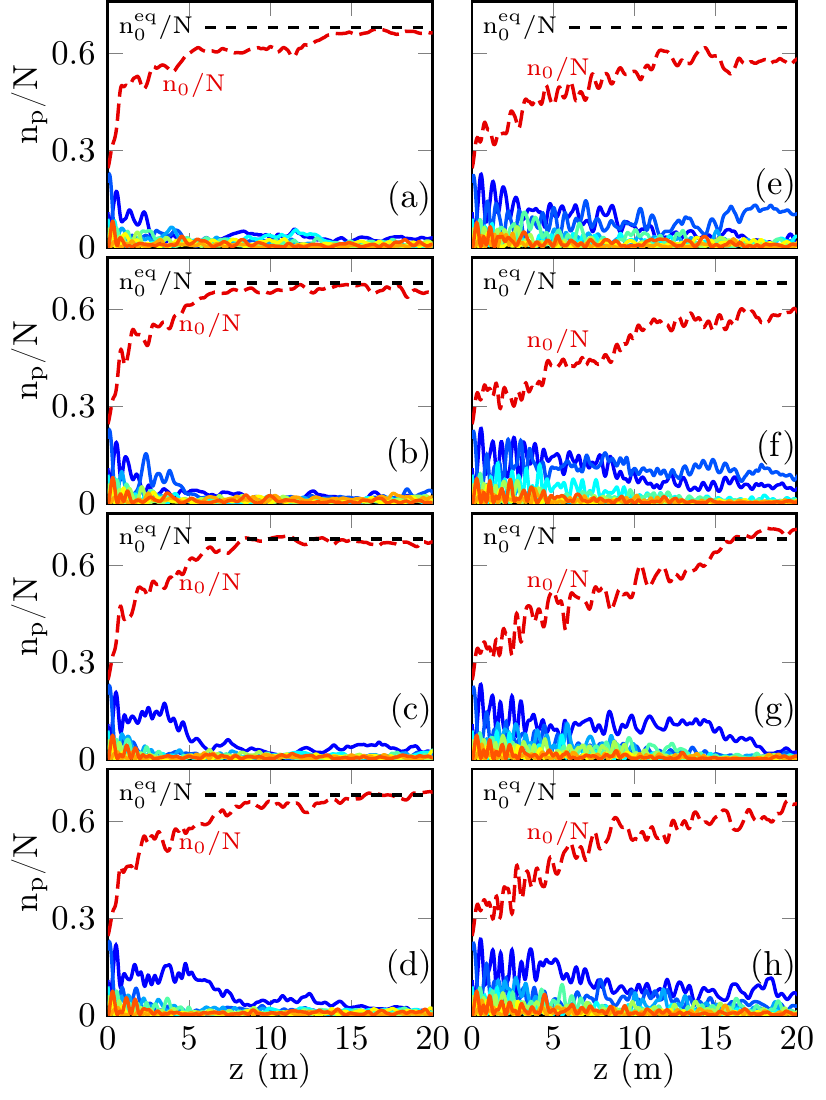}
\caption{Mixed coherent-incoherent regime with mode-decorrelated disorder:
Numerical simulations of the NLS Eq.(\ref{eq:nls_Ap}) showing the evolutions of the modal components $n_p(z)$, for different amounts of the strengths of disorder  $2\pi/\sigma_\beta$, and correlation lengths $l_\beta$:
fundamental mode $p=0$ (red dashed), $p=1$ (dark blue solid), $p=2$ (blue solid), $p=3$ (light blue solid), $p=4$ (cyan solid), $p=5$ (light green solid), $p=6$ (green solid), $p=7$ (yellow solid), $p=8$ (orange solid).
Parameters are:
$2\pi/\sigma_\beta$=5.3m (1st column), $2\pi/\sigma_\beta$=10.6m (2nd column).
1st (top) row: $l_\beta=0.5$m, 2nd row: $l_\beta=1$m,  
3rd row: $l_\beta=3$m, 4th row: $l_\beta=5$m.
As the impact of disorder is reduced, the modal components enter a mixed coherent-incoherent regime of interaction characterized by a (phase-sensitive) oscillatory behavior of the modal components.
At complete thermal equilibrium $n_0^{eq}/N \simeq 0.68$ (dashed black line from Eq.(\ref{eq:n_eq_vs_E})).
The power is $N=19$kW, the initial condition is a coherent Gaussian beam ($N_*=120$ modes, $a=26\mu$m).
}
\label{fig:4}
\end{center}
\end{figure}

\begin{figure}[]
\begin{center}
\includegraphics[width=1\columnwidth]{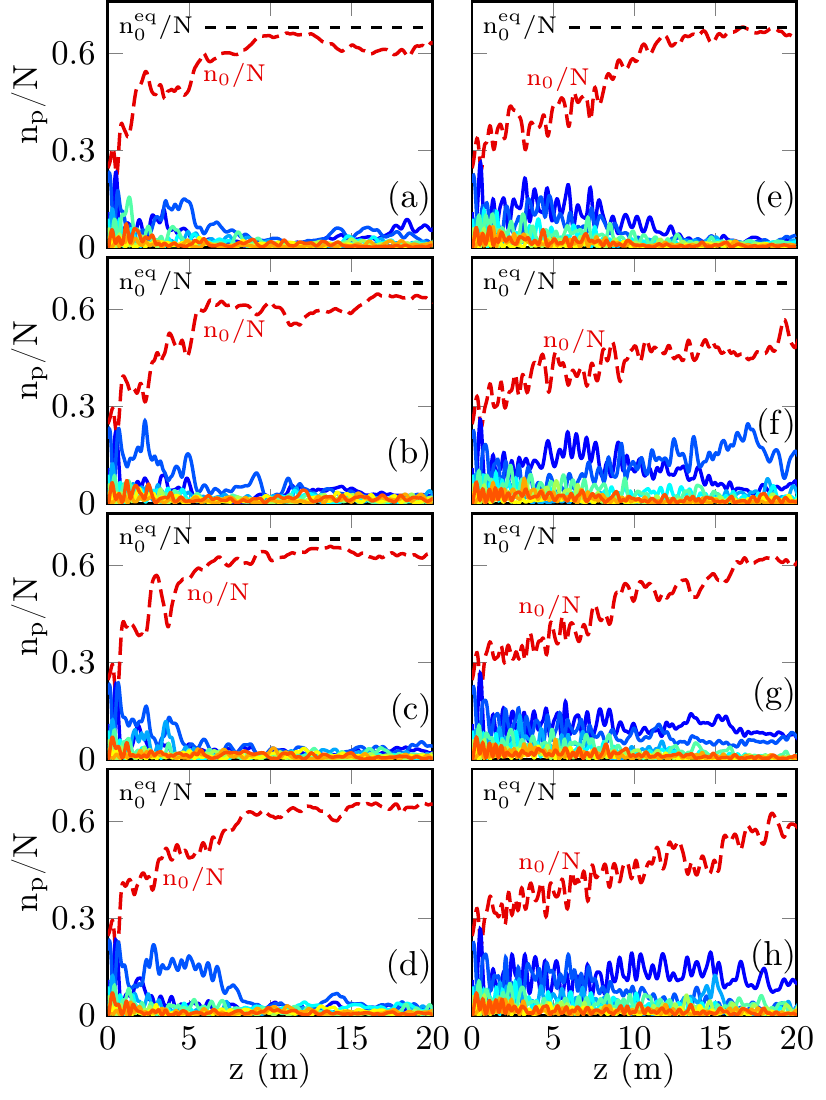}
\caption{Mixed coherent-incoherent regime with partially mode-correlated disorder:
Numerical simulations of the NLS Eq.(\ref{eq:nls_Ap}) showing the evolutions of the modal components $n_p(z)$, for different amounts of the strengths of disorder  $2\pi/\sigma_\beta$, and correlation lengths $l_\beta$.
Parameters and initial condition are the same as in Fig.~\ref{fig:4}, except that a partially mode-correlated disorder has been considered.
}
\label{fig:5}
\end{center}
\end{figure}

\subsection{Mixed coherent-incoherent regime}

We report in this section numerical simulations of the modal NLS Eq.(\ref{eq:nls_Ap}) where disorder and nonlinearity are typically of the same order of magnitude $L_d \sim L_{nl}$.
For completeness, we consider both models of disorder discussed in section \ref{sec:corr_part_corr_dis} where the modes experience a decorrelated or a partially correlated noise.
Figure~\ref{fig:4} reports the results for a decorrelated model of disorder in which we have considered different values of $l_\beta$ and $2\pi/\sigma_\beta$ of the noise.
We note that, at first sight, for relative small values of ($l_\beta, 2\pi/\sigma_\beta$), the global evolutions of the modal components $n_p(z)$ are similar to those reported in the regime where disorder dominates the nonlinearity.
There is however a difference that distinguishes the two regimes.
The rapid fluctuations on the small length scale of disorder $L_d (\ll L_{nl})$ of Figs.~\ref{fig:1}-\ref{fig:2} get smoother in the regime where disorder and nonlinearity are of the same order in Fig.~\ref{fig:4}.
Also remark that a variation of the correlation length $l_\beta$ (different lines in Fig.~\ref{fig:4}) has a marginal impact on the dynamics as compared to the strength of disorder $2\pi/\sigma_\beta$ (columns in Fig.~\ref{fig:4}) -- note that the values of $2\pi/\sigma_\beta$ in Fig.~\ref{fig:4} correspond to refractive index fluctuations of the order $\delta n \sim \sigma_\beta/k_0 \sim 10^{-7}$.

It is interesting to note that, by increasing further the correlation and beating lengths of disorder ($l_\beta, 2\pi/\sigma_\beta$), the system enters a different regime, which is characterized by the presence of pronounced oscillations of the modal components.
This oscillatory behavior reflects the presence of a phase-correlation among the modal components, as it was discussed through the {\it coherent modal regime} of interaction in the absence of any disorder in section \ref{sec:coherent_regime}.
Here, the presence of a moderate disorder is not sufficient to remove such coherent phase-correlation dynamics, so that the modes exhibit a mixed coherent-incoherent regime of interaction.

The same phenomenological behavior about the impact of disorder on the system is observed by considering a partially correlated model of noise, as illustrated in Fig.~\ref{fig:5}.
Interesting to note, for moderate values of disorder the evolutions of the modal components is very similar to that observed for a decorrelated model of disorder (compare the first columns in Fig.~\ref{fig:4} and Fig.~\ref{fig:5}).
The fact that a partial correlation among the modes does not alter the rate of thermalization was already discussed in the regime where disorder dominates the nonlinearity (see section \ref{sec:corr_part_corr_dis}).
Actually, the main difference between the decorrelated and partially correlated models of disorder is observed by further reducing the impact of disorder (i.e., by further increasing $l_\beta$ and $2\pi/\sigma_\beta$).
In this case, the partially correlated noise model leads to a more pronounced oscillatory behavior, a feature that can easily be interpreted by remarking that since the degenerate modes see the same noise, the impact of disorder is  less efficient in breaking the phase-correlations among the modes.
As a result, when one considers the partially mode-correlated noise for large values of $l_\beta$ and  $2\pi/\sigma_\beta$, the mixed coherent-incoherent regime of interaction becomes clearly apparent, and the corresponding oscillatory behavior of the modal components leads to a deceleration of the thermalization process, see the second column of Fig.~\ref{fig:5}.
We note that the simulations in Figs.~\ref{fig:4}-\ref{fig:5} have been realized with the parameters of the recent experiment of Ref.\cite{PRL19} (in particular the same power). 
The simulations then appear consistent with the experimental results -- for instance a propagation length similar to that of the fiber length used in Ref.[41] (i.e. $\sim$12m) is sufficient to evidence a significant process of condensation starting from a coherent initial condition.

We finally comment the impact of a mode-correlated model of disorder.
We have seen in section~\ref{sec:corr_part_corr_dis} that in the regime $L_d \ll L_{nl}$ this model of disorder does not introduce an effective dissipation so that it does not lead to a fast process of condensation.
In the regime $L_d \sim L_{nl}$ the simulations still reveal a persistent oscillatory behavior of the modes by starting from a coherent initial condition (as in Figs.~\ref{fig:4}-\ref{fig:5}), while a significant deceleration of condensation featured by a mixed coherent-incoherent regime has been observed by starting the simulations from a speckle beam.

\begin{figure}[]
\begin{center}
\includegraphics[width=1\columnwidth]{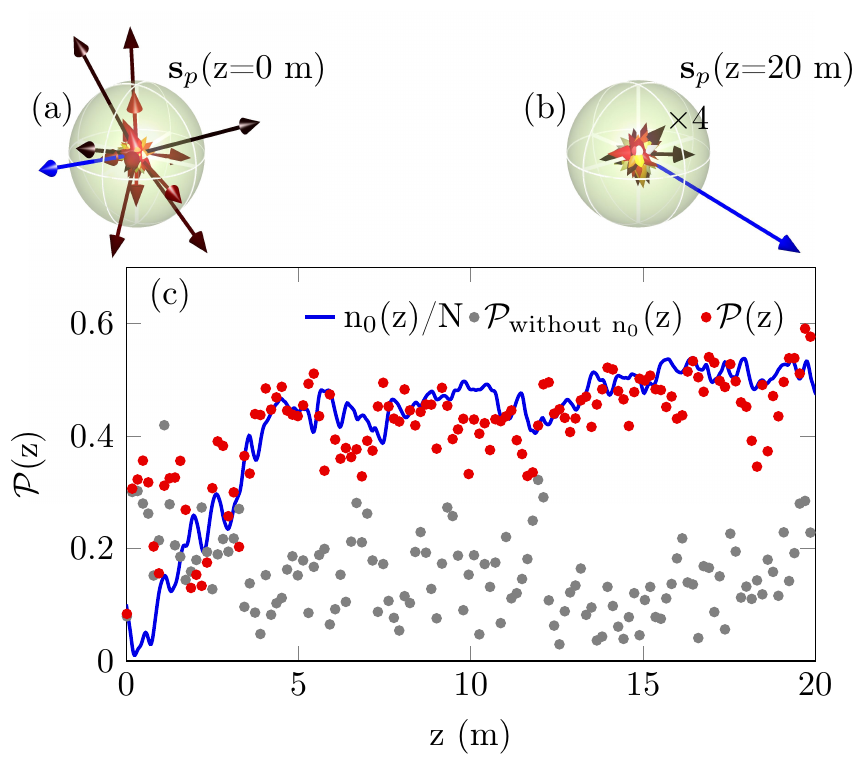}
\caption{
Evolution during the propagation $z$ of the effective degree of polarization ${\cal P}$ computed by taking into account the contribution of the fundamental mode (red points), and without including its contribution (grey points) (c).
Corresponding evolution of the condensate amplitude $n_0(z)$ (blue line) (c).
The initial condition is a speckle beam, so that the Stokes vectors of the modal components are not aligned with each other (a), leading to a small ${\cal P}(z=0)$.
During the propagation, the condensation process entails an effective repolarization of the beam because $|\left<{\bm s}_{0}\right>| \gg |\left<{\bm s}_{p \neq 0}\right>|$ (a-b) -- to improve their visibility the lengths of the Stokes vectors $\left<{\bm s}_{p \neq 0}\right>$ have been multiplied by a factor $\times 4$ in (b).
Parameters: $l_\beta=30$cm, $2\pi/\sigma_\beta=2.1$m, $N=47.5$kW ($N_*=120$ modes, $a=26\mu$m). 
}
\label{fig:dop_1}
\end{center}
\end{figure}

\subsection{Polarization effects}

In this section we discuss the polarization dynamics of the modal components during their nonlinear propagation through the MMF in the presence of disorder effects.
We consider the usual definition of the Stokes vectors integrated over the transverse spatial section of the optical beam
$S^{(0)}(z)=\int |\psi_x(\br,z)|^2+|\psi_y(\br,z)|^2d\br$,
$S^{(1)}(z)=\int |\psi_x(\br,z)|^2-|\psi_y(\br,z)|^2d\br$,
$S^{(2)}(z)=2 \operatorname{Re}\int \psi_x(\br,z)^*\psi_y(\br,z)d\br$,
$S^{(3)}(z)=-2\operatorname{Im}\int \psi_x(\br,z)^*\psi_y(\br,z)d\br$.
By expanding the field over the modes, we have
\begin{subequations}  \label{eq:stokes_vecteurs_modes}
\begin{alignat}{2}
S^{(0)}(z)&=\sum_{p=0}^{N_*-1}|A_{p,x}(z)|^2+|A_{p,y}(z)|^2=\sum_{p=0}^{M-1}s_{p}^{(0)}, \nonumber \\
S^{(1)}(z)&=\sum_{p=0}^{N_*-1}|A_{p,x}(z)|^2-|A_{p,y}(z)|^2=\sum_{p=0}^{M-1}s_{p}^{(1)},\nonumber \\
S^{(2)}(z)&=2 \operatorname{Re}\sum_{p=0}^{N_*-1}A_{p,x}(z)^*A_{p,y}(z)=\sum_{p=0}^{N_*-1}s_{p}^{(2)},\nonumber  \\
S^{(3)}(z)&=-2\operatorname{Im}\sum_{p=0}^{N_*-1}A_{p,x}(z)^*A_{p,y}(z)=\sum_{p=0}^{N_*-1}s_{p}^{(3)}. \nonumber 
\end{alignat}
\end{subequations}
It is important to recall that our model does not account for the temporal dynamics of the optical wave.
Accordingly, we resort to a definition of an effective degree of polarization ${\cal P}(z)=\sqrt{\sum_{j=1}^3\left<S^{(j)}\right>^2}/\left<S_0\right>$, by considering an average ($\left<\cdot\right>$) over the evolution $z-$variable.
As is well-known, caution should be exercised with the notion of degree of polarization, which is inherently related to the underlying averaging procedure.
In this respect, we note that for an averaging length $\Delta z$ much larger than the correlation lengths $l_\beta, 2\pi/\sigma_\beta$, the degree of polarization  vanishes ${\cal P} \simeq 0$.
This results from the theory developed in Appendix~A (section~\ref{appA1}), where we have shown that in the regime $L_d \ll L_{nl}$  the correlations among the orthogonal polarization components are vanishingly small.
To study the correlations $\left<S^{(j)}\right>$, here we consider a moderate value of the spatial averaging $\Delta z \simeq 15$cm.

We report in Fig.~\ref{fig:dop_1} the evolution of ${\cal P}(z)$ starting from a random distribution of the modal components with random phases (`speckle' beam).
Accordingly, the initial Stokes vectors of the modes ${\bm s}_p$ are not aligned with each other, i.e. their random orientation leads to a vanishing sum $\left<{\bm S}\right>=\sum_p\left<{\bm s}_{p}\right> \simeq 0$ and thus ${\cal P}(z=0) \simeq 0$.
As the beam propagates through the MMF, it undergoes a beam-cleaning condensation in which the evolution of ${\cal P}(z)$ follows a behavior similar to that of the condensate fraction $n_0(z)/N$.
This is because the modal distribution gets dominated by the macroscopic population of the fundamental mode:
\begin{eqnarray}
|\left<{\bm s}_{0}\right>| \gg |\left<{\bm s}_{p}\right>| \quad {\rm for} \quad p \neq 0,
\label{eq:s0_sp}
\end{eqnarray}
where $|\cdot|$ denotes the modulus of the vector.
According to (\ref{eq:s0_sp}) the sum over the modal Stokes vectors no longer vanishes, $\left<{\bm S}\right>=\sum_p\left<{\bm s}_{p}\right> \neq 0$. 
Note that the effective degree of polarization does not increase if ${\cal P}(z)$ is computed without including the contribution of the fundamental mode (grey points in Fig.~\ref{fig:dop_1}(c)), which confirms that the repolarization of the beam is a consequence of wave condensation.

\begin{figure}[]
\begin{center}
\includegraphics[width=1\columnwidth]{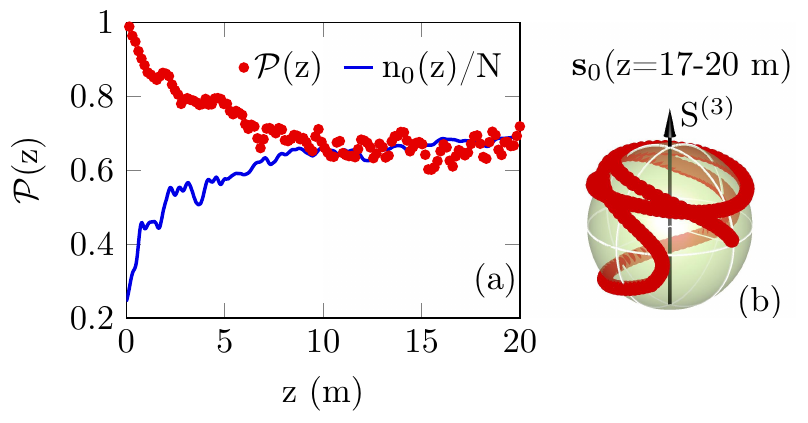}
\caption{
(a) Evolution during the propagation $z$ of the effective degree of polarization ${\cal P}$ (red points), and condensate fraction $n_0(z)/N$ (blue line).
The initial condition is a polarized coherent optical beam, i.e., the Stokes vectors of the modal components are aligned with each other, leading to ${\cal P}(z=0)=1$.
During the propagation, the disorder misaligns the Stokes vectors leading to a partial depolarization of the beam. 
Then the condensation-induced repolarization effect leads to a saturation of the decrease of ${\cal P}(z)$.
(b) Trajectory of the Stokes vector ${\bm s}_{0}(z)$ on the Poincar\'e sphere over the propagation length 17m to 20m, showing  an effect of nonlinear polarization rotation.
Parameters: $l_\beta=5$m, $2\pi/\sigma_\beta=25$m, $N=47.5$kW ($N_*=120$ modes, $a=26\mu$m). 
}
\label{fig:dop_2}
\end{center}
\end{figure}

Let us now consider the case where the launched optical beam is spatially coherent and polarized, as for typical experiments of beam self-cleaning.
The numerical simulation corresponding to this initial condition is reported in Fig.~\ref{fig:dop_2}(a) for a moderate disorder $L_d \gtrsim L_{nl}$.
In this case, all of the Stokes vectors are initially aligned, so that ${\cal P}(z=0)=1$.
For short propagation lengths, the disorder scrambles the phase relation-ship among the modes, thus leading to a misalignment of the Stokes vectors and then to a partial depolarization of the beam.
Subsequently, the effective repolarization induced by condensation leads to a saturation of the decrease of ${\cal P}(z)$.

In this regime of beam cleaning where disorder effects are moderate $L_d \gtrsim L_{nl}$, the fundamental mode is macroscopically populated and its polarization dynamics is not significantly altered by disorder nor by the coupling to other modes.
As a consequence, the polarization dynamics of the fundamental mode exhibits a well-known effect of nonlinear polarization rotation on the Poincar\'e sphere \cite{agrawal_book}, as can be observed in Fig.~\ref{fig:dop_2}(b).
This effect is characterized by a rotation of the Stokes vector in the $(s_0^{(1)},s_0^{(2)})$ plane with an  ellipticity that can remain almost constant over some propagation lengths, $s_0^{(3)} =$const.
Note that this nonlinear polarization rotation of the self-cleaned optical beam has been observed experimentally in Ref.\cite{krupa_polar}.
To interpret their results, the authors of \cite{krupa_polar} invoke the key impact of the temporal profile of the optical pulse injected in the MMF.
Indeed, the frequency rotation of the Stokes vector depends on the instantaneous value of the power, so that different power levels of the injected pulse profile exhibit different rotation speeds of the Stokes vector.
Accordingly, a quantitative comparison between the experimental results of Ref.\cite{krupa_polar} and our numerical results is not possible without including temporal effects in our model, and specifically the temporal profile of the injected optical pulse.

\begin{figure}[]
\begin{center}
\includegraphics[width=.8\columnwidth]{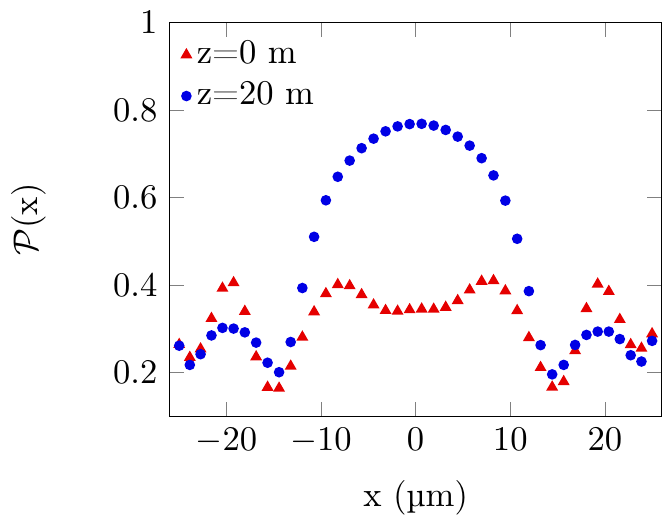}
\caption{
Polarization properties across the transverse surface section of the self-cleaned beam reported in Fig.~\ref{fig:dop_1}: 
Effective degree of polarization ${\cal P}$ computed along the $x-$axis by keeping fixed the $y-$position ($y=0$), at $z=0$m (red triangles), and $z=20$m (blue dots).
As a consequence of wave condensation, the central part of the beam exhibits a significant repolarization process.
Parameters: $l_\beta=30$cm, $2\pi/\sigma_\beta=2.1$m, $N=47.5$kW ($N_*=120$ modes, $a=26\mu$m). 
}
\label{fig:dop_3}
\end{center}
\end{figure}

We followed the experimental procedure of Ref.\cite{krupa_polar} by analyzing the polarization properties in the transverse spatial distribution of the optical beam.
The effective degree of polarization ${\cal P}(x,y=0)$ reported in Fig.~\ref{fig:dop_3} is computed by performing a spatial integration over a diaphragm of same diameter as the fundamental mode of the MMF (9$\mu$m), which is moved across the beam along the $x-$axis by keeping fixed the $y=0$ position.
As expected, the transverse profile of the degree of polarization exhibits a bell-shaped profile indicating a significant repolarization of the self-cleaned beam.
This confirms that the effective repolarization of the beam discussed here results from the process of condensation and the associated macroscopic population of the fundamental mode of the MMF.

\section{Freezing and slowing-down of thermalization}
\label{sec:discussion}

In this section we discuss an important consequence of the discrete nature of wave turbulence in MMFs, namely an effective freezing of the process of thermalization and condensation, which should also explain why optical beam cleaning has not been observed in step-index MMFs. 
In addition, we comment the impact on the turbulent dynamics of a perturbation of the dispersion relation, which is shown to modify the regularization of wave resonances and then the rate of thermalization of the optical beam.

\subsection{Absence of beam-cleaning in step-index fibers: Freezing of thermalization }
\label{sec:step_index}

The theory developed in this article can be applied in principle to different waveguide geometries, such as  step-index MMFs that are characterized by a homogeneous circular potential $V(r)$.
This example is important in that the effect of optical beam self-cleaning has not been observed in step index MMFs.

To discuss this experimental observation, we first note that at variance with a GRIN MMF, in a step-index MMF the eigenvalues $\beta_p$ are not equally spaced and degenerate modes are scarce.
We recall in this respect that thermalization takes place through the excitation of higher-order modes, a feature that essentially occurs by ``nontrivial" resonances, i.e., resonances that involve at least three (or four) non-degenerate modes.
It is important to note that, at variance with a GRIN fiber, in step-index MMF such nontrivial resonances are {\it not exact resonances} verifying $\Delta \omega_{lmnp}=0$.
According to our discussion on discrete vs continuous wave turbulence (see section \ref{sec:cont_wt}-\ref{sec:disc_wt}), this means that incoherent light propagation in step-index MMFs should be described by quasi-resonances through a continuous wave turbulence approach.
However, as will be shown below quasi-resonances are poorly efficient in actual step-index MMFs and many of them verify $|\Delta \omega_{lmnp}| > 1/L_{nl}$, i.e., the frequency mismatch is too large to provide a non-vanishing contribution to the kinetic equation. 

\begin{center}
\begin{tabular}{|c|c|c|c|c|}
\hline
             & $N_{\rm res}^{(1)}$  & $\chi_{\rm eff}^{(1)}$  & $N_{\rm res}^{(2)}$    & $\chi_{\rm eff}^{(2)}$  \\
\hline
Step-index   & 1 376             & 4.7                    & 12 256                & 15                   \\
\hline
GRIN         & 8 369 504        & 630                      & 8 369 504        & 630                    \\
\hline
\end{tabular}
\captionof{table}{Number of quasi-resonances $N_{\rm res}^{(1,2)}$ and corresponding efficiencies $\chi_{\rm eff}^{(1,2)}=\sum_{lmnp} S_{lmnp}^2$ for a step-index MMF for $L_{nl}=1$m and for $L_{nl}=25$cm, respectively.
For a GRIN MMF, $N_{\rm res}$ denote the number of exact resonances.
The significant reduction of  $N_{\rm res}^{(j)}$ and $\chi_{\rm eff}^{(j)}$ for the step-index MMF with respect to the GRIN MMF is responsible for an effective freezing of the process of thermalization and condensation.
}
\label{tab:step_grin}
\end{center}

We illustrate this by comparing the number of non-trivial resonances $N_{\rm res}$ for a GRIN and a step-index MMF -- more specifically we compute the number of exact resonances for the GRIN fiber ($\Delta \omega_{lmnp}=0$), and the number of quasi-resonances for the step-index fiber $|\Delta \omega_{lmnp}| \ll L_{nl}^{-1}$.
The resonances are non-trivial in the sense that they involve at least three different groups of non-degenerate modes. 
We considered fibers with approximatively the same number of modes: $N_*=120$ for the GRIN ($n_0=1.47$, $n_1=1.457$, $a=26\mu$m), $N_*=121$ for the step-index fiber ($n_0=1.4496$, $n_1=1.4462$, $a=37\mu$m).
We also considered different  values of the nonlinear length $L_{nl}=25$cm and $L_{nl}=1$m, and computed the number of quasi-resonances in the step-index fiber with the criterion $|\Delta \omega_{lmnp}| \le L_{nl}^{-1}/10$.
The results are reported in Table~\ref{tab:step_grin}, which show a drastic reduction of the number of resonances and corresponding efficiencies in the step-index fiber as compared to the GRIN fiber.

To summarize, incoherent light propagation in a step-index MMF is not described by a discrete turbulence regime because of the absence of exact non-trivial resonances.
The step-index MMF also exhibits poorly efficient quasi-resonances that essentially freeze the development of a continuous turbulence regime.
Note that the number of quasi-resonances contributing to the kinetic equation can be increased by considering larger radii of step-index MMFs, thus leading to a reduced mode spacing.
In this case an efficient continuous turbulence regime can be established in principle (provided that sufficient power is launched in the fiber).
However, as discussed in section~\ref{sec:cont_wt}, in this case the impact of weak disorder is expected to prevent the conservation of kinetic energy, which would thus inhibit wave condensation.

\subsection{Freezing discrete turbulence and thermalization with specific initial conditions}
\label{sec:LP11}

The wave turbulence theory developed in this paper is relevant when the optical beam populates many modes of the MMF.
This is the case for instance for the experiments reported in \cite{PRL19}, in which the optical beam is passed through a diffuser to degrade its transverse wavefront profile before injection into the MMF.
It has been shown in this case that, by increasing the excitation of modes, the effect of beam self-cleaning is degraded, as described by wave condensation and the equilibrium condensation curve, see \cite{PRL19}.
It is important to stress, however, that there exist particular conditions of beam injection into the MMF that can excite only few modes of the fiber, as it has been recently reported in Refs.\cite{deliancourt_lp11,deliancourt_lpmn}. 
Consider for instance the case where a Gaussian beam with a radius comparable to that of the fundamental mode is injected at perfect normal incidence and exactly at the center of the MMF. 
In this case only few radial modes are excited, while all modes featured by a non-homogeneous azimuthal profile are not excited at all.
Another important example reported in Ref.~\cite{deliancourt_lp11} is the excitation with a Gaussian beam whose incident external angle is adjusted around 2.5 degrees, in order to excite the fiber beyond the numerical aperture of the fundamental mode. 
In this way, the amount of power coupled into the fundamental mode is limited, while a high fraction of power results to be coupled into the LP$_{11}$ mode.
With this specific initial condition, the experiments in Ref.~\cite{deliancourt_lp11} reported a remarkable effect of beam self-cleaning upon power on this preferentially excited mode LP$_{11}$ mode.

This latter observation may appear at first sight in contradiction with the effect of condensation and thermalization to the RJ distribution.
However, as discussed all along this paper, incoherent light propagation in MMFs is described by the discrete wave turbulence regime that is dominated by exact resonances.
Considering the small number of modes excited with the above specified initial conditions, we shall see that the processes of thermalization and condensation result to be essentially frozen. 
Indeed, following the discussion about the absence of beam cleaning in step-index MMFs (section~\ref{sec:step_index}), here the same argument of freezing of thermalization should explain why condensation of power into the fundamental mode is not observed with the specific tilted injection favouring the excitation of the LP$_{11}$ mode \cite{deliancourt_lp11}.
In the same way, for perfect normal incidence injection of a small Gaussian beam, the numerical simulations do not evidence the establishment of a RJ equilibrium state featured by energy equipartition among the modes within the short fiber length used in the experiments.
This is a consequence of the discrete nature of the resonances manifold in MMFs, which  exhibits clusters of resonant mode interactions in relation with finite size effects in discrete wave turbulence 
\cite{nazarenko11,Kartashova98,Zakharov05,Nazarenko06,kartashova08,Kartashova09,Kartashova10,Lvov10,Harris13,
Harper13,Bustamante14,kuksin,Mordant18}.

\begin{center}
\begin{tabular}{|c|c|c|c|c|}
\hline
{\bf mode excitation} & $N_{\rm res}^{(1)}$  & $\chi_{\rm eff}^{(1)}$  & $N_{\rm res}^{(2)}$    & $\chi_{\rm eff}^{(2)}$  \\
\hline
(i)          & 18 380        & 7.4                    & 93 840        & 16                   \\
\hline
(ii)         & 11  600      & 10.7                    & 229 776      & 46                    \\
\hline
(iii)        & 96 840       & 67                      & 1 716 000       & 283                      \\
\hline
(iv)         & 922 944       & 210                     & 8 369 504       & 603                       \\
\hline
\end{tabular}
\captionof{table}{Number of resonances $N_{\rm res}^{(1,2)}$ and corresponding efficiencies $\chi_{\rm eff}^{(1,2)}=\sum_{lmnp} S_{lmnp}^2$ for two different values of kinetic energies $E^{(1)}/E_{\rm crit}=0.32$ (columns 2-3) and $E^{(2)}/E_{\rm crit}=0.54$ (columns 4-5).
The initial conditions (i) and (ii) refer to  particular injection conditions: (i) coherent Gaussian beam at perfect normal incidence; (ii) coherent Gaussian beam with a specific tilt angle favouring the excitation of the LP$_{11}$ mode (see Ref.\cite{deliancourt_lp11}). 
The initial conditions (iii) and (iv) refer to generic incoherent (speckle-like) beams with exponential (iii), Lorentzian (iv), spectral distributions in mode space.
The number of resonances and their efficiencies are considerably smaller for the particular excitations (i) and (ii) as compared to the generic states (iii)-(iv).
Since the energy $E^{(1)}$ (or $E^{(2)}$) is kept fixed, the corresponding `amount of disorder' is the same for all initial conditions (i)-(iv): 
The significant reduction of  $N_{\rm res}^{(j)}$ and $\chi_{\rm eff}^{(j)}$ for the particular states (i)-(ii) 
reflects the peculiar modal excitation due to the specific injection conditions into the MMF.
}
\label{tab:lp11}
\end{center}

We have computed the number of non-trivial resonances $N_{\rm res}$ (involving at least three groups of non-degenerate modes) and their corresponding efficiencies $\chi_{\rm eff}=\sum_{lmnp} S_{lmnp}^2$ for the particular initial conditions discussed above, namely: (i) Gaussian beam at normal incidence; (ii) Gaussian beam with a specific tilt angle (2.5 degrees) favouring the excitation of the LP$_{11}$ mode.
This computation has been realized for two different values of the kinetic energy $E^{(1)}$ and $E^{(2)}$, corresponding to a Gaussian beam with FWHM=2 FWHM$_0$ for $E^{(1)}$, and FWHM=3 FWHM$_0$ for $E^{(2)}$, FWHM$_0$ being the full-width-half-maximum of the fundamental mode.
The results are reported in Table~\ref{tab:lp11} for the values of the energies $E^{(1)}/E_{\rm crit}=0.32$ (columns 2-3) and $E^{(2)}/E_{\rm crit}=0.54$ (columns 4-5), where $E_{\rm crit}=NV_0(1+2\beta_0/V_0)/2$ denotes the critical energy of the transition to condensation \cite{PRA11b}.
We have compared these results with a generic incoherent (speckle-like) beam, which is characterized by an exponential or a Lorentzian spectral distribution in mode space (see (iii)-(iv) in Table~\ref{tab:lp11}).
A resonance is retained in the computation whenever the modal amplitudes are larger than some threshold value ($n_p \ge 0.1\%$ in Table~\ref{tab:lp11}, with $\sum_p n_p=1$). 
The main result is that $N_{\rm res}$ and  $\chi_{\rm eff}$ decrease in a substantial way for the particular initial conditions (i)-(ii) where a small Gaussian beam is injected either at perfect normal incidence or with a specific tilt angle favouring the excitation of the LP$_{11}$ mode.
It is important to stress that this comparison is realized for the {\it same kinetic energy $E$}, i.e., the same `amount of disorder' in the initial condition.
Accordingly, the significant reduction of $N_{\rm res}$ and  $\chi_{\rm eff}$ for the particular initial conditions (i)-(ii) reveals the specificity of such modal excitation as compared to the generic initial conditions (iii)-(iv).
We note that such specific initial conditions are commented by the authors of Refs.\cite{deliancourt_lp11,deliancourt_lpmn}, who state that beam-cleaning on the LP$_{11}$ mode requires ``a lengthy and difficult to exactly reproduce manual procedure, i.e., with a tricky adjustment of tilted launching conditions". 

To conclude this discussion, we have seen in section~\ref{sec:coherent_regime} that typical MMFs used in beam-cleaning experiments behave as a dynamical system with a limited number of degrees of freedom, in relation with the discrete nature of  wave turbulence in MMFs.
When many modes are excited through generic initial conditions in the presence of a significant amount of structural disorder ($L_d \ll L_{nl},z$), the system exhibits a well developed discrete turbulence regime that is accurately described by the discrete kinetic equation derived here.
On the other hand, for specific initial conditions characterized by the excitation of a small number of modes, then the discrete structure of the resonance manifold of MMFs can freeze the thermalization and thus the condensation processes within the short fiber lengths considered in the experiments (5-8m in \cite{deliancourt_lp11,deliancourt_lpmn}).
The derived kinetic equations should not be relevant to describe this regime of few-mode interaction over small propagation lengths.
In particular, the impact of disorder may be considered as perturbative in this regime where the system can exhibit a partially phase-sensitive coherent interaction, which may probably be described by 
the tools developed to study finite size effects in turbulence, such as discrete and mesoscopic wave turbulence and the associated clusters of resonant mode interactions 
\cite{nazarenko11,Kartashova98,Zakharov05,Nazarenko06,kartashova08,Kartashova09,Kartashova10,Lvov10,Harris13,
Harper13,Bustamante14,kuksin,Mordant18}.
Such a partially coherent few-mode interaction regime can be mastered by a fine tuning of the transverse profile of the laser excitation \cite{deliancourt_lp11,deliancourt_lpmn}.
In the recent experiments \cite{deliancourt_lpmn} beam cleaning of many low-order modes has been reported owing to an ingenious feedback-induced adaptive profiling of the transverse wavefront phase of the coherent beam excitation (also see \cite{silberberg17}).
More precisely, owing to a feedback loop, the  transverse wavefront phase of the injected beam is adjusted by an iterative procedure so as to force the beam to convergence toward a pre-established target mode at the fiber output.
This adaptively controlled cleaning of low-order modes is of different nature than the spontaneous phenomenon of condensation on the fundamental mode resulting from the natural process of thermalization toward the equilibrium distribution.


\subsection{Corrections of the dispersion relation}
\label{sec:b_p}

The wave turbulence approach developed in this article is based on the NLS equation with an ideal parabolic potential and the (linear) dispersion relation $\beta_p=\beta_0(p_x+p_y+1)$.
Several factors introduce perturbations to this expression of the dispersion relation, which we write in the form ${\tilde \beta}_p=\beta_p + b_p$, where the perturbation $b_p$ is a function of $(p_x,p_y)$ with $b_0/\beta_0 \ll 1$.
An example of perturbation is provided by the well known fact that a GRIN MMF usually exhibits deviations from the ideal parabolic shape, i.e., $V(\br) \sim |\br|^{\nu}$ with an exponent that deviates from $\nu=2$.
The general expression of the eigenvalue is rather complicated and of the form ${\tilde \beta}_p \propto (1+p_x+p_y)^{2\nu/(\nu+2)}$ \cite{kahn09}.
Considering a deviation of a few percents from $\nu=2$ \cite{kahn05}, one has $b_0/\beta_0 \sim 5\times 10^{-3}$ with usual parameters of beam cleaning experiments.
An other example is the leading order correction due to angular dispersion effects in the Helmholtz equation, $b_p=(\beta_0^2/(2 k_0 n_0))(1+p_x+p_y)^{2}$, which gives $b_0/\beta_0 \sim 2\times 10^{-4}$ 
with usual beam-cleaning experimental parameters.
In addition, we can notice that the truncation of the parabolic refractive index profile due to the presence of the fiber cladding introduces significant perturbations of the higher-order eigenvalues (see Fig.~5 in Ref.\cite{PRA11b}). 
It is also important to note that the standard deviation of the fluctuations of the structural disorder of the MMF due to imperfections and external perturbations (term ${\bf D}_p {\bm A}_p$ in the modal NLS Eq.(\ref{eq:nls_Ap})) may be of the same order as the correction of the dispersion relation.
In this respect, we also recall that a bending of the fiber introduces a correction in the propagation constant of the order $\simeq a/R_c$, where $a$ is the fiber radius and $R_c$ the radius of the bending ($a/R_c \simeq 10^{-4}$ with $a\simeq25\mu$m and $R_c\simeq 25$cm).

Let us discuss on the possible impact of perturbations of the dispersion relations.
In this respect, resonances that are exact at leading order ($\Delta \omega_{lmnp} =\beta_l+\beta_m-\beta_n-\beta_p=0$) exhibit a residual non-resonant contribution, i.e., $\Delta {\tilde \omega_{lmnp}} = {\tilde \beta}_l+{\tilde \beta}_m-{\tilde \beta}_n-{\tilde \beta}_p=\Delta b_{lmnp}$ with  $\Delta b_{lmnp}=b_l+b_m-b_n-b_p$.
We derive the kinetic equation accounting for the correction on the dispersion relation with the above assumption  $L_d = 1/\Delta \beta \ll L_{nl} < L_{kin}^{disor}$.
The result of the convolution integral Eq.(\ref{eq:conv_int}) is approximated by 
\begin{eqnarray}
{J}_{lmnp}^{(j)} \simeq \gamma \left< {Y}_{lmnp}^{(j)}\right> 
\frac{i8\Delta \beta-\Delta b_{lmnp}}{\Delta b_{lmnp}^2+(8\Delta \beta)^2} \delta^K(\Delta \omega_{lmnp}).
\label{eq:I_lmnp_damp_Bp}
\end{eqnarray}
Proceeding as in section~\ref{sec:weak_disorder}, we obtain the discrete kinetic equation 
\begin{eqnarray}
\nonumber
\partial_z n_p(z) &=&  \frac{4 \gamma^2 \overline{\Delta \beta}}{3} \sum_{l,m,n} \frac{ \delta^K(\Delta \omega_{lmnp})}{\Delta b_{lmnp}^2+\overline{\Delta \beta}^2}   |S_{lmnp}|^2 M_{lmnp}({\bm n}) \quad  \\
&&
+  \,  \frac{32\gamma^2 \overline{\Delta \beta}}{9}  \sum_l  \frac{ \delta^K(\Delta \omega_{lp})}{\Delta b_{lp}^2+\overline{\Delta \beta}^2}
 |  s_{lp}({\bm n}) |^2 (n_l-n_p)  \quad 
\label{eq:kin_np_dsp_corr}
\end{eqnarray}
where we recall that $s_{lp}({\bm n})=\sum_{m'} S_{lm'm'p} n_{m'}$, $M_{lmnp}({\bm n})=  n_l n_m n_p+n_l n_m n_n -  n_n n_p n_m -n_n n_p n_l$, with $\Delta b_{lp}=b_l-b_p$ and $\overline{\Delta \beta}=8\Delta \beta$.
As already commented through Eq.(\ref{eq:kin_contin_dis}), the Lorentzian distribution reflects the finite bandwidth of the four-wave resonances due to the effective dissipation $\Delta \beta$. 
Accordingly, the kinetic Eq.(\ref{eq:kin_np_dsp_corr}) conserves $E=\sum_p \beta_p n_p(z)$,  but not ${\tilde E}=\sum_p {\tilde \beta}_p n_p(z)$.
In addition, the conservation of the power $N=\sum_p n_p(z)$ and the $H-$theorem of entropy growth for ${\cal S}(z)=\sum_p \log\big(n_p(z)\big)$ describe a relaxation to $n_p^{eq}=T/(\beta_p-\mu)$, i.e., the same equilibrium as in the absence of the correction on the dispersion relation ($b_p=0$). 
The main difference is that in the regime $b_0 \sim \Delta \beta$, the correction $b_p$ leads to a deceleration of the rate of thermalization and condensation.
On the other hand, in the regime $\Delta \beta \gg b_0$, the Lorentzian distribution can be simplified 
$\overline{\Delta \beta}/\big(\Delta b_{lmnp}^2+\overline{\Delta \beta}^2 \big)
\rightarrow 1/\overline{\Delta \beta}$,  
and the kinetic Eq.(\ref{eq:kin_np_dsp_corr}) exactly recovers the previous  kinetic Eq.(\ref{eq:kin_np_disc}), i.e. the correction of the dispersion relation $b_p$ does not affect the rate of thermalization.

\begin{figure}[]
\begin{center}
\includegraphics[width=1\columnwidth]{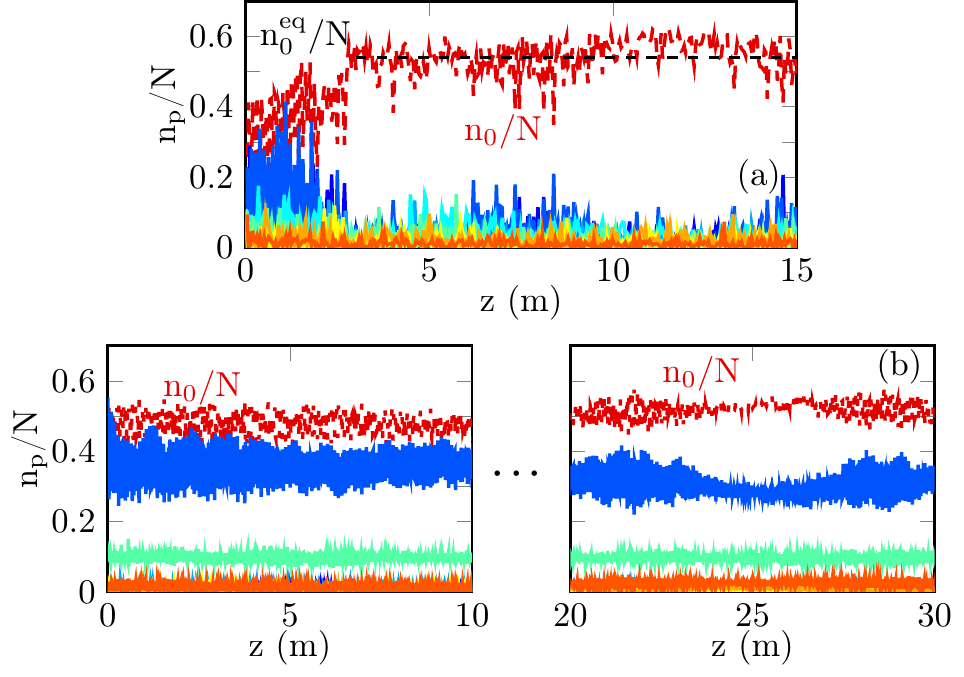}
\caption{
Numerical simulations of the `continuous' NLS Eq.(\ref{eq:nls}) showing the evolutions of the modal populations $n_p/N$: 
fundamental mode $p=0$ (red dashed), $p=1$ (dark blue solid), $p=2$ (blue solid), $p=3$ (light blue solid), $p=4$ (cyan solid), $p=5$ (light green solid), $p=6$ (green solid), $p=7$ (yellow solid), $p=8$ (orange solid).
(a) Starting from a `speckle' beam with random phase among the modes, the beam exhibits a condensation process: $n_0(z)/N$ relaxes to the theoretical equilibrium value $n_0^{eq}/N \simeq 0.54$ (dashed black line from Eq.(\ref{eq:n_eq_vs_E})).
(b) Starting from a coherent initial condition, the modal components exhibit an oscillatory behavior, similar to that observed for the discretized modal NLS Eq.(\ref{eq:modes_A_00}), see Fig.~\ref{fig:0}(a).
At complete thermal equilibrium, the fundamental mode would reach the condensate fraction $n_0^{eq}/N \simeq 0.8$.
The power is $N=47.5$kW ($N_*=66$ modes, $a=15 \mu$m).}
\label{fig:contin_nls}
\end{center}
\end{figure}

Let us now discuss the impact of a correction on the dispersion relation in the absence of disorder.
This situation is relevant for direct numerical simulations of the (continuous) NLS Eq.(\ref{eq:nls}), where the truncation of the parabolic potential $V(\br)$ introduces a perturbation on the dispersion relation for higher-order modes (see Fig.~5 in Ref.\cite{PRA11b}).
At variance with simulations of the (discretized) modal NLS Eq.(\ref{eq:modes_A_00}) without disorder that do not evidence wave condensation even for large propagation lengths (see Fig.~\ref{fig:0} and the thermalization length scale $L_{kin}^{ord}\sim \beta_0 L_{nl}^2/{\bar S^2}_{lmnp}$), the simulations of the NLS Eq.(\ref{eq:nls}) show a rather rapid condensation process in the presence of an initial condition with random phases among the modes, as illustrated in Fig.~\ref{fig:contin_nls}.
This rapid condensation can be explained by the perturbation on the dispersion relation due to the truncation of the parabolic potential, which modifies the regularization of wave resonances by removing the mode degeneracies, $L_{kin}^{pert} \sim b_p L_{nl}^2/{\bar S^2}_{lmnp}$, see Appendix~B.
However, at variance with experiments of beam cleaning, this estimation of $L_{kin}^{pert}$ assumes that the initial condition exhibits random phases among the modes, a feature of fundamental importance for the applicability of the wave turbulence theory \cite{nazarenko11,chibbaro17,chibbaro18}.
In addition, this estimation assumes that the perturbation $b_p$ is maintained fixed, while in a real experiment such a perturbation may not be controlled all over the fiber length due to inherent imperfections and external perturbations, and the corresponding fluctuations contribute to the disorder term in the NLS equation, as discussed through the kinetic Eq.(\ref{eq:kin_np_dsp_corr}) here above.

We finally notice that the refractive index profile of the MMF can be tailored to optimize mode properties.
For instance, one can generalize the GRIN parabolic profile by using exponents strongly deviating from 2, the higher that exponent, the closer the profile would be to a step-index profile.
Since the effect of beam cleaning does not occur in step-index fibers (see section~\ref{sec:step_index}), it would be interesting to study the degradation of wave condensation in the discrete turbulence regime by gradually increasing the exponent of the refractive index profile, in line with the experiments initiated  in \cite{dupiol18}.


\section{Conclusion}
\label{sec:conclusion}

In summary, we have discussed experiments of beam self-cleaning where long ($\sim$ns) pulses are injected in relative short multimode fiber lengths ($\sim 10$m), for which the dominant contribution of disorder originates from polarization random fluctuations (weak disorder) \cite{ho14}.
On the basis of the wave turbulence theory, we have derived kinetic equations describing the nonequilibrium evolution of random waves in a regime where disorder dominates nonlinear effects ($L_d \ll L_{nl}$).
The theory revealed that the presence of a conservative weak disorder introduces an effective dissipation in the system whose resonance broadening prevents the conservation of the energy, which inhibits the effect of condensation in the usual continuous wave turbulence approach.
On the other hand, we have shown that usual experiments of beam-cleaning are not described by the continuous wave turbulence theory, but instead by a discrete wave turbulence approach.
In this discrete turbulence regime only exact resonances contribute to the kinetic equation, which is no longer sensitive to the effect of dissipation-induced resonance broadening. 
Accordingly, the discrete kinetic equation conserves the energy, which re-establishes the process of wave condensation.
The main result is that the effective dissipation induced by disorder modifies the regularization of such discrete resonances, which leads to an acceleration of the rate of thermalization and condensation.

In order to improve our understanding of beam cleaning experiments, we have considered different models of weak disorder in MMFs.
The theory shows that when all  modes experience the same (mode-correlated) noise, the dissipation induced by disorder vanishes and the system no longer exhibits a fast process of condensation. 
However, even a relative small decorrelation among the noise experienced by the modes is sufficient to re-establish a disorder-induced acceleration of condensation. 
The simulations are in quantitative agreement with the theory without using adjustable parameters in the regime where disorder dominates nonlinear effects ($L_d \ll L_{nl}$). 
However, the impact of weak disorder due to polarization fluctuations in beam cleaning experiments is expected to be of the same order as nonlinear effects \cite{ho14}.
We have thus considered the impact of a moderate disorder ($L_d \gtrsim L_{nl}$). 
In this regime, the phase-correlations among the modes are not completely suppressed and the system enters a mixed coherent-incoherent regime. 
Accordingly, the modal components exhibit an oscillatory behavior that slows down the thermalization process, though the optical beam still exhibits a fast condensation process that is consistent with the experimental results reported in Ref.\cite{PRL19}. 
In addition, the analysis of polarization effects revealed that optical beam cleaning is responsible for an effective partial repolarization of the central part of the beam. 
This property was observed experimentally in Ref.\cite{krupa_polar} and it can be interpreted as a natural consequence of the condensation-induced macroscopic population of the fundamental mode of the MMF. 
To summarize, by considering the dominant contribution of weak disorder originating in polarization fluctuations, our theory and simulations provide a qualitative understanding of the effect of optical beam self-cleaning, in particular when a large number of modes are excited into the MMF as in the recent experiments reported in Ref.\cite{PRL19}.
On the other hand, the discrete nature of wave turbulence in MMFs is responsible for a freezing of thermalization and condensation when a small number of modes are excited.
Such an effective freezing of condensation also explains why optical beam cleaning has not been observed in step-index MMFs.

It is important to recall that, at variance with the experiments of beam cleaning where sub-nanosecond pulsed are injected into the MMF, we considered in this work a purely spatial NLS equation to model the propagation of the beam through the fiber. 
This is justified by the fact that in the nanosecond regime temporal dispersion effects can be neglected in first approximation. 
Then although our theory and simulations describe the essential mechanism underlying beam cleaning condensation, it is clear that they cannot provide a quantitative description of the experimental results, which would require a detailed analysis of the temporal averaging effect inherent to the pulsed regime considered in the experimental measurements.

We have also considered the impact of strong disorder at the leading order linear regime, which revealed that random mode coupling among degenerate modes enforces the thermalization toward the Rayleigh-Jeans equilibrium distribution.
We note in this respect that, according to  the  simulations reported in \cite{pod19}, strong disorder only weakly affects the evolutions of the modal components as compared to the significant impact of weak disorder discussed in this work.
Nevertheless, it would be important to extend the theory developed for weak disorder to strong disorder in order to study the interplay of random mode coupling and nonlinearity, a feature that will be considered in future works.


From a broader perspective, the wave turbulence theory discussed in this work can be extended to study spatio-temporal effects in MMFs
 \cite{wrightOE15,renninger13,wright15_0,wright15,liu16,krupa16,wright16,wright17,laegsgaard18,turitsyn18,conforti18}.
The development of a spatio-temporal theory would also be important to study complex incoherent behaviors, such as the formation of incoherent shocks \cite{PRL13,NC15}, or the generation of supercontinuum radiation in MMFs \cite{wright15_0,christodoulides_oe17,krupa_ol16,lopez_ol16}, in relation with the wave turbulence kinetic approach developed to study supercontinuum generation in single-mode fibers \cite{PRA09,PRA13,PRA17}.
The discrete wave turbulence approach discussed in this work can also be extended to study the impact of disorder on turbulence cascades \cite{nazarenko11,Newell_Rumpf,proment12,fisch18}, from both the theoretical and experimental points of views.
Extension of the theory to consider other forms of nonlinear effects would also be interesting, such as saturable or nonlocal nonlinearities in relation with atomic vapors beam cleaning experiments \cite{labeyrie03}.
In addition, there is a growing interest for experimental demonstrations of superfluid light flows in bulk materials \cite{vocke16,michel18,glorieux18}.
The experiments of condensation in MMFs would allow to study the nucleation of superfluid vortices induced by a rotating confining potential (along the `time' $z-$variable) in manufactured multimode fibers, in analogy with rotating trapped BECs \cite{stringari}.

The acceleration of the process of thermalization reported in this work is also relevant to the notion of prethermalization to out of equilibrium states \cite{berges04,langen16,larre16,larre18,PRL18}, which is attracting a growing interest in different research communities, including long range interacting systems with fast relaxation towards quasi-stationary states~\cite{Ruffo2009,Morigi2014,PRL11}, or one dimensional (nearly) integrable (quantum) systems \cite{OE11,randoux16,Schmiedmayer2012,Schmiedmayer2016,gasenzer19}.
From a broader perspective, the present work can contribute to the challenging question of spontaneous organization of coherent states in nonlinear disordered (turbulent) systems \cite{silberberg08,segev,cherroret15,leonetti14,schirmacher18,ermann15,delre17}, in relation with the paradigm of statistical light-mode dynamics (glassy behaviors) and complexity in random lasers \cite{conti11,leuzzi15}.

\section{Acknowledgements}

The authors are grateful to A. Tonello, P. B\'ejot, S. Gu\'erin, V. Couderc, and A. Barth\'elemy for fruitful discussions.
We acknowledge financial support from the French ANR under Grant No. ANR-19-CE46-0007 (project ICCI),
iXcore research foundation, EIPHI Graduate School (Contract No. ANR-17-EURE-0002), French program ``Investissement d'Avenir," Project No. ISITE-BFC-299 (ANR-15 IDEX-0003);  H2020 Marie Sklodowska-Curie Actions (MSCA-COFUND) (MULTIPLY Project No. 713694). Calculations were performed using HPC resources from DNUM CCUB (Centre de Calcul, Universit\'e de Bourgogne).


\bigskip


\begin{widetext}
\section{Appendix A: Derivation of the discrete kinetic equation with weak disorder}
\label{app:weak_disorder}

To study correlations among the modes, we derive an equation for the moments of the $2\times 2$ matrix $\left<{\bm A}_p^* {\bm A}_q^T\right>(z)$.
The computation for distinct modes ($p \neq q$) was reported in the Supplemental of Ref.\cite{PRL19}.
The computation for the correlations within a specific mode ($p=q$) is more delicate and it is detailed here below.

\subsection{Modal correlations}
\label{appA1}

The second moments satisfy:
\begin{eqnarray*}
\partial_z \left< {\bm A}_p^* {\bm A}_p^T\right> = i \left< {\bf D}_p^* {\bm A}_p^* {\bm A}_p^T\right>
-i \left< {\bm A}_p^* {\bm A}_p^T {\bf D}_p^T \right> 
- i\gamma \big< {\bm G}_{pp}({\bm A})\big>  .
\end{eqnarray*}
where ${\bm G}_{pq}({\bm A}(z))={\bm P}_p({\bm A})^* {\bm A}_q^T(z)
- {\bm A}_p^* {\bm P}_q({\bm A})^T (z)$.
According to the Furutsu-Novikov theorem:
\begin{eqnarray*}
 \left<  \nu_{p,j} \bsigma_j^* {\bm A}_p^* {\bm A}_p^T\right> &=& 
 \int_0^z \left< \frac{\delta \big( \bsigma_j^* {\bm A}_p^* {\bm A}_p^T(z) \big)}
 {\delta \nu_{p,j}(z') } \right>   \sigma^2_\beta  {\cal R}\Big( \frac{z-z'}{l_\beta}\Big) dz'
\end{eqnarray*}
The variational derivative can be computed by following \cite{konotop}.
For $z>z'$ it is the solution of 
\begin{eqnarray}
\nonumber
\partial_z  \frac{\delta \big( {\bm A}_p^* {\bm A}_p^T(z) \big)}
 {\delta \nu_{p,j}(z') } = i  {\bf D}_p^*  \frac{\delta \big( {\bm A}_p^* {\bm A}_p^T(z) \big)}
 {\delta \nu_{p,j}(z') } -i  \frac{\delta \big( {\bm A}_p^* {\bm A}_p^T(z) \big)}
 {\delta \nu_{p,j}(z') }  {\bf D}_p^T 
 - i\gamma \frac{\delta\big( {\bm G}_{pp}({\bm A}(z))\big)}{\delta \nu_{p,j}(z') }   
\label{eq:varder1}
\end{eqnarray}
starting from
$$
 \frac{\delta \big( {\bm A}_p^* {\bm A}_p^T(z) \big)}
 {\delta \nu_{p,j}(z') }  \mid_{z=z'}= 
 i \bsigma_j^*  {\bm A}_p^* {\bm A}_p^T(z) - i {\bm A}_p^* {\bm A}_p^T(z) \bsigma_j^T.
$$
We need to know the form of the variational derivative for  $z'<z$ and  $|z-z'| = O(l_\beta)$.
As $\sigma_\beta l_\beta \ll1 $ and $l_\beta \ll L_{nl}$,  all terms in the right-hand side of the 
differential equation (\ref{eq:varder1}) satisfied by the variational derivative are negligeable for $|z-z'| = O(l_\beta)$,
so that the leading-order expression of the  variational derivative  for $z'<z$ and  $|z-z'| = O(l_\beta)$ is
\begin{eqnarray}
 \frac{\delta \big(  {\bm A}_p^* {\bm A}_p^T(z) \big)}
 {\delta \nu_{p,j}(z') } = i \bsigma_j^*  {\bm A}_p^* {\bm A}_p^T(z') 
-i   {\bm A}_p^* {\bm A}_p^T(z') \bsigma_j^T,
\end{eqnarray}
and therefore
\begin{eqnarray*}
 \left<  \nu_{p,j} \bsigma_j^* {\bm A}_p^* {\bm A}_p^T\right> = 
 i 
 \int_0^z \left< \bsigma_j^*\bsigma_j^*  {\bm A}_p^* {\bm A}_p^T(z') \right>   \sigma^2_\beta  {\cal R}\Big( \frac{z-z'}{l_\beta}\Big) dz' 
-i 
 \int_0^z \left< \bsigma_j^* {\bm A}_p^* {\bm A}_p^T(z') \bsigma_j^T \right>   \sigma^2_\beta  {\cal R}\Big( \frac{z-z'}{l_\beta}\Big) dz'.
\end{eqnarray*}
For $j=0$ this is zero and for $j=1,2,3$ this can be approximated by (using $l_\beta \ll z$):
\begin{eqnarray*}
 \left<  \nu_{p,j} \bsigma_j^* {\bm A}_p^* {\bm A}_p^T\right> &=& 
\frac{ i \Delta \beta }{2} 
\big(
\left< \bsigma_j^*\bsigma_j^*  {\bm A}_p^* {\bm A}_p^T \right>  
-  \left< \bsigma_j^* {\bm A}_p^* {\bm A}_p^T  \bsigma_j^T \right> \big) 
,
\end{eqnarray*}
and we find
\begin{eqnarray*}
\partial_z \left< {\bm A}_p^* {\bm A}_p^T\right> = -  
 \Delta \beta 
 \Big( 3  \left< {\bm A}_p^* {\bm A}_p^T \right> 
-\sum_{j=1}^3 \bsigma_j^* \left< {\bm A}_p^* {\bm A}_p^T \right> \bsigma_j^T \Big)     
- i\gamma \big< {\bm G}_{pp}({\bm A})\big>  .
\end{eqnarray*}
The mean $2\times 2$ matrix $\left< {\bm A}_p^* {\bm A}_p^T\right> $ is Hermitian and therefore it can be expanded as
$$
\left< {\bm A}_p^* {\bm A}_p^T\right> =  w_p(z) \bsigma_0 + \sum_{j=1}^3 w_{p,j}(z) \bsigma_j .
$$
The real-valued functions $w_p$ and $(w_{p,j})_{j=1}^3$ satisfy
\begin{eqnarray}
\partial_z w_p &=& - \gamma \bPi_0 \big\{\big< i {\bm P}_p({\bm A})^* {\bm A}_p^T
- i {\bm A}_p^* {\bm P}_p({\bm A})^T \big> \big\},
\label{eq:wp}\\
\partial_z w_{p,j} &=& 
- 4\Delta \beta   w_{p,j} 
- \gamma \bPi_j \big\{\big< i {\bm G}_{pp}({\bm A}) \big>\big\}, 
\label{eq:wpj}
\end{eqnarray}
where $\bPi_j {\bf W}=$ coefficient of the decomposition of the Hermitian matrix ${\bf W}$ on $\bsigma_j$. In particular, $\bPi_0 {\bf W} = {\rm Tr}({\bf W})/2 $ so that
$$
\bPi_0 \big\{\big< i {\bm G}_{pp}({\bm A})\big> \big\} 
= - {\rm Im} \big<   {\bm P}_p({\bm A})^\dag {\bm A}_p \big>   .
$$
The coefficients $w_{p,j} $  for $j=1,2,3$ satisfy the damped equations
$$
\partial_z w_{p,j} =
- 4\Delta \beta   w_{p,j} 
- \gamma \bPi_j \big\{\big< i {\bm G}_{pp}({\bm A}) \big>\big\}.
$$
They are of the form
\begin{eqnarray*}
w_{p,j}(z) = w_{p,j}(0) \exp\big(-4 \Delta \beta   z\big) 
- \gamma \int_0^z\exp\big(-4 \Delta \beta   (z-z')\big) 
 \bPi_j \big\{\big< i {\bm G}_{pp}({\bm A}) \big>\big\}(z') dz'  .
\label{express:wpj1}
\end{eqnarray*}
As we have  $z,L_{nl} \gg 1/\Delta \beta$, 
the initial condition is forgotten in (\ref{express:wpj1}) and the second term in the right-hand side can be simplified
and we get for $j=1,2,3$ 
\begin{eqnarray}
w_{p,j}
&=& - \frac{ \gamma}{4 \Delta \beta}  
\bPi_j \big\{  \big< i {\bm G}_{pp}({\bm A}(z)) \big>\big\} .
\label{eq:w_pj_0}
\end{eqnarray}
Using the assumption $L_d=1/\Delta \beta \ll L_{nl}$,  we have
\begin{equation}
w_{p,j} \simeq 0
\label{eq:mom2pq2b}
\end{equation}
to leading order.

\subsection{Closure of the moments equations}
\label{appA2}

The diagonal modal components $w_p$ in (\ref{eq:wp}) satisfy the undamped equation
\begin{eqnarray*}
\partial_z w_p = 
\gamma {\rm Im} \big<   {\bm P}_p({\bm A})^\dag {\bm A}_p \big>  .
\end{eqnarray*}
With the expression (\ref{def:nlterm0}) of the nonlinear term, we get that the mode occupancies $w_p$
satisfy the coupled equations (\ref{eq:pzwp1}-\ref{def:Xp2}).
We now derive the equations governing the evolutions of the fourth-order moments $\left< X_p^{(1)}\right> $
and $\left<  {X}_p^{(2)}\right>$ given by Eqs.(\ref{def:Xp1}-\ref{def:Xp2}).

\subsubsection{Computation of the term $\left<  {X}_p^{(1)}\right> $:}

We first write the equation satisfied by the product of four vector fields:
\begin{eqnarray}
\partial_z ({\bm A}_l^\dag {\bm A}_m^*) ( {\bm A}_n^T {\bm A}_p)&=&
i(\beta_l+\beta_m-\beta_n-\beta_p) ({\bm A}_l^\dag {\bm A}_m^*) ({\bm A}_n^T {\bm A}_p)
+ i ({\bm A}_l^\dag {\bf D}_l^\dag {\bm A}_m^*) ( {\bm A}_n^T {\bm A}_p)\\
\nonumber
&&
+ i ({\bm A}_l^\dag {\bf D}_m^* {\bm A}_m^*) ( {\bm A}_n^T {\bm A}_p)
- i ({\bm A}_l^\dag {\bm A}_m^*) ( {\bm A}_n^T {\bf D}_n^T {\bm A}_p)
- i ({\bm A}_l^\dag {\bm A}_m^*) ( {\bm A}_n^T {\bf D}_p {\bm A}_p)\\
\nonumber
&&
+ i \gamma {Y}_{lmnp}^{(1)} ,  
\label{eq:4f_Xp1}\\
\nonumber
{Y}_{lmnp}^{(1)}&=& -\sum_{l',m',n'} S_{l'm'n'l}^* \Big[ \frac{1}{3} ({\bm A}_{l'}^\dag {\bm A}_{m'}^*) ({\bm A}_{n'}^T {\bm A}_m^*)
+\frac{2}{3}({\bm A}_{n'}^T {\bm A}_{m'}^*) ({\bm A}_{l'}^\dag {\bm A}_m^*)\Big]  ({\bm A}_n^T{\bm A}_p) \\
\nonumber
&&-\sum_{l',m',n'} S_{l'm'n'm}^* \Big[ \frac{1}{3} ({\bm A}_{l'}^\dag {\bm A}_{m'}^*) ({\bm A}_{l}^\dag {\bm A}_{n'})
+\frac{2}{3}({\bm A}_{n'}^T {\bm A}_{m'}^*) ( {\bm A}_{l}^\dag {\bm A}_{l'}^*)\Big]  ({\bm A}_n^T {\bm A}_p) \\
\nonumber
&&+\sum_{l',m',n'} S_{l'm'n'n} \Big[ \frac{1}{3}  ({\bm A}_{l'}^T {\bm A}_{m'})
({\bm A}_{n'}^\dag {\bm A}_p)
+\frac{2}{3}(  ( {\bm A}_{n'}^\dag {\bm A}_{m'}) ({\bm A}_{l'}^T {\bm A}_p) \Big] ({\bm A}_{l}^\dag {\bm A}_{m}^*)\\
&&+\sum_{l',m',n'} S_{l'm'n'p} \Big[ \frac{1}{3}  ({\bm A}_{l'}^T {\bm A}_{m'})
({\bm A}_{n}^T {\bm A}_{n'}^*)
+\frac{2}{3} ( {\bm A}_{n'}^\dag {\bm A}_{m'}) ({\bm A}_{n}^T {\bm A}_{l'}) \Big]({\bm A}_{l}^\dag {\bm A}_{m}^*) .
\label{express:tildeY1}
\end{eqnarray}
We take the expectation and we apply the Gaussian summation rule to the sixth-order moments in the expression of $\left< {Y}_{lmnp}^{(1)}\right>$:
\begin{eqnarray}
\nonumber
\left< {Y}_{lmnp}^{(1)}\right> &=& \frac{16}{3} S_{lmnp} \big( w_l w_m w_p+w_lw_mw_n-w_nw_pw_m-w_nw_pw_l\big)  
+ \frac{16}{3} \delta^K_{mp} s_{ln}({\bm w}) w_p (w_l-w_n) \\
&&+ \frac{16}{3} \delta^K_{mn} s_{lp}({\bm w}) w_n (w_l-w_p)
+ \frac{16}{3} \delta^K_{lp} s_{mn}({\bm w}) w_p (w_m-w_n)+ \frac{16}{3} \delta^K_{ln} s_{mp}({\bm w}) w_n (w_m-w_p) , 
\label{eq:Y_lmnp_1}\\
s_{ln}({\bm w})  &=& \sum_{n'}S_{ln'n'n} w_{n'}.
\label{eq:s_nl_1}
\end{eqnarray}

We now extend the procedure of Sec.~\ref{appA1} to the computation of fourth-order modes considered here. 
Making use of the Furutsu-Novikov theorem, and considering the general case $l\neq m\neq n\neq p$ in the regime $\sigma_\beta l_\beta \ll 1 $ and $l_\beta \ll L_{nl}$, we obtain
\begin{eqnarray}
\partial_z \left< ({\bm A}_l^\dag {\bm A}_m^*) ( {\bm A}_n^T {\bm A}_p) \right> &=&
\big( - 8  \Delta \beta + i  (\beta_l+\beta_m-\beta_n-\beta_p)  \big)
\left<({\bm A}_l^\dag {\bm A}_m^*) ({\bm A}_n^T {\bm A}_p) \right>   
+ i \gamma \left< {Y}_{lmnp}^{(1)} \right>,
\label{eq_app:J_lmnp_1}
\end{eqnarray}
whose solution has the form
\begin{eqnarray}
\nonumber
\left< ({\bm A}_l^\dag {\bm A}_m^*) ( {\bm A}_n^T {\bm A}_p) \right>(z) &=& 
\left< ({\bm A}_l^\dag {\bm A}_m^*) ( {\bm A}_n^T {\bm A}_p) \right>(0) \exp\big( 
\big( - 8  \Delta \beta + i  (\beta_l+\beta_m-\beta_n-\beta_p)  \big)z\big)\\
&&
+ i \gamma \int_0^z\left< {Y}_{lmnp}^{(1)}\right> (z') \exp\big( 
\big( - 8  \Delta \beta + i  (\beta_l+\beta_m-\beta_n-\beta_p)  \big)(z-z')\big) dz'  .
\end{eqnarray}
These equations correspond to those reported in Eq.(\ref{eq:4th_order_moment}) and Eq.(\ref{eq:mom4pq}) for the fourth-order moment $J_{lmnp}^{(1)}$.


In the other cases when (at least) two indices are equal, i.e., the fourth-order moments involve degenerate modes, the calculation shows that there is still a damping in the moment equation, although the damping factor can be different from $8 \Delta \beta$.
However we will neglect this change because: (i)  these terms are negligible in the triple sum and when the number of modes is large ($N_* \gg 1$), and (ii) this change only affects the multiplicative coefficient $8\Delta \beta$  in  (\ref{eq_app:J_lmnp_1}).
Finally we give the expression of $\left< X_p^{(1)}\right>$ in the discrete wave turbulence regime where $\beta_0 \gg \Delta \beta$. Collecting all terms we obtain
\begin{eqnarray}
\nonumber
\left< X_p^{(1)}\right> &=& \frac{2 \gamma}{3 \Delta \beta} \sum_{l,m,n}
 |S_{lmnp}|^2 \delta^K (\beta_l+\beta_m- \beta_n - \beta_p)
 \big( w_l w_m w_p+w_lw_mw_n-w_nw_pw_m-w_nw_pw_l\big)  \\
&&
+    \frac{4\gamma}{3 \Delta \beta}  \sum_l  |  s_{lp}({\bm w}) |^2 \delta^K (\beta_l- \beta_p) (w_l-w_p) .
\label{eq:momX_p1}
  \end{eqnarray}

\subsubsection{Computation of the term $\left< {X}_p^{(2)}\right> $:}

The equation satisfied by the product of four vector fields reads
\begin{eqnarray}
\nonumber
\partial_z ({\bm A}_n^T {\bm A}_m^*) ( {\bm A}_l^\dag {\bm A}_p)&=&
i(\beta_l+\beta_m-\beta_n-\beta_p) ({\bm A}_n^T {\bm A}_m^*) ({\bm A}_l^\dag {\bm A}_p)
+ i ({\bm A}_n^T  {\bm A}_m^*) ( {\bm A}_l^\dag {\bf D}_l^\dag {\bm A}_p)\\
\nonumber
&&
+ i ({\bm A}_n^T {\bf D}_m^* {\bm A}_m^*) ( {\bm A}_l^\dag  {\bm A}_p)
- i ({\bm A}_n^T {\bf D}_n^T {\bm A}_m^*) ( {\bm A}_l^\dag  {\bm A}_p)
- i ({\bm A}_n^T {\bm A}_m^*) ( {\bm A}_l^\dag {\bf D}_p {\bm A}_p)\\
&&
+ i \gamma {Y}_{lmnp}^{(2)} ,  
\label{eq:4f_Xp2}\\
\nonumber
{Y}_{lmnp}^{(2)}&=& -\sum_{l',m',n'} S_{l'm'n'l}^* 
\Big[ \frac{1}{3} ({\bm A}_{l'}^\dag {\bm A}_{m'}^*) ({\bm A}_{n'}^T {\bm A}_p)
+\frac{2}{3}({\bm A}_{n'}^T {\bm A}_{m'}^*) ({\bm A}_{l'}^\dag {\bm A}_p)\Big] 
 ({\bm A}_m^\dag{\bm A}_n) \\
\nonumber
&&-\sum_{l',m',n'} S_{l'm'n'm}^* \Big[ 
\frac{1}{3} ({\bm A}_{l'}^\dag {\bm A}_{m'}^*)
 ({\bm A}_{n'}^T {\bm A}_{n})
+\frac{2}{3}({\bm A}_{n'}^T {\bm A}_{m'}^*) 
( {\bm A}_{l'}^\dag {\bm A}_n)\Big]  
({\bm A}_l^\dag {\bm A}_p) \\
\nonumber
&&+\sum_{l',m',n'} S_{l'm'n'n} \Big[ \frac{1}{3} ({\bm A}_{l'}^T {\bm A}_{m'})
({\bm A}_{n'}^\dag {\bm A}_m^*)
+\frac{2}{3}  ( {\bm A}_{n'}^\dag {\bm A}_{m'}) ({\bm A}_{l'}^T {\bm A}_m^*) \Big]  ({\bm A}_l^\dag {\bm A}_p)\\
&&+\sum_{l',m',n'} S_{l'm'n'p} \Big[ \frac{1}{3} 
 ({\bm A}_{l'}^T {\bm A}_{m'}) ({\bm A}_{n'}^\dag {\bm A}_{l}^*)
+\frac{2}{3}  ( {\bm A}_{n'}^\dag {\bm A}_{m'}) ({\bm A}_{l'}^T {\bm A}_{l}^*) \Big] ({\bm A}_m^\dag {\bm A}_n).
\label{express:tildeY}
\end{eqnarray}
We take the expectation 
and we apply the Gaussian summation rule to the sixth-order moments in the expression of $\left< {Y}_{lmnp}^{(2)}\right>$:
\begin{eqnarray}
\nonumber
\left< {Y}_{lmnp}^{(2)}\right> &=& \frac{16}{3} S_{lmnp} \big( w_l w_m w_p+w_lw_mw_n-w_nw_pw_m-w_nw_pw_l\big) 
+ \frac{16}{3} \delta^K_{mp} s_{ln}({\bm w}) w_p (w_l-w_n) \\
\nonumber
&&+ \frac{32}{3} \delta^K_{mn} s_{lp}({\bm w}) w_n (w_l-w_p)
+ \frac{32}{3} \delta^K_{lp} s_{mn}({\bm w}) w_p (w_m-w_n)+ \frac{16}{3} \delta^K_{ln} s_{mp}({\bm w}) w_n (w_m-w_p) .
\label{eq:Y_lmnp_2}
\end{eqnarray}
If $l\neq m\neq n\neq p$, then Furutsu-Novikov formula gives
\begin{eqnarray}
\nonumber
\partial_z \left< ({\bm A}_n^T {\bm A}_m^*) ( {\bm A}_l^\dag {\bm A}_p)\right> &=&
\big( -8 \Delta \beta  + i(\beta_l+\beta_m-\beta_n-\beta_p)\big)
\left<({\bm A}_n^T {\bm A}_m^*) ( {\bm A}_l^\dag {\bm A}_p)\right>   
 + i \gamma \left< {Y}_{lmnp}^{(2)}\right>  .
\end{eqnarray}
This equation corresponds to that reported in Eq.(\ref{eq:4th_order_moment}) for the fourth-order moment $J_{lmnp}^{(2)}$.

Following the same procedure, we have also derived the following equation
\begin{eqnarray}
\partial_z \left< ({\bm A}_l^T {\bm A}_l^*) ( {\bm A}_m^\dag {\bm A}_m) \right> =0,
\label{eq:4_l_m}
\end{eqnarray}
for any $l,m$. 
This implies in particular that the variance of the intensity fluctuations of each modal component $p$ is preserved $\left< |{\bm A}_p|^4\right> =$const, i.e., Gaussian statistics is preserved during the propagation.\\
Note that Eq.(\ref{eq:4_l_m}) is undamped, but this does not affect our results. 
Indeed, 
$ \left< ({\bm A}_l^T {\bm A}_l^*) ( {\bm A}_m^\dag {\bm A}_m) \right>$ is real-valued
so that  it does not contribute when it is substituted  into Eq.(\ref{def:Xp2}) because $S_{lmml}$ is real-valued as well.

By neglecting the small corrections that appear in  the cases when (at least) two indices are equal (i.e. fourth-order modes involving degenerate modes), we obtain in the discrete wave turbulence regime ($\beta_0 \gg \Delta \beta$):
\begin{eqnarray}
\nonumber
\left< {X}_p^{(2)}\right> &=& \frac{2 \gamma}{3   {\Delta \beta} } \sum_{l,m,n}
\delta^K (\beta_l+\beta_m- \beta_n - \beta_p)  |S_{lmnp}|^2 
 \big( w_l w_m w_p+w_lw_mw_n-w_nw_pw_m-w_nw_pw_l\big) \\
&&
+   \frac{2\gamma}{  {\Delta \beta} } \sum_l 
\delta^K (\beta_l - \beta_p)  |  s_{lp}({\bm w}) |^2 (w_l-w_p)  .
\label{eq:momX_p2}
  \end{eqnarray}
By replacing the expressions of $\left< {X}_p^{(j)}\right>$ ($j=1,2$) given in (\ref{eq:momX_p1}) and (\ref{eq:momX_p2}) into the equation for the evolution of the modal components (\ref{eq:pzwp1}), we obtain the discrete kinetic Eq.(\ref{eq:kin_np_disc}).



\subsection{Impact of a correlated noise model of disorder on the kinetic equation}
\label{appA3}

\subsubsection{Computation of the moment $\left< ({\bm A}_l^\dag {\bm A}_m^*) ( {\bm A}_n^T {\bm A}_p) \right>$}

In the model of correlated disorder, all modes experience the same noise, ${\bf D}_n\equiv {\bf D} = \sum_{j=0}^3 \nu_j \bsigma_j$. 
The equation (\ref{eq:4f_Xp1}) for the evolution of the product of four fields now reads
\begin{eqnarray}
\nonumber
\partial_z ({\bm A}_l^\dag {\bm A}_m^*) ( {\bm A}_n^T {\bm A}_p)&=&
i(\beta_l+\beta_m-\beta_n-\beta_p) ({\bm A}_l^\dag {\bm A}_m^*) ({\bm A}_n^T {\bm A}_p)
+ i ({\bm A}_l^\dag {\bf D}^\dag {\bm A}_m^*) ( {\bm A}_n^T {\bm A}_p)\\
\nonumber
&&
+ i ({\bm A}_l^\dag {\bf D}^* {\bm A}_m^*) ( {\bm A}_n^T {\bm A}_p)
- i ({\bm A}_l^\dag {\bm A}_m^*) ( {\bm A}_n^T {\bf D}^T {\bm A}_p)
- i ({\bm A}_l^\dag {\bm A}_m^*) ( {\bm A}_n^T {\bf D} {\bm A}_p)\\
&&
+ i \gamma {Y}_{lmnp}^{(1)} .
\end{eqnarray}
We note that ${\bm A}_l^\dag {\bf D}^\dag {\bm A}_m^* + {\bm A}_l^\dag {\bf D}^* {\bm A}_m^* = 2\sum_{j\in \{1,3\}}
\nu_j {\bm A}_l^\dag \bsigma_j {\bm A}_m^*$.
We follow the procedure outlined above in sections \ref{appA1}-\ref{appA2}.
Using the Furutsu-Novikov theorem and assuming $\sigma_\beta l_\beta \ll 1 $ and $l_\beta \ll L_{nl}$ we obtain
$$
\frac{\delta {\bm A}_l^\dag \bsigma_j {\bm A}_m^* {\bm A}_n^T {\bm A}_p(z)}{\delta \nu_j(z')}
=2i
\Big[
{\bm A}_l^\dag {\bm A}_m^* {\bm A}_n^T  {\bm A}_p(z') 
-
{\bm A}_l^\dag \bsigma_j {\bm A}_m^* {\bm A}_n^T\bsigma_j   {\bm A}_p(z') \big]
\exp\big[ i (\beta_l+\beta_m-\beta_n-\beta_p)(z-z')\big]
$$
and therefore
\begin{eqnarray}
\nonumber
\partial_z \left< ({\bm A}_l^\dag {\bm A}_m^*) ( {\bm A}_n^T {\bm A}_p) \right> &=&
\big( - 4  \Delta \beta + i  (\beta_l+\beta_m-\beta_n-\beta_p)  \big)
\left<({\bm A}_l^\dag {\bm A}_m^*) ({\bm A}_n^T {\bm A}_p) \right>   
+ i \gamma \left< {Y}_{lmnp}^{(1)} \right>  \\
&& +2 \Delta \beta \sum_{j\in \{1,3\}}  \left< ({\bm A}_l^\dag \bsigma_j {\bm A}_m^*) ( {\bm A}_n^T  \bsigma_j{\bm A}_p) \right>.
\label{eq:4m_corr_disor}
\end{eqnarray}
The last term can also be written as:
$$
 \sum_{j\in \{1,3\}}  \left< ({\bm A}_l^\dag \bsigma_j {\bm A}_m^*) ( {\bm A}_n^T  \bsigma_j{\bm A}_p) \right>
 =
  \left< ({\bm A}_l^\dag {\bm A}_n) ( {\bm A}_m^\dag {\bm A}_p) \right>
+
  \left< ({\bm A}_l^\dag \bsigma_2 {\bm A}_n) ( {\bm A}_m^\dag \bsigma_2 {\bm A}_p) \right>.
$$
The analysis reveals that the terms involving the dissipation that are proportional to $\Delta \beta$ in (\ref{eq:4m_corr_disor}) essentially vanish.
Indeed, using the factorizability property of statistical Gaussian fields to split the fourth-order moments into products of second-order moments, then we get
$\left<({\bm A}_l^\dag {\bm A}_m^*) ({\bm A}_n^T {\bm A}_p) \right>  = \frac{1}{2} w_l w_m ( \delta^K_{ln}\delta^K_{mp}+\delta^K_{lp}\delta^K_{mn})$
and
$\sum_{j\in \{1,3\}}  \left< ({\bm A}_l^\dag \bsigma_j {\bm A}_m^*) ( {\bm A}_n^T  \bsigma_j{\bm A}_p) \right>
= w_l w_m ( \delta^K_{ln}\delta^K_{mp}+\delta^K_{lp}\delta^K_{mn})$.
This shows that, to leading order,  the terms involving the dissipation indeed
compensate with each other.

\subsubsection{Computation of the moment $\left< ({\bm A}_n^T {\bm A}_m^*) ( {\bm A}_l^\dag {\bm A}_p)\right>$}

The equation (\ref{eq:4f_Xp2}) for the evolution of the product of four fields now reads
\begin{eqnarray}
\nonumber
\partial_z ({\bm A}_n^T {\bm A}_m^*) ( {\bm A}_l^\dag {\bm A}_p)&=&
i(\beta_l+\beta_m-\beta_n-\beta_p) ({\bm A}_n^T {\bm A}_m^*) ({\bm A}_l^\dag {\bm A}_p)
+ i ({\bm A}_n^T  {\bm A}_m^*) ( {\bm A}_l^\dag {\bf D}^\dag {\bm A}_p)\\
\nonumber
&&
+ i ({\bm A}_n^T {\bf D}^* {\bm A}_m^*) ( {\bm A}_l^\dag  {\bm A}_p)
- i ({\bm A}_n^T {\bf D}^T {\bm A}_m^*) ( {\bm A}_l^\dag  {\bm A}_p)
- i ({\bm A}_n^T {\bm A}_m^*) ( {\bm A}_l^\dag {\bf D} {\bm A}_p)\\
&&
+ i \gamma {Y}_{lmnp}^{(2)} .
\end{eqnarray}
Since the random matrix is Hermitian (${\bf D}^\dag={\bf D}$), the expression can be reduced
\begin{eqnarray}
\nonumber
\partial_z ({\bm A}_n^T {\bm A}_m^*) ( {\bm A}_l^\dag {\bm A}_p)&=&
i(\beta_l+\beta_m-\beta_n-\beta_p) ({\bm A}_n^T {\bm A}_m^*) ({\bm A}_l^\dag {\bm A}_p)
+ i \gamma {Y}_{lmnp}^{(2)} .  
\end{eqnarray}
The fourth-order moment $\left< ({\bm A}_n^T {\bm A}_m^*) ( {\bm A}_l^\dag {\bm A}_p)\right>$ then evolves as in the absence of any disorder ($\Delta \beta=0$).

\subsection{Impact of a partially correlated noise model of disorder on the kinetic equation}
\label{appA4}

In the partially correlated model of disorder, all degenerate modes experience the same realization of disorder.
Groups of degenerate modes with the same reduced eigenvalue $\beta$ are indexed by $p$, the modes within the $p$th group are indexed by $(p,j)$, and the  linear polarization components of the $(p,j)$-th mode are ${\bm A}_{pj}$.
The structural disorder induces a linear random coupling between the modes of different groups, as described by the $2\times 2$ matrix ${\bf D}_p(z)= \sum_{l=0}^3 \nu_{p,l}(z) \bsigma_l$ where $p$ labels the mode group number, i.e., degenerate modes that belong to the same group experience the same noise through the random process $\nu_{p,l}(z)$ (this notation should not be confused with the decorrelated model of disorder where $p$ labels individual modes).
The evolutions of the modal components are governed by 
$$
i \partial_z {\bm A}_{pj} = \beta_p {\bm A}_{pj}
+ {\bf D}_p (z) {\bm A}_{pj} 
- \gamma  {\bm P}_{pj}({\bm A}).
$$
We look at the second-order moments $\left< {\bm A}_{pj}^* {\bm A}_{ql}^T\right>$.
We follow the procedure outlined above in section~\ref{appA1} by using the Furutsu-Novikov theorem with $\sigma_\beta l_\beta \ll 1 $ and $l_\beta \ll L_{nl}$.

Considering non-degenerate modes ($p\neq q$), we obtain 
$$
\left< {\bm A}_{pj}^* {\bm A}_{ql}^T\right>(z) =
\frac{i \gamma}{4\Delta \beta-i(\beta_p-\beta_q)}
 \big< {\bm P}_{pj}({\bm A})^* {\bm A}_{ql}^T
- {\bm A}_{pj}^* {\bm P}_{ql}({\bm A})^T \big>  .
$$
In the regime $L_d = 1/\Delta \beta \ll L_{nl}$ the correlation is vanishing small, $\left< {\bm A}_{pj}^* {\bm A}_{ql}^T\right>(z) \simeq 0$, as in the model of decorrelated disorder.\\
Let us now consider correlations among the orthogonal polarization components of a mode ($p = q$ and $j=l$), then we find as before
$$
\left< {\bm A}_{pj}^* {\bm A}_{pj}^T\right> (z)=  w_{p,jj}(z) \bsigma_0 .
$$

We consider correlations among distinct degenerate modes ($p = q$ and $j\neq l$).
We define the two Hermitian matrices 
${\bf W}_{p,jl} = \frac{1}{2} \big( \left< {\bm A}_{pj}^* {\bm A}_{pl}^T\right>
+\left< {\bm A}_{pl}^* {\bm A}_{pj}^T\right>\big)$ and $\tilde{\bf W}_{p,jl} = \frac{i}{2} \big( \left< {\bm A}_{pj}^* {\bm A}_{pl}^T\right>
-\left< {\bm A}_{pl}^* {\bm A}_{pj}^T\right>\big)$
and then by carrying out the same calculations as in the case $p = q$ and $j=l$,
we find 
$$
\left< {\bm A}_{pj}^* {\bm A}_{pl}^T\right>(z) =  w_{p,jl}(z) \bsigma_0 .
$$
The coefficients $w_{p,jl}$ satisfy
\begin{eqnarray*}
\partial_z w_{p,jl} &=&  \frac{\gamma}{2} {\rm Im} \big<   {\bm P}_{pj}({\bm A})^\dag {\bm A}_{pl} + {\bm P}_{pl}({\bm A})^\dag {\bm A}_{pj} \big> =
\frac{1}{6}  \gamma \left< X_{p,jl}^{(1)}+ X_{p,lj}^{(1)}\right> 
+ \frac{1}{3}  \gamma \left<  {X}_{p,jl}^{(2)} +{X}_{p,lj}^{(2)} \right> ,\\
X_{p,jl}^{(1)} &=&{\rm Im}\Big\{ \sum_{p_1j_1,p_2j_2,p_3j_3} S_{pl,p_1j_1,p_2j_2,p_3j_3}^* ({\bm A}_{p_1j_1}^\dag {\bm A}_{p_2j_2}^*) ({\bm A}_{p_3j_3}^T {\bm A}_{pj}) \Big\}  ,\\
{X}_{p,jl}^{(2)} &=&{\rm Im}\Big\{ \sum_{p_1j_1,p_2j_2,p_3j_3} S_{pl,p_1j_1,p_2j_2,p_3j_3}^* ({\bm A}_{p_3j_3}^T {\bm A}_{p_2j_2}^*) ({\bm A}_{p_1j_1}^\dag {\bm A}_{pj}) \Big\}  .
\end{eqnarray*}
Let us look at the fourth-order moments $\left< ({\bm A}_{p_1j_1}^\dag {\bm A}_{p_2j_2}^*) ( {\bm A}_{p_3j_3}^T{\bm A}_{p_4j_4})\right>$ or 
$\left< ({\bm A}_{p_1j_1}^T {\bm A}_{p_2j_2}^*) ( {\bm A}_{p_3j_3}^\dag {\bm A}_{p_4j_4})\right>$. \\
There are different types of such fourth-order moments that depend on the specific modes that they involve. Almost all of them satisfy an evolution equation with damping proportional to $\Delta \beta$, 
with different coefficients in front of $\Delta \beta$ that depend on the number of equal indices.
These terms are of the same form as those obtained by considering the decorrelated model of disorder, see section~\ref{appA2}.
There are special cases when $p_j$ are equal by pairs (e.g. $p_1=p_2$ and $p_3=p_4$) where 
$$
\partial_z \left< ({\bm A}_{p_1j_1}^T {\bm A}_{p_1j_2}^*) ( {\bm A}_{p_3j_3}^\dag {\bm A}_{p_3j_4})\right>=
i \gamma {Y}_{p_3j_3,p_1j_2,p_1j_1,p_3j_4}^{(2)},
$$
which shows that there is no damping. From the expression of ${Y}^{(2)}$ (see Eq.(\ref{express:tildeY})), and the evaluation of its expectation in terms of the $w_{p,jl}$ according to the Gaussian rule for sixth-order moments, such terms induce a reversible exchange of energy between the modes within each group.


\section{Appendix B: Kinetic equations without disorder}
\label{app:no_disorder}

\subsection{Continuous and discrete wave turbulence regimes}

The continuous kinetic equation describing weak turbulence in the presence of a parabolic potential was derived in Ref.\cite{PRA11b}, see Eq.(17-18).
The presence of the factor $1/\beta_0^6$ in the kinetic equation is due to the introduction of a continuous frequency variable $\kappa=\beta_0(p_x,p_y)$, as discussed through the continuous kinetic Eq.(\ref{eq:kin_contin_dis}) above.
The characteristic scale of evolution of the modal components then reads $L_{kin}^{ord} \sim \beta_0 L_{nl}^2/{\bar S_{lmnp}^2}$, as given in Eq.(\ref{eq:accel_thermal}).

Let us now consider the discrete wave turbulence regime without disorder.
By ignoring polarization effects (${\bm A}_p \to A_{p,1}$), the set of closed equations for the second- and fourth-order moments read \cite{PRA11b}:
\begin{eqnarray}
\partial_z n_p &=& - \gamma i \sum_{l,m,n} S_{lmnp}  \tilde{J}_{lmnp} \exp( i\Delta {\omega}_{lmnp} z) + \gamma i \sum_{l,m,n} S_{lmnp}^*  \tilde{J}_{lmnp}^* \exp(- i\Delta {\omega}_{lmnp} z) 
\label{eq:no_dis_1}\\
\partial_z \tilde{J}_{lmnp} &=& 2 i \gamma S_{lmnp}^* M_{lmnp}({\bm n})  \exp(- i\Delta {\omega}_{lmnp} z) 
\label{eq:no_dis_2} 
\end{eqnarray}
where 
$M_{lmnp}({\bm n})=n_l n_m n_p+n_l n_m n_n -  n_n n_p n_m -n_n n_p n_l$, 
${J}_{lmnp} (z)=\left< A_l^* A_m^* A_n A_p\right>$ is the fourth order moment with $\tilde{J}_{lmnp}=J_{lmnp} \exp(- i\Delta \omega_{lmnp} z)$ and $\Delta {\omega}_{lmnp}=\beta_l+\beta_m-\beta_n-\beta_p$ ($\beta_p=\beta_0(1+p_x+p_y)$).
Note that for the sake of simplicity and clarity, we do not write in Eq.(\ref{eq:no_dis_1}) the contribution that leads to the second term in the rhs of the kinetic equations (\ref{eq:kin_contin_dis2}) or (\ref{eq:kin_np_disc}), whose main effect is to enforce energy equipartition within groups of degenerate modes.

Recalling that $\beta_0 L_{nl} \gg 1$, we make use of the standard homogenization theorem to obtain the effective equations  for $z \sim L_{nl}$~:
\begin{eqnarray}
\partial_z n_p &=&  i\gamma  \sum_{l,m,n} \delta^K(\Delta \omega_{lmnp})  \big(-S_{lmnp}  \tilde{J}_{lmnp}   + S_{lmnp}^*  \tilde{J}_{lmnp}^* \big)  \\
\partial_z \tilde{J}_{lmnp} &=& 2 i \gamma S_{lmnp}^* M_{lmnp}({\bm n})  \delta^K(\Delta \omega_{lmnp})  
\end{eqnarray}
where we recall that $\delta^K(\Delta \omega_{lmnp})=1$ if $\Delta \omega_{lmnp}=0$, and zero otherwise.
Accordingly, one obtains an equation for the modal components that is formally reversible in $z$.
This equation is valid for propagation lengths of the order of $L_{nl}$. 
For longer propagation lengths, corrective terms must be taken into account \cite{kurzweil87}, showing that the characteristic length of evolution of the modal components $n_p(z)$ scales as in the continuous case $L_{kin}^{ord} \sim \beta_0 L_{nl}^2/{\bar S_{lmnp}^2}$ (up to corrective prefactors).


\subsection{Impact of a perturbation of the dispersion relation}

In this paragraph we consider the impact of a perturbation of the dispersion relation ${\tilde \beta}_p= \beta_p + b_p$ (with $b_p/\beta_p \ll 1$) on the kinetic equation without disorder.
This issue is addressed in section~\ref{sec:b_p} in the presence of disorder.
The equations for the second- and fourth-order moments (\ref{eq:no_dis_1}-\ref{eq:no_dis_2}) take the form
\begin{eqnarray}
\partial_z n_p &=& 2 \gamma \sum_{l,m,n} {\rm Im}[S_{lmnp}  {J}_{lmnp}(z)]
\label{eq:n_m_disc1}\\
{J}_{lmnp} (z) &=& {J}_{lmnp} (0) \exp( i\Delta {\tilde \omega}_{lmnp} z) +2 i \gamma \int_0^z S_{lmnp}^* M_{lmnp}({\bm n}(\zeta))  \exp( i\Delta {\tilde \omega}_{lmnp} (z-\zeta)) d\zeta   
\label{eq:n_m_disc2}
\end{eqnarray}
where $\Delta {\tilde \omega_{lmnp}} = {\tilde \beta}_l+{\tilde \beta}_m-{\tilde \beta}_n-{\tilde \beta}_p$. In the following we neglect the initial condition $J_{lmnp}(0)$. 
This can be justified when the initial condition exhibits random phases among the modes, but not when the initial condition is a coherent beam as in usual experiments of beam self-cleaning.
Resonances that are exact at leading order ($\Delta \omega_{lmnp} =\beta_l+\beta_m-\beta_n-\beta_p=0$) exhibit a residual non-resonant contribution, i.e., $\Delta {\tilde \omega_{lmnp}} = {\tilde \beta}_l+{\tilde \beta}_m-{\tilde \beta}_n-{\tilde \beta}_p=\Delta b_{lmnp}$ with  $\Delta b_{lmnp}=b_l+b_m-b_n-b_p$.
Since $\beta_0 L_{nl} \gg 1$ only leading order exact resonances can contribute:
\begin{eqnarray}
\partial_z n_p = 4 \gamma^2 \sum_{l,m,n} |S_{lmnp}|^2  \delta^K(  \Delta {\omega}_{lmnp} )  
\int_0^z \cos (\Delta b_{lmnp} (z-\zeta)) M_{lmnp}({\bm n}(\zeta))   d\zeta .
\label{eq:n_m_disc3}
\end{eqnarray}
If the resonances are approximate, then the quartets $\{l,m,n,p\}$ such that $|\Delta b_{lmnp}| L^{pert}_{kin} \sim 1$ contribute to the convolution in (\ref{eq:n_m_disc3}), where $L^{pert}_{kin} $  is the characteristic evolution length of the moments in the presence of the perturbation $b_p$. 
Those quartets for which $|\Delta b_{lmnp}| L^{pert}_{kin} \gg 1$ average out to zero. 
Those quartets for which $|\Delta b_{lmnp}| L^{pert}_{kin} \ll 1$ should give rise to reversible equations but we neglect them because we assume here that there are many more quasi-resonances than resonances. 
Since quasi-resonances are more important than exact resonances we can take the continuous limit with $\sin(\Delta b_{lmnp}z)/\Delta b_{lmnp} \to \pi \delta(\Delta b_{lmnp})$ for $z \gg \pi/{\rm min}(\Delta b_{lmnp})$, where $\delta(x)$ denotes the Dirac distribution.
Note that the passage to the continuous limit can be justified for a perturbation that removes mode degeneracies, such as the truncation of the parabolic potential (see Fig.~5 in \cite{PRA11b}), but not for the perturbation induced by the Helmholtz equation that does not remove degeneracies, see section~\ref{sec:b_p}.
According to the continuous kinetic equation, the characteristic evolution length of the kinetics scales as $L^{pert}_{kin} \sim b_p L_{nl}^2/{\bar S_{lmnp}^2}$.



\section{Appendix C: Derivation of the discrete kinetic equation with strong disorder}
\label{app:strong_disorder}

\subsection{First order moments}

The model of strong disorder is defined from the $2N_* \times 2N_*$ random matrix ${\bf D}$ in Eq.(\ref{eq:D_sd}). 
The matrices ${\bf H}^{qr} $, $ {\bf K}^{qr} $, $ {\bf J}^{q} $ are defined by:
$$
( {\bf H}^{qr} )_{jl} = \frac{1}{\sqrt{2}}
 \left\{
\begin{array}{ll}
1 & \mbox{ if } (j,l)=(q,r)\\ 
1 & \mbox{ if } (j,l)=(r,q) \\
0 & \mbox{otherwise}
\end{array}
\right. 
\quad
( {\bf K}^{qr} )_{jl} = \frac{1}{\sqrt{2}}
 \left\{
\begin{array}{ll}
i & \mbox{ if } (j,l)=(q,r)\\ 
-i & \mbox{ if } (j,l)=(r,q) \\
0 & \mbox{otherwise}
\end{array}
\right. 
\quad
( {\bf J}^{q} )_{jl} =  
 \left\{
\begin{array}{ll}
1 & \mbox{ if } (j,l)=(q,q)\\
0 & \mbox{otherwise}
\end{array}
\right.
$$
We have the identity
\begin{equation}
\sum_{q<r} \big( ( {\bf H}^{qr} )^2+( {\bf K}^{qr} )^2\big)+\sum_q ( {\bf J}^{q} )^2 = 2N_* {\bf I}.
\label{eq:invH}
\end{equation}

In these conditions $i (\bbeta+{\bf D}(z))$ generates a shifted unitary Brownian motion ${\bf U}(z)$, 
$$
\partial_z {\bf U} = i (\bbeta+{\bf D}(z)) {\bf U}, \quad \quad {\bf U}(z=0)={\bf I},
$$
which is a random diffusion on the group of unitary matrices
whose stationary distribution is the uniform (Haar) measure.
We have $\| {\bf U}(z) {\bm A}_0 \|=\| {\bm A}_0\|$ for any constant vector ${\bm A}_0$.

The evolution of the average of the unitary matrix ${\bf U}$ is governed by
\begin{eqnarray*}
\frac{d \EE[{\bf U}(z)]}{dz}  = i\bbeta \EE[ {\bf U}(z)]+
  \frac{i}{\sqrt{2N_*}} \Big\{ \sum_{q<r}
g_{qr} \big( {\bf H}^{qr}  \EE[ {\nu}_{qr}(z) {\bf U}(z)]
+
 {\bf K}^{qr}  \EE[ {\mu}_{qr}(z) {\bf U}(z)] \big)
+
\sum_q g_q  {\bf J}^q \EE[ {\eta}_{q}(z) {\bf U}(z)] \Big\}.
\end{eqnarray*}
We follow the same procedure as in Appendix~A to evaluate the averages by using the Furutsu-Novikov theorem:
\begin{eqnarray}
\nonumber
 \EE \left[  \nu_{qr} (z){\bf U}(z) \right] &=& \int_0^z \EE \left[ \frac{\delta  {\bf U}(z)  }
 {\delta \nu_{qr}(z') } \right]   \sigma^2_\beta  {\cal R}\Big( \frac{z-z'}{l_\beta}\Big) dz' .
 \end{eqnarray}
The variational derivative satisfies for $z>z'$:
 $$
 \partial_z  \frac{\delta  {\bf U}(z)  }
 {\delta \nu_{qr}(z') } = i \bbeta  \frac{\delta  {\bf U}(z)  }
 {\delta \nu_{qr}(z') } + {\bf D}(z)  \frac{\delta  {\bf U}(z)  }
 {\delta \nu_{qr}(z') }  ,
 $$
 with
 $$
 \frac{\delta  {\bf U}(z)  }
 {\delta \nu_{qr}(z') } \mid_{z=z'} = \frac{i}{\sqrt{2N_*}} g_{qr} {\bf H}^{qr} {\bf U}(z')  .
 $$
Therefore, if $\sigma_\beta l_\beta \ll 1 $, we get the simplified equation
 \begin{eqnarray*}
  \frac{\delta  {\bf U}(z)  }
 {\delta \nu_{qr}(z') } &=& \frac{i}{\sqrt{2N_*}} g_{qr}
 \exp\big( i\bbeta (z-z') \big)
  {\bf H}^{qr} {\bf U}(z'),\\
 \Big(
 \frac{i}{\sqrt{2N_*}}
 {\bf H}^{qr}   \frac{\delta  {\bf U}(z)  }
 {\delta \nu_{qr}(z') } \Big)_{jl}&=& - \frac{1}{4N_*} g_{qr}^2
 \Big( \delta^K_{qj} \exp\big( i\beta_r (z-z') \big) + \delta^K_{rj} \exp\big( i\beta_q (z-z') \big)\Big)  {U}_{jl} (z') \\
 &=& - \frac{1}{4N_*} g_{qr}^2
 \Big( \delta^K_{qj} \exp\big( i (\beta_r-\beta_q) (z-z') \big) +\delta^K_{rj}
 \exp\big( i (\beta_q-\beta_r) (z-z') \big)\Big)  {U}_{jl} (z) ,
\end{eqnarray*}
 and
 \begin{eqnarray}
 \nonumber
 \Big(
 \frac{i}{\sqrt{2N_*}}
 {\bf H}^{qr}  \EE \left[  \nu_{qr} (z){\bf U} (z) \right] \Big)_{jl}
 & =&   - \frac{1}{2N} g_{qr}^2
 \int_0^z  \sigma^2_\beta  {\cal R}\Big( \frac{z-z'}{l_\beta}\Big)
   \Big( \delta^K_{qj} \exp\big( i(\beta_r-\beta_q) (z-z') \big)  \\
 && \quad  + \delta^K_{rj} \exp\big( i(\beta_q-\beta_r) (z-z') \big)\Big)  
 dz' U_{jl}(z) .
\nonumber
\end{eqnarray}
We get similarly
 \begin{eqnarray}
 \nonumber
 \Big(
 \frac{i}{\sqrt{2N_*}}
 {\bf K}^{qr}  \EE \left[  \mu_{qr} (z){\bf U} (z) \right] \Big)_{jl}
 & =&   - \frac{1}{4N_*} g_{qr}^2
 \int_0^z  \sigma^2_\beta  {\cal R}\Big( \frac{z-z'}{l_\beta}\Big)
   \Big(\delta^K_{qj} \exp\big( i(\beta_r-\beta_q) (z-z') \big)  \\
 && \quad  +\delta^K_{rj} \exp\big( i(\beta_q-\beta_r) (z-z') \big)\Big)  
 dz' U_{jl}(z) ,
\nonumber
\\
 \nonumber
 \Big(
 \frac{i}{\sqrt{2N_*}}
 {\bf J}^{q}  \EE \left[  \eta_{q} (z){\bf U} (z) \right] \Big)_{jl}
 & =&   - \frac{1}{2N_*} g_{q}^2
 \int_0^z  \sigma^2_\beta  {\cal R}\Big( \frac{z-z'}{l_\beta}\Big)
 dz' {\bf 1}_q(j) U_{jl}(z) .
\nonumber
\end{eqnarray}

Therefore, using $z \gg l_\beta$,
\begin{eqnarray}
\frac{d \EE[{\bf U}(z)]}{dz}  &=&
i\bbeta  \EE[{\bf U}(z)]
 -  {\bf Q}   \EE[{\bf U}(z)]     ,
\label{eq:sd_avU}
\end{eqnarray}
with ${\bf Q}$ the diagonal matrix with diagonal coefficients 
\begin{eqnarray}
Q_{mm} = \frac{\Delta \beta}{2N_*}
\int_0^\infty  \sigma^2_\beta  {\cal R}(s)
\Big( g_m^2 
+ \sum_{p> m} g_{mp}^2 \exp\big( i(\beta_p-\beta_m) l_\beta s \big) +\sum_{p<m} g_{pm}^2 
\exp\big( i(\beta_p-\beta_m)  l_\beta s\big)  ds
\Big) .
\label{eq:sd_avU_Q}
\end{eqnarray}
The real part of $Q_{mm}$ is positive (because it is the sum of Fourier coefficients of ${\cal R}$,
which are non-negative by Bochner's theorem), which means that the mean wave mode amplitudes decay exponentially with the decay rate $1/Q_{mm}$.

\subsection{Second order moments}

We can proceed in the same way for the second-order moment.
Let us denote $a_p(z)=({\bf U}(z) {\bm A}_0)_p$.
We have
$$
\frac{da_n}{dz} =  i\beta_n a_n+
\frac{i}{\sqrt{4N_*}} \Big\{ \sum_{q<n}
g_{qn}  {\nu}_{qn} a_q
+
\sum_{q>n}
g_{nq}  {\nu}_{nq} a_q
- i 
 \sum_{q<n}
g_{qn}  {\mu}_{qn} a_q
+i
\sum_{q>n}
g_{nq}  {\mu}_{nq} a_q
+\sqrt{2} g_n \eta_n a_n   \Big\} ,
$$
and therefore
\begin{eqnarray*}
\frac{d|a_n|^2}{dz} &=& 
\frac{i}{\sqrt{2N}} \Big\{ \sum_{q<n}
g_{qn}  {\nu}_{qn} (a_q {a_n}^*-a_n {a_q}^*)
+
\sum_{q>n}
g_{nq}  {\nu}_{nq} (a_q {a_n}^*-a_n {a_q}^*)
\\
&&- i 
 \sum_{q<n}
g_{qn}  {\mu}_{qn} (a_q {a_n}^*+a_n {a_q}^*)
+i
\sum_{q>n}
g_{nq}  {\mu}_{nq} (a_q {a_n}^*+a_n {a_q}^*)   \Big\} .
\end{eqnarray*}
By Furutsu-Novikov formula, we have for $q<n$
\begin{eqnarray}
\nonumber
\EE \left[ \nu_{qn} (z)a_n {a_q}^* (z) \right] &=& \int_0^z \EE \left[ \frac{\delta  a_n {a_q}^*(z)  }
 {\delta \nu_{qn}(z') } \right]   \sigma^2_\beta  {\cal R}\Big( \frac{z-z'}{l_\beta}\Big) dz' .
\end{eqnarray}
 The variational derivative $\frac{\delta  a_n {a_q}^*(z)  }
 {\delta \nu_{qn}(z') }$, for $q<n$, satisfies for $z>z'$:
 $$
 \partial_z   \frac{\delta  a_n {a_q}^*(z)  }
 {\delta \nu_{qn}(z') }  = i (\beta_n-\beta_q)
  \frac{\delta  a_n {a_q}^*(z)  }
 {\delta \nu_{qn}(z') }  + \cdots  ,
 $$
 with
 $$
 \frac{\delta  a_n {a_q}^*(z)  }
 {\delta \nu_{qn}(z') }  \mid_{z=z'} = \frac{i}{\sqrt{4N_*}} g_{qn} 
 \big(|a_q|^2-|a_n|^2 \big)(z')  .
 $$
 Therefore, if $\sigma_\beta l_\beta \ll 1 $, then we get
 $$
 \frac{\delta  a_n {a_q}^*(z)  }
 {\delta \nu_{qn}(z') } 
  = \frac{i}{\sqrt{4N_*}} g_{qn} 
 \big(|a_q|^2-|a_n|^2 \big)(z')
 \exp\big( i(\beta_n-\beta_q) (z-z') \big)  ,
 $$ 
and we have similar expressions for the other terms.
Consequently, $w_m=\EE[|a_m|^2]$ satisfy Eq.(\ref{eq:sd_w_p}):
\begin{eqnarray}
\frac{d w_m}{dz} &=& 2 \sigma_\beta^2 \sum_{p=0}^{2N_*-1}  g_{mp}^2 \int_0^z \big( {w_p} - {w_m}\big)(z') {\cal R}\Big(\frac{z-z'}{l_\beta}\Big) 
\cos\big( (\beta_p-\beta_m) (z-z')\big) dz'.
\label{eq:w_m_app}
\end{eqnarray}



\end{widetext}



\end{document}